\documentclass[12pt]{article}
\usepackage{amsmath,amssymb,amsfonts}
\usepackage{color,graphicx,cite,soul}
\usepackage{array,booktabs,longtable}
\usepackage{caption,microtype,url}

\newcommand{\lsim}
{\;\raisebox{-.3em}{$\stackrel{\displaystyle <}{\sim}$}\;}
\newcommand{\gsim}
{\;\raisebox{-.3em}{$\stackrel{\displaystyle >}{\sim}$}\;}

\newcommand\Code[1]{\ensuremath{\texttt{#1}}}
\newcommand\Var[1]{\ensuremath{\mathit{#1}}}
\newcommand\Vi{\Var{i}}
\newcommand\Vj{\Var{j}}

\newcommand\al{\alpha}

\newcommand\tb{\tan\beta}
\newcommand\TB{t_\beta}

\newcommand\CBA{c_{\beta - \alpha}}
\newcommand\SBA{s_{\beta - \alpha}}

\newcommand\LP{\left(}
\newcommand\RP{\right)}
\newcommand\LB{\left[}
\newcommand\RB{\right]}
\newcommand\LV{\left\{}
\newcommand\RV{\right\}}

\newcommand\ReDiag{\mathop{%
  \raise .5pt\hbox{[}%
  \widetilde{\mathrm{Re}}%
  \raise .5pt\hbox{]}}}
\newcommand\ReOffDiag{\mathop{%
  \raise .5pt\hbox{$\llbracket$}%
  \widetilde{\mathrm{Re}}%
  \raise .5pt\hbox{$\rrbracket$}}}

\newcommand\DRbar{\ensuremath{\smash{\overline{\mathrm{DR}}}}}
\newcommand\MSbar{\ensuremath{\overline{\mathrm{MS}}}}
\newcommand\matr[1]{\mathbf{#1}}

\newcommand\cMl{{\cal M}_{\text{1-loop}}}
\newcommand\cL{{\cal L}}

\newcommand\SW{s_\mathrm{w}}
\newcommand\CW{c_\mathrm{w}}
\newcommand\MW{M_W}
\newcommand\MZ{M_Z}
\newcommand\Mh{M_h}

\newcommand\MHp{M_{H^\pm}}

\newcommand\mt{m_t}

\newcommand\At{A_t}

\newcommand\Sf{{\tilde f}}

\renewcommand\stop{\tilde t}

\newcommand\sbot{\tilde b}

\newcommand\dTB{\delta\TB}
\newcommand\ino[1]{\tilde\chi_{#1}}

\newcommand\chapm[1]{\ino{#1}^\pm}

\newcommand\cha{\chapm}
\newcommand\mcha[1]{m_{\chapm{#1}}}

\newcommand\neu[1]{\ino{#1}^0}
\newcommand\mneu[1]{m_{\neu{#1}}}

\newcommand\refeq[1]{Eq.~(\ref{#1})}

\newcommand\refta[1]{Tab.~\ref{#1}}
\newcommand\refse[1]{Sect.~\ref{#1}}

\newcommand\citere[1]{Ref.~\cite{#1}}
\newcommand\citeres[1]{Refs.~\cite{#1}}

\newcommand\uscore{\symbol{95}}
\newcommand\eg{e.g.\ }
\newcommand\ie{i.e.\ }

\newcommand{\CP}{{\cal CP}}

\newcommand{\onel}{one-loop}
\newcommand{\tev}{\,\, \mathrm{TeV}}
\newcommand{\gev}{\,\, \mathrm{GeV}}
\newcommand{\mev}{\,\, \mathrm{MeV}}

\newcommand{\eehh}{e^+e^- \to h_i h_j}
\newcommand{\eehZ}{e^+e^- \to h_i Z}
\newcommand{\eehga}{e^+e^- \to h_i \ga}

\newcommand\FA{\texttt{FeynArts}}
\newcommand\FC{\texttt{FormCalc}}
\newcommand\LT{\texttt{LoopTools}}
\newcommand\FH{\texttt{FeynHiggs}}
\newcommand\FT{\texttt{FeynTools}}
\newcommand\FHXS{\texttt{FeynHiggsXS}}

\newcommand\HDECAY{\texttt{HDECAY}}

\newcommand\iab{\ensuremath{\mbox{ab}^{-1}}}

\newcommand\mh[1]{m_{h_{#1}}}

\newcommand{\Scs}{$\mathcal S$}

\def\order#1{\ensuremath{{\cal O}(#1)}}
\def\reffi#1{\mbox{Fig.~\ref{#1}}}
\def\reffis#1{\mbox{Figs.~\ref{#1}}}

\def\ga{\gamma}
\def\de{\delta}
\def\la{\lambda}

\def\phiAt{\varphi_{\At}}

\def\phimu{\varphi_{\mu}}

\def\phiMz{\varphi_{M_2}}

\definecolor{Orange}{named}{Orange}
\definecolor{Purple}{named}{Purple}
\definecolor{Lightblue}{cmyk}{0.9,0.1,0.1,0.3}
\definecolor{dgelborange}{cmyk}{0.,0.3,0.5, 0.}
\definecolor{Lila}{rgb}{0.5,0.,1}

\graphicspath{{figs/}}

\evensidemargin \oddsidemargin
\allowdisplaybreaks
\begin{document}
\thispagestyle{empty}

\def\thefootnote{\fnsymbol{footnote}}

\begin{flushright}
\mbox{}
\end{flushright}

\vspace{0.5cm}

\begin{center}

{\large\sc 
{\bf Neutral Higgs Boson Production at \boldmath{$e^+e^-$} Colliders}} 

\vspace{0.4cm}

{\large\sc {\bf in the Complex MSSM: A Full One-Loop Analysis}}

\vspace{1cm}

{\sc
S.~Heinemeyer$^{1,2}$%
\footnote{email: Sven.Heinemeyer@cern.ch}%
~and C.~Schappacher$^{3}$%
\footnote{email: schappacher@kabelbw.de}%
\footnote{former address}%
}

\vspace*{.7cm}

{\sl
$^1$Instituto de F\'isica de Cantabria (CSIC-UC), E-39005 Santander,  Spain

\vspace*{0.1cm}

$^2$Instituto de F\'isica Te\'orica, (UAM/CSIC), Universidad
  Aut\'onoma de Madrid,\\ Cantoblanco, E-28049 Madrid, Spain
\vspace*{0.1cm}

$^3$Institut f\"ur Theoretische Physik,
Karlsruhe Institute of Technology, \\
D--76128 Karlsruhe, Germany

}

\end{center}

\vspace*{0.1cm}

\begin{abstract}
\noindent
For the search for additional Higgs bosons in the Minimal Supersymmetric 
Standard Model (MSSM) as well as for future precision analyses in the 
Higgs sector a precise knowledge of their production properties is mandatory.
We evaluate the cross sections for the neutral Higgs boson production 
at $e^+e^-$ colliders in the MSSM with complex parameters (cMSSM). 
The evaluation is based on a full one-loop calculation of the production 
channels $e^+e^- \to h_i Z,\, h_i \ga,\, h_i h_j$ $(i,j = 1,2,3)$, including 
soft and hard QED radiation.  
The dependence of the Higgs boson production cross sections on the relevant 
cMSSM parameters is analyzed numerically.  We find sizable contributions 
to many cross sections.  They are, depending on the production
channel, roughly of 10-20\% of the tree-level results, but can go up to
50\% or higher.  The full one-loop contributions are important for a future 
linear $e^+e^-$ collider such as the ILC or CLIC.  There are plans to 
implement the evaluation of the Higgs boson production cross sections into 
the code \FH.
\end{abstract}


\def\thefootnote{\arabic{footnote}}
\setcounter{page}{0}
\setcounter{footnote}{0}

\newpage


\section{Introduction}
\label{sec:intro}

The discovery of a new particle with a mass of about $125 \gev$
in the Higgs searches at the Large Hadron Collider (LHC), which has been
announced by ATLAS~\cite{ATLASdiscovery} and CMS~\cite{CMSdiscovery}, 
marks the culmination of an effort that has been 
ongoing for almost half a century and opens a new era of particle physics. 
Within the experimental and theoretical uncertainties the properties of
the newly discovered particle measured 
so far are in agreement with a Higgs boson as it is predicted in the
Standard Model (SM)~\cite{Moriond15}.

The identification of the underlying physics of the discovered new
particle and the exploration of the mechanism of electroweak symmetry breaking
will clearly be a top priority in the future program of particle physics. 
The most frequently studied realizations are the Higgs mechanism within the 
SM and within the Minimal Supersymmetric Standard Model
(MSSM)~\cite{mssm,HaK85,GuH86}. 
Contrary to the case of the SM, in the MSSM two Higgs doublets are required.
This results in five physical Higgs bosons instead of the single Higgs
boson in the SM.  In lowest order these are the light and heavy 
$\CP$-even Higgs bosons, $h$ and $H$, the $\CP$-odd Higgs boson, 
$A$, and two charged Higgs bosons, $H^\pm$. Within the MSSM with complex
parameters (cMSSM), taking higher-order corrections into account, the
three neutral Higgs bosons mix and result in the states 
$h_i$ ($i = 1,2,3$)~\cite{mhiggsCPXgen,Demir,mhiggsCPXRG1,mhiggsCPXFD1}.
The Higgs sector of the cMSSM is described at the tree-level by two
parameters: 
the mass of the charged Higgs boson, $\MHp$, and the ratio of the two
vacuum expectation values, $\tb \equiv \TB = v_2/v_1$.
Often the lightest Higgs boson, $h_1$ is identified~\cite{Mh125} with 
the particle  discovered at the LHC~\cite{ATLASdiscovery,CMSdiscovery} 
with a mass around $\sim 125\gev$~\cite{MH125}.
If the mass of the charged Higgs boson is assumed to be larger than 
$\sim 200\gev$ the four additional Higgs bosons are roughly mass
degenerate, $\MHp \approx \mh2 \approx \mh3$ and referred to as the
``heavy Higgs bosons''. 
Discovering one or more of the additional Higgs bosons would be an
unambiguous sign of physics beyond the SM and could yield important 
information about their possible supersymmetric origin.

If supersymmetry (SUSY) is realized in nature and the charged
Higgs boson mass is $\MHp \lesssim 1.5\tev$, then the additional Higgs 
bosons could be detectable at the LHC~\cite{ATLAS-HA,CMS-HA} 
(including its high luminosity upgrade, HL-LHC; see \citere{holzner} 
and references therein). This would yield some initial data on the 
extended Higgs sector.  Equally important, the additional Higgs bosons
could also be produced at a future linear $e^+e^-$ collider such as the
ILC~\cite{ILC-TDR,teslatdr,ilc,LCreport} or CLIC~\cite{CLIC,LCreport}. 
(Results on the combination of LHC and ILC results can be found in 
\citere{lhcilc}.) 
At an $e^+e^-$ linear collider several production modes for the neutral
cMSSM Higgs bosons are possible, 
\begin{align}
e^+e^- &\to h_i Z,\,
            h_i \ga,\,
            h_i h_j,\,
            h_i \nu \bar\nu,\,
            h_i e^+e^-,\,
            h_i t \bar{t},\, 
            h_i b \bar{b},\,
            \ldots \qquad (i,j = 1,2,3)\,. \notag
\end{align}

In the case of a discovery of additional Higgs bosons a subsequent
precision measurement of their properties will be crucial to determine
their nature and the underlying (SUSY) parameters. 
In order to yield a sufficient accuracy, one-loop corrections to the 
various Higgs boson production and decay modes have to be considered.
Full one-loop calculations in the cMSSM for various Higgs boson decays
to SM fermions, scalar fermions and charginos/neutralinos have been
presented over the last years~\cite{hff,HiggsDecaySferm,HiggsDecayIno}. 
For the decay to SM fermions see also \citeres{hff0,deltab,db2l}.
Decays to (lighter) Higgs bosons have been evaluated at the full
one-loop level in the cMSSM in \citere{hff}; see also \citeres{hhh,hAA}.
Decays to SM gauge bosons (see also \citere{hVV-WH}) can be evaluated 
to a very high precision using the full SM one-loop 
result~\cite{prophecy4f} combined with the appropriate effective 
couplings~\cite{mhcMSSMlong}.
The full one-loop corrections in the cMSSM listed here together with
resummed SUSY corrections have been implemented into the code 
\FH~\cite{feynhiggs,mhiggslong,mhiggsAEC,mhcMSSMlong,Mh-logresum}.
Corrections at and beyond the one-loop level in the MSSM with real
parameters (rMSSM) are implemented into the code 
\HDECAY~\cite{hdecay,hdecay2}.
Both codes were combined by the LHC Higgs Cross Section Working Group to
obtain the most precise evaluation for rMSSM Higgs boson decays to SM
particles and decays to lighter Higgs bosons~\cite{YR3}.

The most advanced SUSY Higgs boson production calculations at the LHC 
are available via the code \texttt{SusHi}~\cite{sushi}, which are, however, 
so far restricted to the rMSSM.  However, particularly relevant are 
higher-order corrections also for the Higgs boson production at $e^+e^-$
colliders, where a very high accuracy in the Higgs property determination 
is anticipated~\cite{LCreport}. 
In this paper we concentrate on the neutral Higgs boson production at 
$e^+e^-$ colliders in association with a SM gauge boson or another 
cMSSM Higgs boson, i.e.\ we calculate $(i,j = 1,2,3)$,
\begin{align}
\label{eq:eehh}
&\sigma(\eehh) \,, \\
\label{eq:eehZ}
&\sigma(\eehZ) \,, \\
\label{eq:eehga}
&\sigma(\eehga) \,.
\end{align}
The processes $e^+e^- \to h_i h_i$ and $\eehga$ are purely loop-induced. 
The evaluation of the channels (\ref{eq:eehh}) -- (\ref{eq:eehga}) 
is based on a full one-loop calculation, \ie including electroweak (EW) 
corrections, as well as soft and hard QED radiation. 

Results for the cross sections (\ref{eq:eehh}) -- (\ref{eq:eehga}) have 
been obtained over the last two decades. 
A first (nearly) full calculation of the production channels (\ref{eq:eehh}) 
and (\ref{eq:eehZ}) in the rMSSM was presented in \citere{HiggsProd-org}
(leaving out only a detailed evaluation of the initial state radiation).%
\footnote{
  A corresponding computer code is available at 
  \texttt{www.feynhiggs.de}.
}
A tree-level evaluation of the channels (\ref{eq:eehh}) and
(\ref{eq:eehZ}) in the cMSSM was presented in \citere{HiggsProd-cMSSM-tree}, 
where higher-order corrections were included via effective couplings.
Higher-order corrections to the channels (\ref{eq:eehh}) and (\ref{eq:eehZ}) 
in the cMSSM were given in \citere{hff}, where the third generation 
(s)fermion contributions to the production vertex as well as Higgs boson 
propagator corrections were taken into account.
Another full one-loop calculation of $e^+e^- \to h Z$ was given in 
\citere{HiggsProd-1L}.
The production of two equal Higgs bosons in $e^+e^-$ collisions in the
rMSSM, where only box-diagrams contribute, were presented in
\citeres{HiggsProd-hh} and further discussed in \citere{HiggsProd-hh2}.
Finally, the channel (\ref{eq:eehga}) in the rMSSM was evaluated in 
\citere{HiggsProd-gah}.  A short numerical comparison with the literature 
will be given in \refse{sec:comparisons}.

\medskip

In this paper we present a full one-loop calculation for neutral Higgs boson 
production at $e^+e^-$ colliders in association with a SM gauge boson or 
another cMSSM Higgs boson, taking into account soft and hard QED radiation. 
In \refse{sec:renorm} we very briefly review the renormalization of the
relevant sectors of the cMSSM.  
Details about the calculation can be found in \refse{sec:calc}. 
In \refse{sec:comparisons} various comparisons with results from other
groups are given. The numerical results for all production channels 
(\ref{eq:eehh}) -- (\ref{eq:eehga}) are presented in \refse{sec:numeval}.
The conclusions can be found in \refse{sec:conclusions}.  
There are plans to implement the evaluation of the production cross sections
into the Fortran code 
\FH~\cite{feynhiggs,mhiggslong,mhiggsAEC,mhcMSSMlong,Mh-logresum}.


\subsection*{Prolegomena}

We use the following short-hands in this paper:
\begin{itemize}

\item \FT\ $\equiv$ \FA\ + \FC\ + \LT\ + \FH.

\item $\SW \equiv \sin\theta_W$, $\CW \equiv \cos\theta_W$.

\item $\SBA \equiv \sin(\beta-\alpha)$, 
      $\CBA \equiv \cos(\beta-\alpha)$, 
      $\TB \equiv \tb$.

\end{itemize}
They will be further explained in the text below.


\section{The complex MSSM}
\label{sec:renorm}

The cross sections (\ref{eq:eehh}) -- (\ref{eq:eehga}) are calculated 
at the one-loop level, including soft and hard QED radiation, see the 
next section.  This requires the simultaneous renormalization of the 
Higgs and gauge boson sector as well as the fermion sector of the cMSSM.  
We give a few relevant details about these sectors and their 
renormalization. More information can be found in 
\citeres{HiggsDecaySferm,HiggsDecayIno,MSSMCT,SbotRen,Stop2decay,%
Gluinodecay,Stau2decay,LHCxC,LHCxN,LHCxNprod}.

The renormalization of the Higgs and gauge-boson sector follow strictly 
\citere{MSSMCT} and references therein (see especially \citere{mhcMSSMlong}). 
This defines in particular the counterterm $\dTB$,
as well as the counterterms for the $Z$~boson mass, $\de\MZ^2$, and for the 
sine of the weak mixing angle, $\de\SW$ 
(with $\SW = \sqrt{1 - \CW^2} = \sqrt{1 - \MW^2/\MZ^2}$, where $\MW$ and 
$\MZ$ denote the $W$~and $Z$~boson masses, respectively).

The renormalization of the fermion sector is described in detail in 
\citere{MSSMCT} and references therein.  For simplification we use 
the \DRbar\ renormalization for all three generations of down-type 
quarks \textit{and} leptons, in the notation of \citere{MSSMCT}:
\begin{align*}
  \Code{UVMf1[4,\,\uscore] = UVDivergentPart} &\qquad
  \text{\DRbar\ renormalization for $m_d$, $m_s$, $m_b$} \\
  \Code{UVMf1[2,\,\uscore] = UVDivergentPart} &\qquad 
  \text{\DRbar\ renormalization for $m_e$, $m_\mu$, $m_\tau$}
\end{align*}


\section{Calculation of diagrams}
\label{sec:calc}

In this section we give some details about the calculation of the
tree-level and higher-order corrections to the production of 
Higgs bosons in $e^+e^-$ collisions. 
The diagrams and corresponding amplitudes have been obtained with 
\FA\ (version 3.9) \cite{feynarts}, using the MSSM model file (including 
the MSSM counterterms) of \citere{MSSMCT}. 
The further evaluation has been performed with \FC\ (version 8.4) and 
\LT\ (version 2.12) \cite{formcalc}.
The Higgs sector quantities (masses, mixings, $\hat{Z}$~factors, etc.) 
have been evaluated using
\FH~\cite{feynhiggs,mhiggslong,mhiggsAEC,mhcMSSMlong,Mh-logresum}
(version 2.11.0).


\subsection{Contributing diagrams}
\label{sec:diagrams}

\begin{figure}
\begin{center}
\framebox[15cm]{\includegraphics[width=0.11\textwidth]{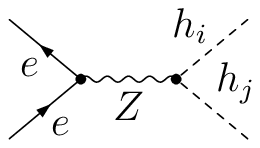}} \\
\framebox[15cm]{\includegraphics[width=0.73\textwidth]{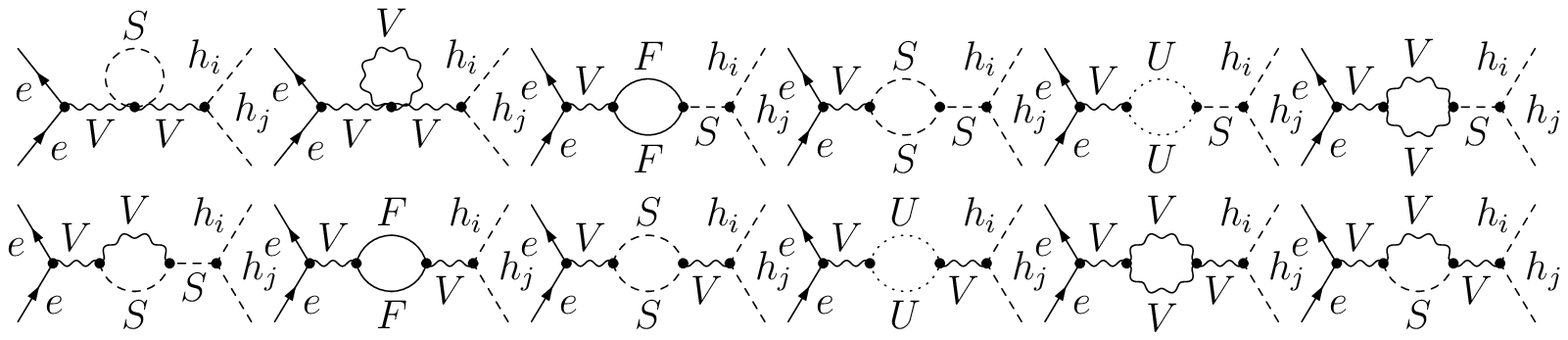}} \\
\framebox[15cm]{\includegraphics[width=0.73\textwidth]{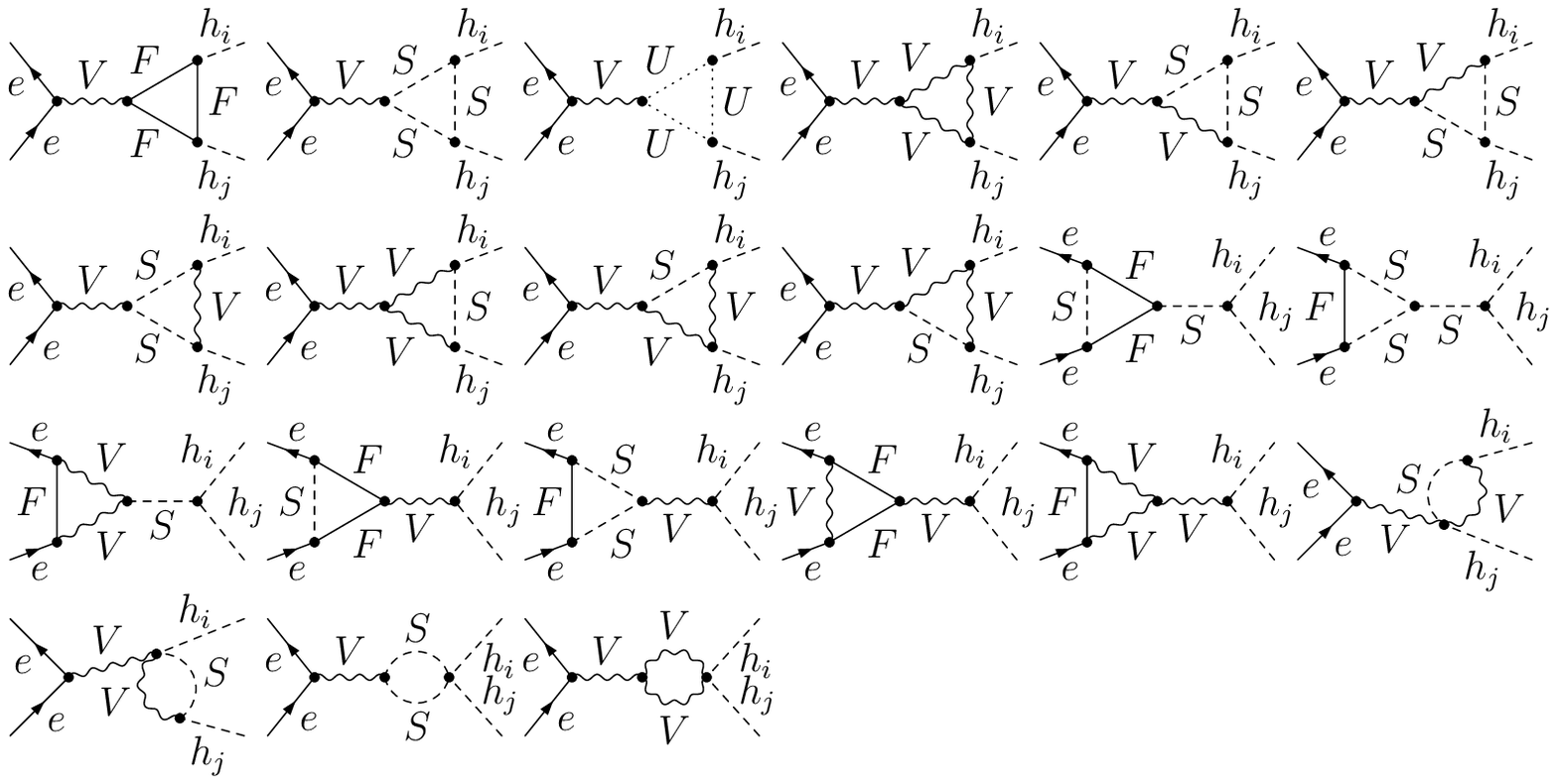}} \\ 
\framebox[15cm]{\includegraphics[width=0.73\textwidth]{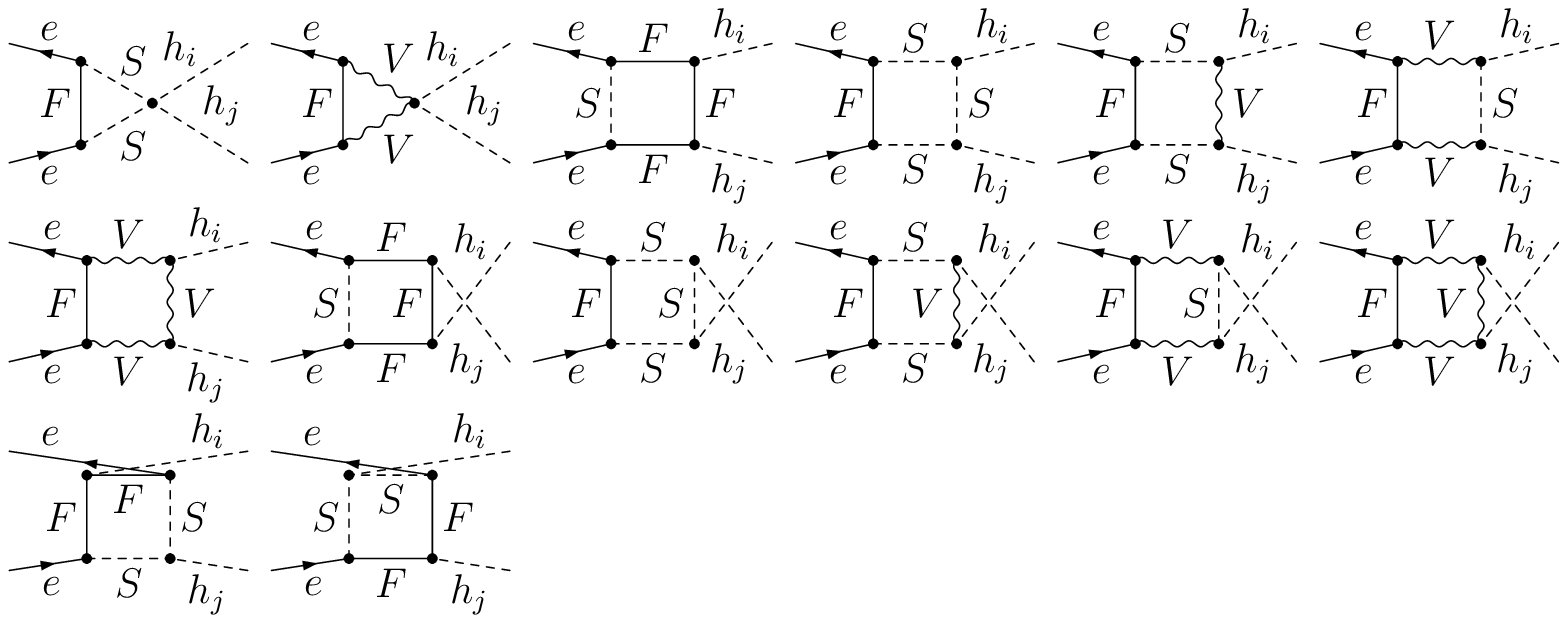}} \\
\framebox[15cm]{\includegraphics[width=0.65\textwidth]{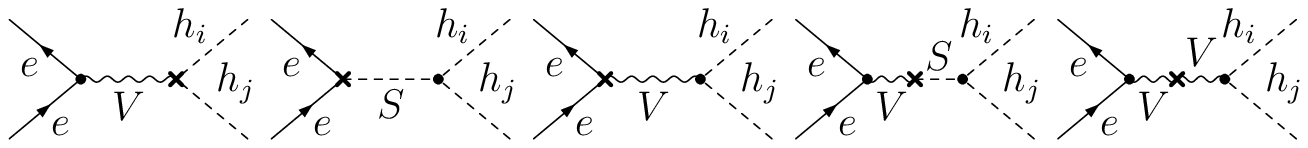}}
\caption{
  Generic tree, self-energy, vertex, box and counterterm diagrams 
  for the process $\eehh$ ($i,j = 1,2,3$).  
  $F$ can be a SM fermion, chargino or neutralino; 
  $S$ can be a sfermion or a Higgs/Goldstone boson; 
  $V$ can be a $\ga$, $Z$ or $W^\pm$. 
  It should be noted that electron-Higgs couplings are neglected.  
}
\label{fig:hhdiagrams}
\end{center}
\end{figure}

\begin{figure}
\begin{center}
\framebox[14.5cm]{\includegraphics[width=0.11\textwidth]{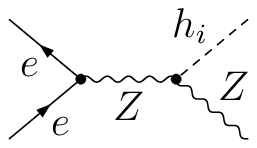}}\\
\framebox[14.5cm]{\includegraphics[width=0.73\textwidth]{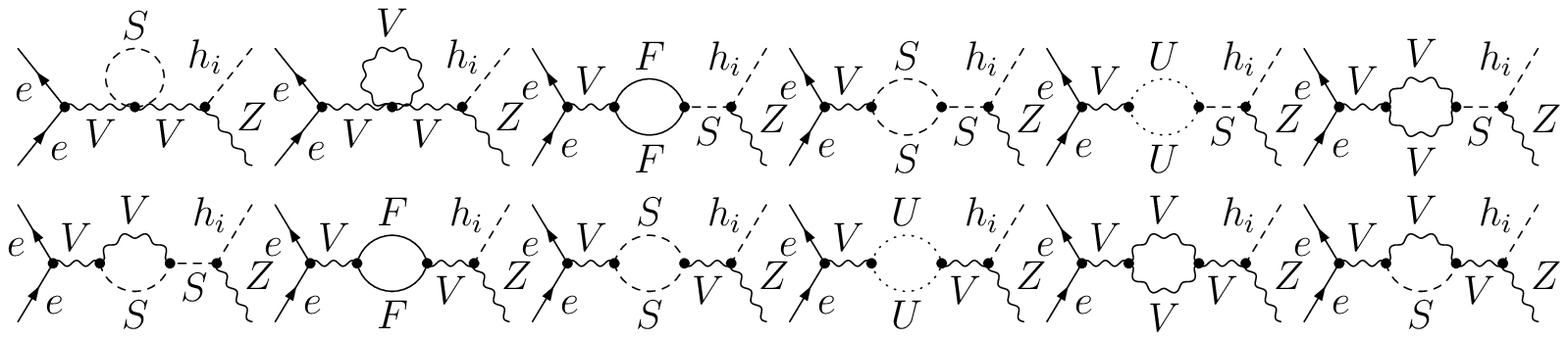}}\\
\framebox[14.5cm]{\includegraphics[width=0.73\textwidth]{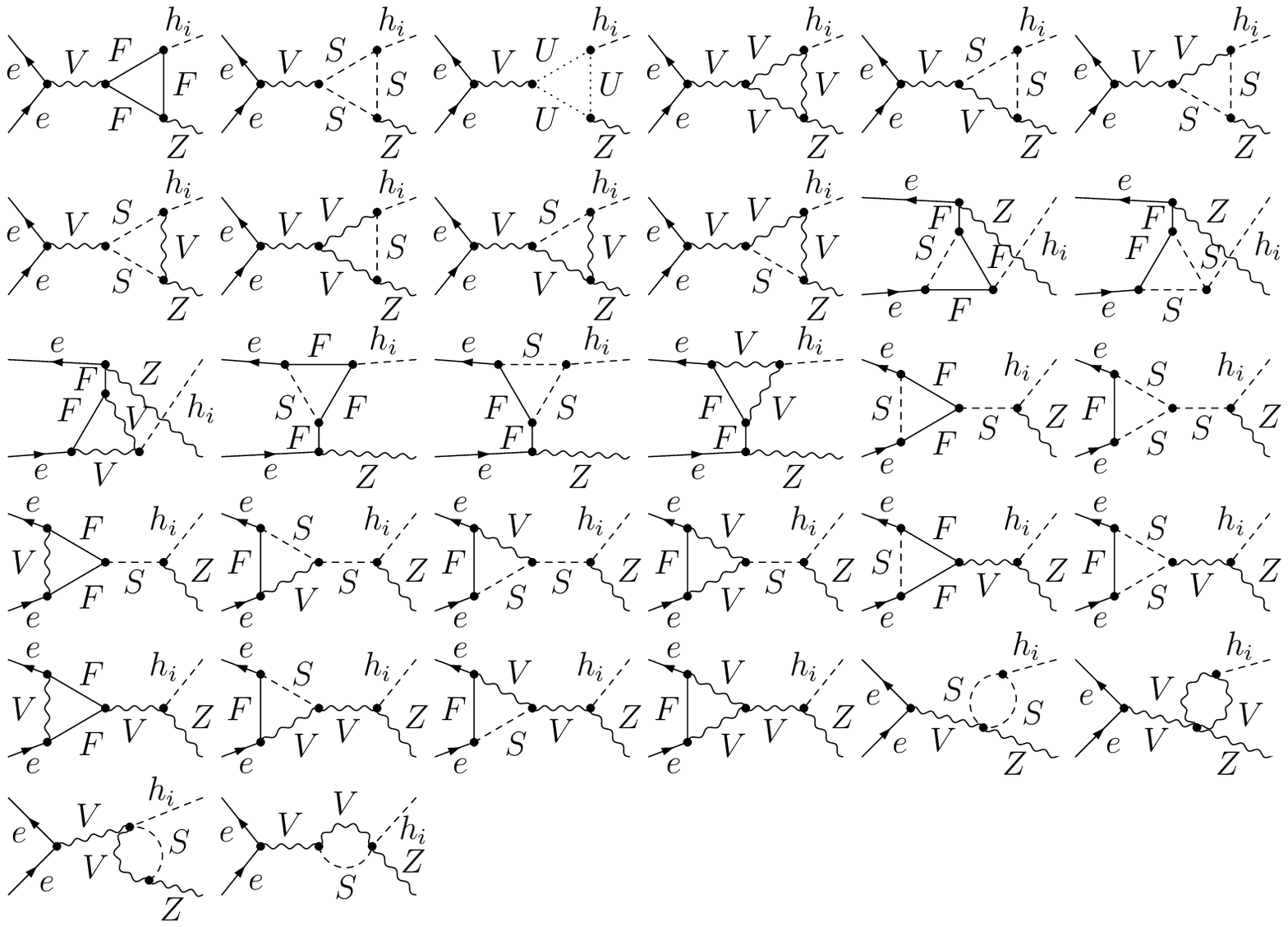}}\\
\framebox[14.5cm]{\includegraphics[width=0.73\textwidth]{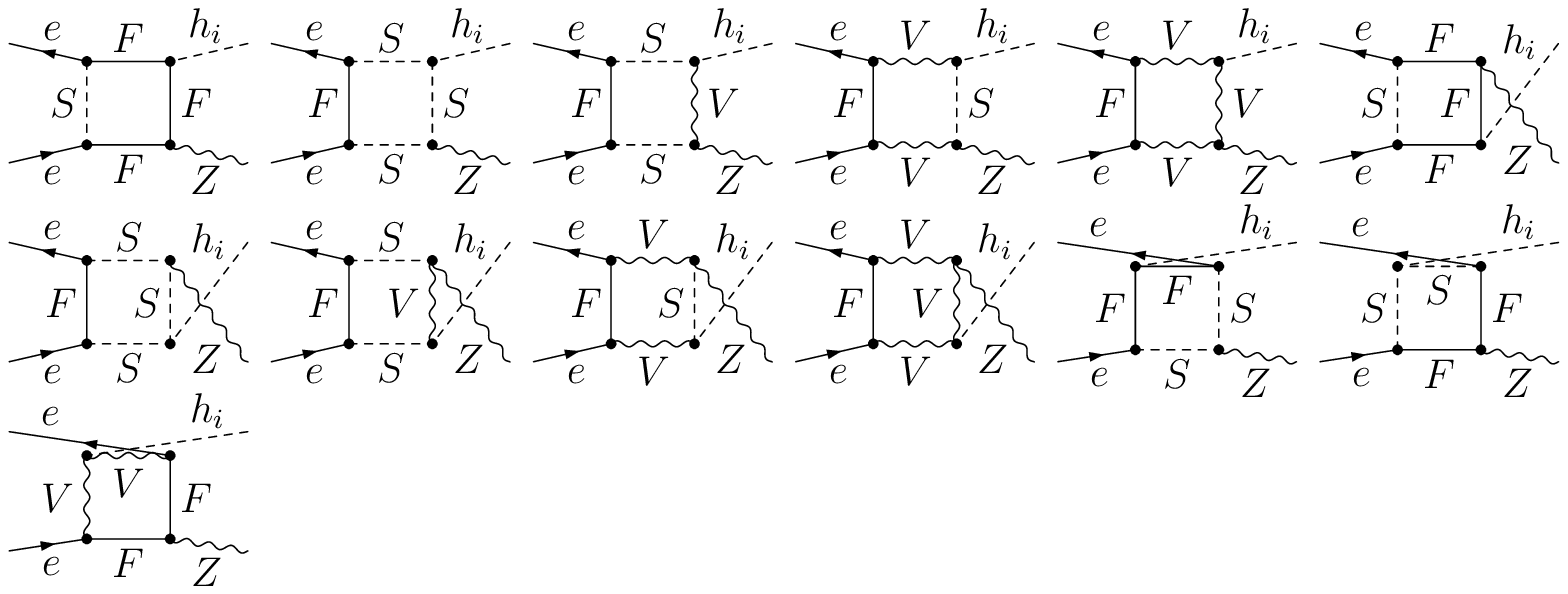}}\\
\framebox[14.5cm]{\includegraphics[width=0.71\textwidth]{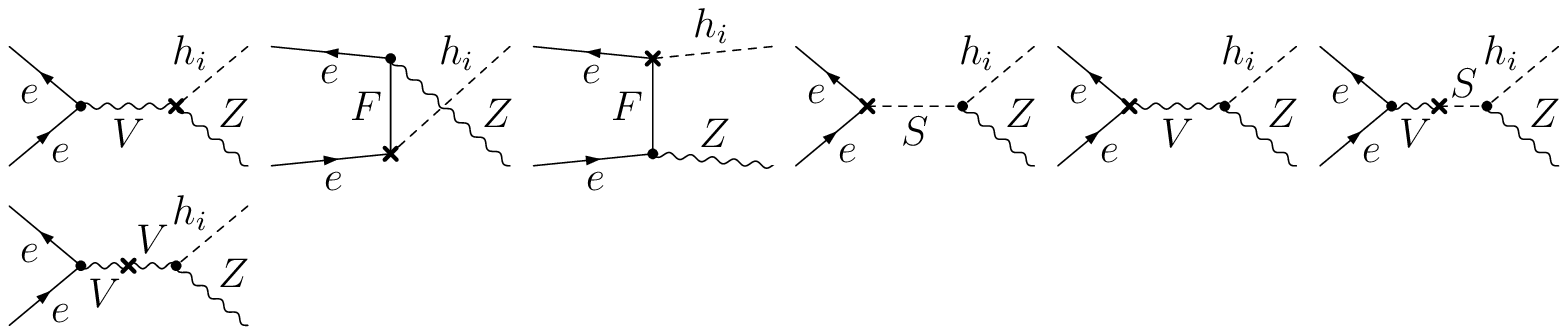}}
\caption{
  Generic tree, self-energy, vertex, box and counterterm diagrams 
  for the process $\eehZ$ ($i = 1,2,3$). 
  $F$ can be a SM fermion, chargino or neutralino; 
  $S$ can be a sfermion or a Higgs/Goldstone boson; 
  $V$ can be a $\ga$, $Z$ or $W^\pm$. 
  It should be noted that electron-Higgs couplings are neglected.  
}
\label{fig:hZdiagrams}
\end{center}
\end{figure}

\begin{figure}[t]
\begin{center}
\framebox[15cm]{\includegraphics[width=0.73\textwidth]{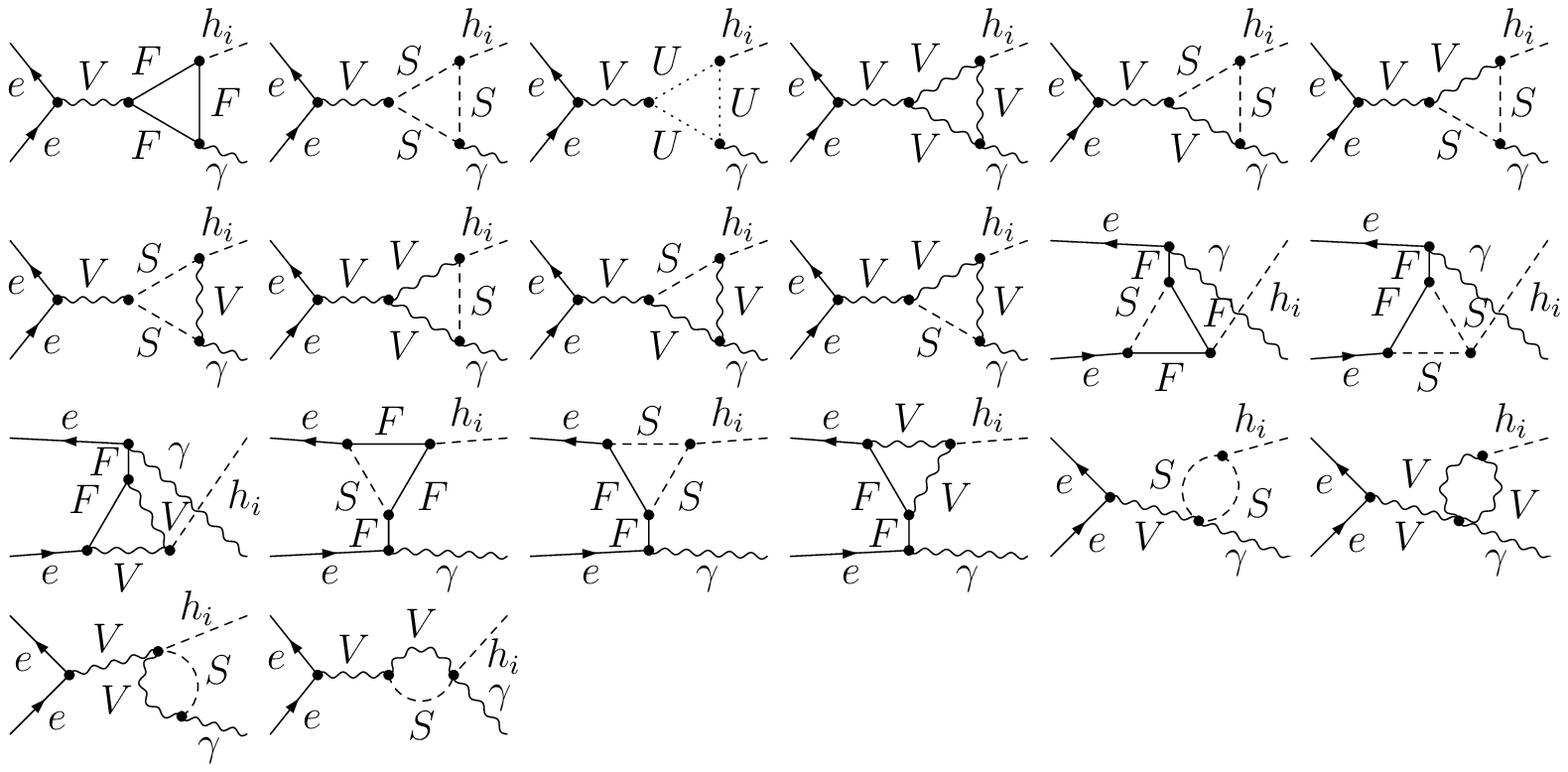}}\\
\framebox[15cm]{\includegraphics[width=0.73\textwidth]{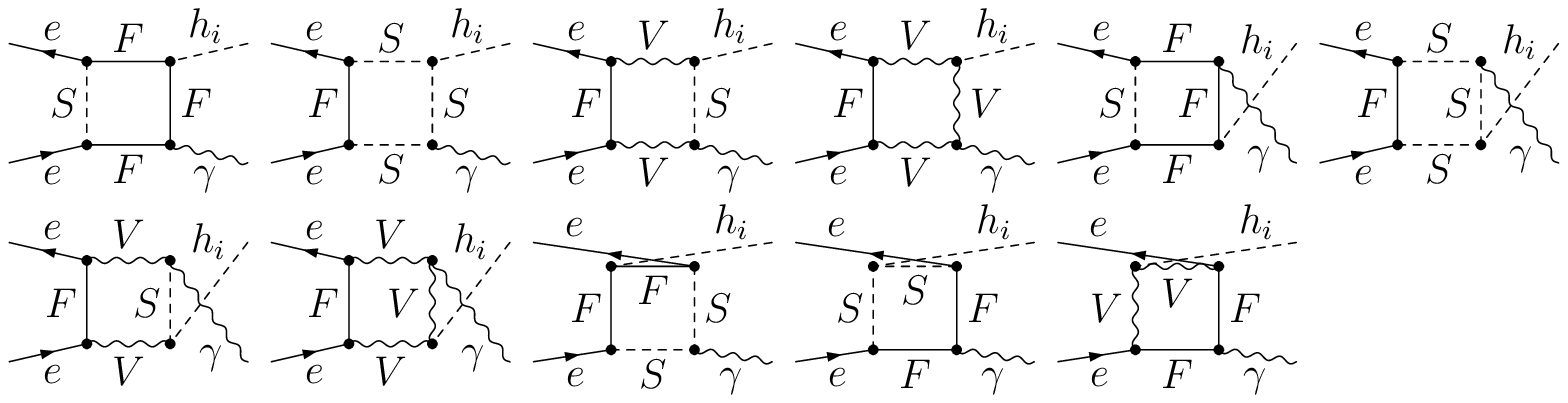}}\\
\framebox[15cm]{\includegraphics[width=0.35\textwidth]{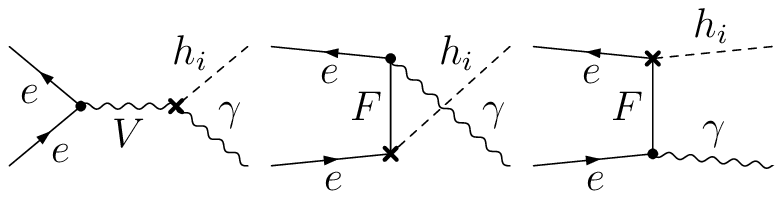}}
\caption{
  Generic vertex, box and counterterm diagrams for the (loop induced) 
  process $\eehga$ ($i = 1,2,3$). 
  $F$ can be a SM fermion, chargino or neutralino; 
  $S$ can be a sfermion or a Higgs/Goldstone boson; 
  $V$ can be a $\ga$, $Z$ or $W^\pm$. 
  It should be noted that electron-Higgs couplings are neglected.  
}
\label{fig:hgadiagrams}
\end{center}
\end{figure}

Sample diagrams for the process $\eehh$ ($i,j = 1,2,3$) are shown in 
\reffi{fig:hhdiagrams}, for the process $\eehZ$ ($i = 1,2,3$) are 
shown in \reffi{fig:hZdiagrams} and for the process $\eehga$ 
$(i = 1,2,3$) in \reffi{fig:hgadiagrams}.
Not shown are the diagrams for real (hard and soft) photon radiation. 
They are obtained from the corresponding tree-level diagrams by 
attaching a photon to the electrons/positrons.
The internal particles in the generically depicted diagrams in 
\reffis{fig:hhdiagrams} -- \ref{fig:hgadiagrams} are labeled as follows: 
$F$ can be a SM fermion $f$, chargino $\cha{c}$ or neutralino 
$\neu{n}$; $S$ can be a sfermion $\Sf_s$ or a Higgs (Goldstone) boson 
$h_i, H^\pm$ ($G, G^\pm$); $U$ denotes the ghosts $u_V$;
$V$ can be a photon $\ga$ or a massive SM gauge boson, $Z$ or $W^\pm$. 
We have neglected all electron-Higgs couplings via the \FA\ 
command \cite{feynarts}
\begin{align*}
\Code{Restrictions -> {NoElectronHCoupling}}
\end{align*}
and terms proportional to the electron mass \Code{ME}
(and the squared electron mass \Code{ME2}) via the \FC\ command 
\cite{formcalc}
\begin{align*}
\Code{Neglect[ME] = Neglect[ME2] = 0}
\end{align*}
which allows \FC\ to replace \Code{ME} by zero whenever this is safe, 
i.e.\ except when it appears in negative powers or in loop integrals.
We have verified numerically that these contributions are indeed totally 
negligible.  For internally appearing Higgs bosons no higher-order
corrections to their masses or couplings are taken into account; 
these corrections would correspond to effects beyond one-loop order.%
\footnote{
  We found that using loop corrected Higgs boson masses 
  in the loops leads to a UV divergent result.
}
For external Higgs bosons, as discussed in \citere{mhcMSSMlong}, the 
appropriate $\hat{Z}$~factors are applied and on-shell (OS) masses 
(including higher-order corrections) are used~\cite{mhcMSSMlong}, 
obtained with 
\FH~\cite{feynhiggs,mhiggslong,mhiggsAEC,mhcMSSMlong,Mh-logresum}.

Also not shown are the diagrams with a $Z$/Goldstone--Higgs boson
self-energy contribution on the external Higgs boson leg. 
They appear in $\eehh$, with a $Z/G$--$h_{i,j}$ transition 
and have been calculated explicitly as far as they are \textit{not}
proportional to the electron mass.  It should be noted that for the
process $\eehZ$ all these contributions are proportional to 
the electron mass and have consistently be neglected.

Furthermore, in general, in \reffis{fig:hhdiagrams} -- \ref{fig:hgadiagrams}
we have omitted diagrams with self-energy type corrections of external 
(on-shell) particles.  While the contributions from the real parts of the 
loop functions are taken into account via the renormalization constants 
defined by OS renormalization conditions, the contributions coming from 
the imaginary part of the loop functions can result in an additional (real) 
correction if multiplied by complex parameters.  In the analytical and 
numerical evaluation, these diagrams have been taken into account via the 
prescription described in \citere{MSSMCT}. 

Within our one-loop calculation we neglect finite width effects that 
can help to cure threshold singularities.  Consequently, in the close 
vicinity of those thresholds our calculation does not give a reliable
result.  Switching to a complex mass scheme \cite{complexmassscheme} 
would be another possibility to cure this problem, but its application 
is beyond the scope of our paper.

For completeness we show here the tree-level cross section formulas:
\begin{align}
\label{eehhTree}
\sigma_{\text{tree}}(\eehh) &= \frac{\pi\, \alpha^2}{96\, s}
       \LP \frac{8\, \SW^4 - 4\, \SW^2 + 1}{\SW^4\,\CW^4}\RP\,
       \frac{\la^{3/2}(1,\mh{i}^2/s,\mh{j}^2/s)}{(1-\MZ^2/s)^2}\,
\times \notag \\
&\mathrel{\phantom{=}}
       \left| \hat{Z}_{j3} (\CBA\, \hat{Z}_{i1} - \SBA\, \hat{Z}_{i2}) -
       \hat{Z}_{i3} (\CBA\, \hat{Z}_{j1} - \SBA\, \hat{Z}_{j2}) \right|^2\,, \\
\label{eehZTree}
\sigma_{\text{tree}}(\eehZ)  &= 
       \frac{\pi\, \alpha^2}{96\, s} 
       \LP \frac{8\, \SW^4 - 4\, \SW^2 + 1}{\SW^4\, \CW^4} \RP\,
       \frac{\la(1,\mh{i}^2/s,\MZ^2/s) + 12\,\MZ^2/s }{(1-\MZ^2/s)^2}\, 
\times \notag \\
&\mathrel{\phantom{=}}
       \la^{1/2}(1,\mh{i}^2/s,\MZ^2/s)\,
       \left| \SBA\, \hat{Z}_{i1} + \CBA\, \hat{Z}_{i2} \right|^2\,,
\end{align}
where $i,j = 1,2,3$ ($i \neq j$)
and $\la(x,y,z) = (x - y - z)^2 - 4yz$ denotes the two-body 
phase space function.  The $Z$-factor matrix is given by 
$\hat{Z}_{ij} \equiv \Code{ZHiggs[\Vi,\,\Vj]}$, see \citere{MSSMCT} 
(and \citere{mhcMSSMlong}) and is calculated by \FH.


\subsection{Ultraviolet divergences}

As regularization scheme for the UV divergences we have used constrained 
differential renormalization~\cite{cdr}, which has been shown to be 
equivalent to dimensional reduction~\cite{dred} at the \onel\ 
level~\cite{formcalc}. 
Thus the employed regularization scheme preserves SUSY~\cite{dredDS,dredDS2}
and guarantees that the SUSY relations are kept intact, \eg that the gauge 
couplings of the SM vertices and the Yukawa couplings of the corresponding 
SUSY vertices also coincide to \onel\ order in the SUSY limit. 
Therefore no additional shifts, which might occur when using a different 
regularization scheme, arise. All UV divergences cancel in the final result.%
\footnote{
  It should be noted that some processes are UV divergent if the 
  electron mass is neglected (see \refse{sec:diagrams}).  The full 
  processes including the terms proportional to the electron mass 
  are, of course, UV finite.  Dropping the divergence, the numerical 
  difference between the two calculations was found to be negligible. 
  Therefore we used the (faster) simplified code with neglected 
  electron mass for our numerical analyses below.
}


\subsection{Infrared divergences}

Soft photon emission implies numerical problems in the phase space 
integration of radiative processes.  The phase space integral diverges 
in the soft energy region where the photon momentum becomes very small,
leading to infrared (IR) singularities.  Therefore the IR divergences from 
diagrams with an internal photon have to cancel with the ones from the 
corresponding real soft radiation.  We have included the soft photon contribution 
via the code already implemented in \FC\ following the description given 
in \citere{denner}.  The IR divergences arising from the diagrams involving 
a photon are regularized by introducing a photon mass parameter, $\la$. 
All IR divergences, \ie all divergences in the limit $\la \to 0$, cancel 
once virtual and real diagrams for one process are added. 
We have (numerically) checked that our results do not depend on $\la$.

\begin{figure}[t!]
\centering
\includegraphics[width=0.49\textwidth,height=7.5cm]{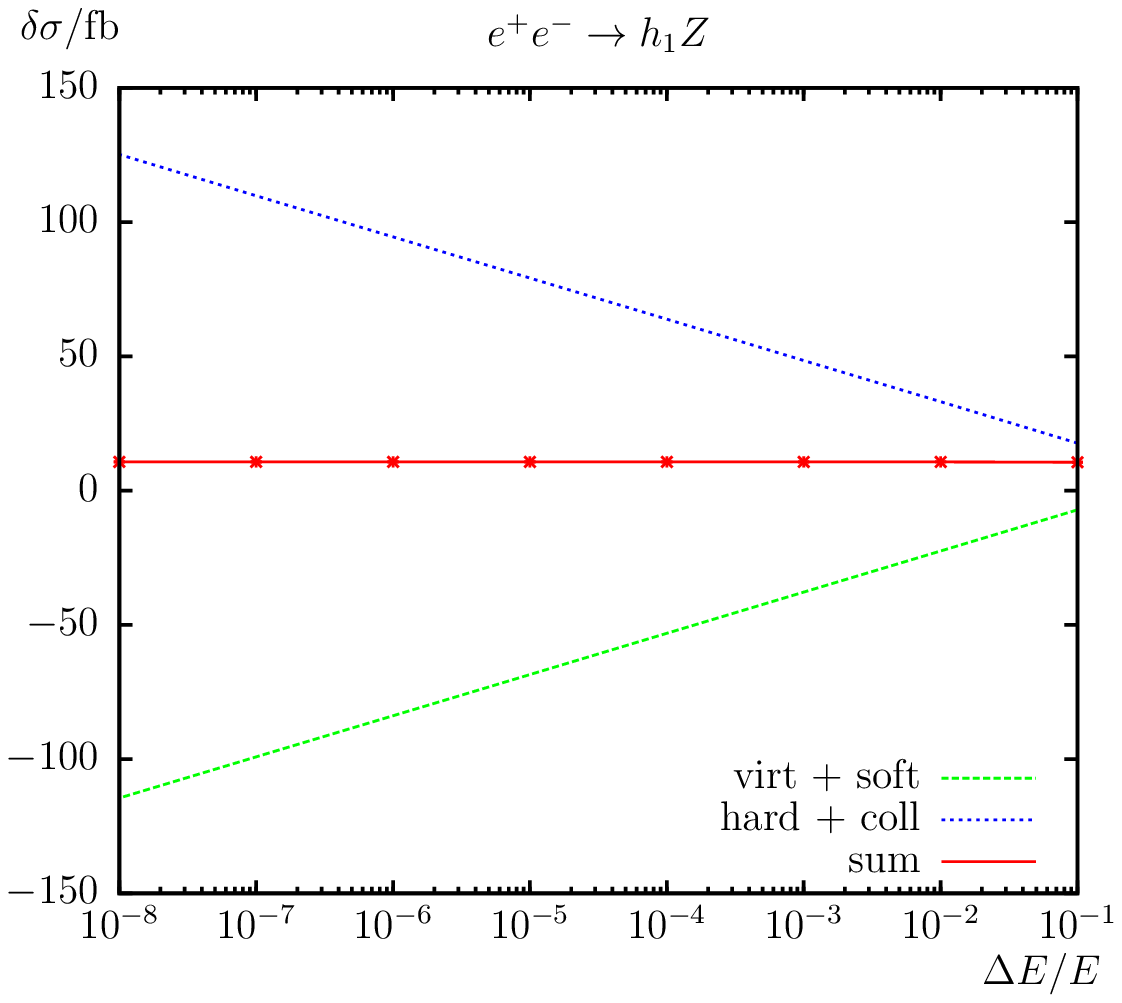}
\vspace{1em}
\begin{minipage}[b]{0.4\textwidth}
\centering
\begin{tabular}[b]{lr}
\toprule
$\Delta E/E$ & $\delta\sigma$/fbarn \\
\midrule
$10^{-1}$ & $10.58 \pm 0.01$ \\
$10^{-2}$ & $10.68 \pm 0.02$ \\
$10^{-3}$ & $10.69 \pm 0.03$ \\
$10^{-4}$ & $10.69 \pm 0.04$ \\
$10^{-5}$ & $10.70 \pm 0.05$ \\
$10^{-6}$ & $10.69 \pm 0.07$ \\
$10^{-7}$ & $10.70 \pm 0.08$ \\
$10^{-8}$ & $10.73 \pm 0.09$ \\
\bottomrule
\end{tabular}
\vspace{2em}
\end{minipage}
\includegraphics[width=0.49\textwidth,height=7.5cm]{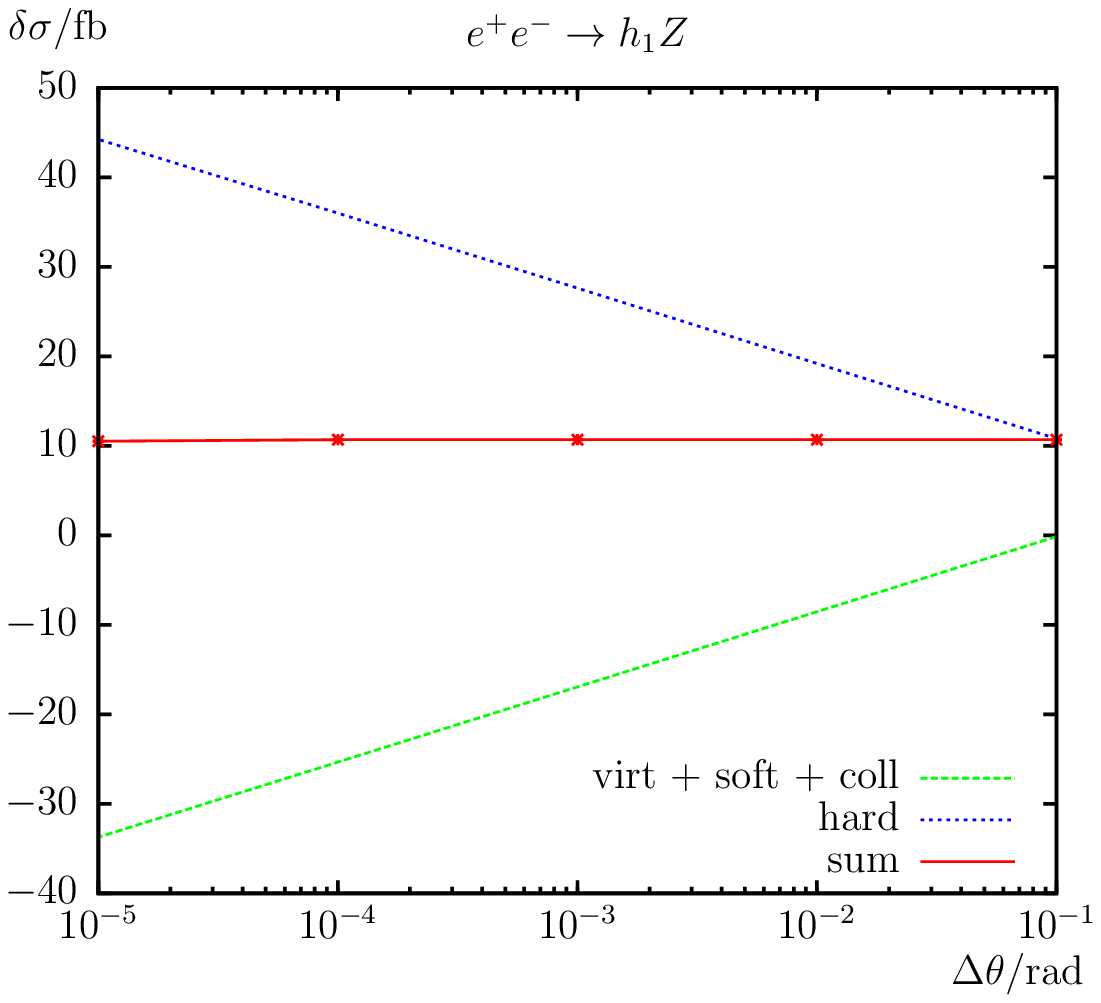}
\begin{minipage}[b]{0.4\textwidth}
\centering
\begin{tabular}[b]{lr}
\toprule
$\Delta \theta$/rad & $\delta\sigma$/fbarn \\
\midrule
$10^{ 0}$ & $10.42 \pm 0.04$ \\
$10^{-1}$ & $10.69 \pm 0.04$ \\
$10^{-2}$ & $10.69 \pm 0.03$ \\
$10^{-3}$ & $10.70 \pm 0.03$ \\
$10^{-4}$ & $10.69 \pm 0.04$ \\
$10^{-5}$ & $10.53 \pm 0.04$ \\
\bottomrule
\end{tabular}
\vspace{2em}
\end{minipage}
\caption{\label{fig:coll} 
  Phase space slicing method.  The different contributions to the loop 
  corrections $\delta\sigma(e^+e^- \to h_1 Z)$ at $\sqrt{s} = 500\gev$
  with fixed $\Delta \theta/\text{rad} = 10^{-2}$ (upper plot) and
  fixed $\Delta E/E = 10^{-3}$ (lower plot).
}
\end{figure}

We have also numerically checked that our results do not depend on 
$\Delta E = \delta_s E = \delta_s \sqrt{s}/2$ defining the energy 
cut that separates the soft from the hard radiation.  As one can see
from the example in the upper plot of \reffi{fig:coll} this holds for 
several orders of magnitude.  Our numerical results below have been 
obtained for fixed $\delta_s = 10^{-3}$.

\subsection{Collinear divergences}

Numerical problems in the phase space integration of the radiative 
process arise also through collinear photon emission. Mass singularities 
emerge as a consequence of the collinear photon emission off massless
particles.  But already very light particles (such as \eg electrons)
can produce numerical instabilities.

There are several methods for the treatment of collinear singularities. 
In the following, we give a very brief description of the so-called
\textit{phase space slicing} (PSS) \textit{method}~\cite{slicing}, 
which we adopted.
The treatment of collinear divergences is not (yet) implemented in \FC, 
and therefore we have developed and implemented the code necessary for 
the evaluation of collinear contributions.

In the PSS method, the phase space is divided into
regions where the integrand is finite (numerically stable) and 
regions where it is divergent (or numerically unstable).
In the stable regions the integration is performed numerically, whereas
in the unstable regions it is carried out (semi-) analytically using 
approximations for the collinear photon emission.

The collinear part is constrained by the angular cut-off parameter 
$\Delta\theta$, imposed on the angle between the photon and the
(in our case initial state) electron/positron.

The differential cross section for the collinear photon radiation 
off the initial state $e^+e^-$ pair corresponds to a convolution
\begin{align}
\text{d}\sigma_{\text{coll}}(s) = \frac{\alpha}{\pi} \int_0^{1-\delta_s} \text{d}z\,
  \text{d}\sigma_{\text{tree}}(\sqrt{z s}) \LV \LB 2\, \ln \LP 
  \frac{\Delta \theta \sqrt{s}}{2\, m_e} \RP - 1 \RB P_{ee}(z) + 1 - z \RV\,,
\end{align}
with $P_{ee}(z) = (1 + z^2)/(1 - z)$
denoting the splitting function of a photon from the initial $e^+e^-$ pair.
The electron momentum is reduced (because of the radiated photon) by 
the fraction $z$ such that the center-of-mass frame of the hard process 
receives a boost.  The integration over all possible factors $z$ is 
constrained by the soft cut-off $\delta_s = \Delta E/E$, to prevent 
over-counting in the soft energy region.

We have (numerically) checked that our results do not depend on 
$\Delta\theta$ over several orders of magnitude; see the example in 
the lower plot of \reffi{fig:coll}.  Our numerical results below have 
been obtained for fixed $\Delta \theta/\text{rad} = 10^{-2}$.

The one-loop corrections of the differential cross section are 
decomposed into the virtual, soft, hard and collinear parts as follows:
\begin{align}
\text{d}\sigma_{\text{loop}} = 
  \text{d}\sigma_{\text{virt}}(\lambda) + 
  \text{d}\sigma_{\text{soft}}(\lambda, \Delta E) + 
  \text{d}\sigma_{\text{hard}}(\Delta E, \Delta\theta) + 
  \text{d}\sigma_{\text{coll}}(\Delta E, \Delta\theta)\,.
\end{align}
The hard and collinear parts have been calculated via the Monte Carlo
integration algorithm \texttt{Vegas} as implemented in the \texttt{CUBA} 
library \cite{cuba} as part of \FC.


\section{Comparisons}
\label{sec:comparisons}

In this section we present the comparisons with results from other groups 
in the literature for neutral Higgs boson production in $e^+e^-$ collisions.
Most of these comparisons were restricted to the MSSM with real parameters.
The level of agreement of such comparisons (at one-loop order) depends 
on the correct transformation of the input parameters from our 
renormalization scheme into the schemes used in the respective literature, 
as well as on the differences in the employed renormalization schemes as such.
In view of the non-trivial conversions and the large number of comparisons 
such transformations and/or change of our renormalization prescription is 
beyond the scope of our paper.

\begin{itemize}

\item
A numerical comparison with the program \FHXS\ \cite{HiggsProd-org} can be 
found in \refta{tab:fhxs3}.
We have neglected the initial state radiation
and diagrams with photon exchange, as done in \citere{HiggsProd-org}.
In \refta{tab:fhxs3} ``self'', ``self+vert'' and ``full'' 
denotes the inclusion of only self-energy corrections, self-energy plus 
vertex corrections or the full calculation including box diagrams.
The comparison for the production of the light Higgs boson is rather 
difficult, due to the different \FH\ versions.  
As input parameters we used our scenario \Scs; see \refta{tab:para} 
below.  (We had to change only $A_{t,b,\tau}$ to $A_{t,b} = 1500$ and 
$A_{\tau} = 0$ to be in accordance with the input options of \FHXS).
It can be observed that the level of agreement for the ``self+vert''
calculation is mostly at the level of 5\% or better. However, the box
contributions appear to go in the opposite direction for the first 
three cross sections in the two calculations.  This hints towards a 
problem in the box contributions in \citere{HiggsProd-org}, where the 
box contributions were obtained independently from the rest of the 
loop corrections, whereas using \FT\ all corrections are evaluated 
together in an automated way.  It should be noted that a self-consistent 
check with the program \FHXS\ gave good agreement with 
\citere{HiggsProd-org} as expected (with tiny differences due to slightly 
different SM input parameters).

\begin{table}[t!]
\caption{\label{tab:fhxs3}
  Comparison of the one-loop corrected Higgs production cross sections 
  (in fb) with \FHXS\ at $\sqrt{s} = 1000\gev$ and $\MHp = 310.86$ and 
  higgsmix = 3 as input in \FH.\\
  \FT:   $m_h = 123.17\gev$, $m_H = 300.00\gev$, $m_{A} = 301.70\gev$.\\
  \FHXS: $m_h = 118.68\gev$, $m_H = 301.84\gev$, $m_{A} = 300.00\gev$.}
\centering
\begin{tabular}{lrrrrrrrr}
\toprule  
 & \multicolumn{3}{c}{\FHXS} & & \multicolumn{3}{c}{\FT} \\ 
\cmidrule{2-4} \cmidrule{6-8}
Process & full & self+vert & self & & full & self+vert & self \\
\midrule
$e^+e^- \to h_1 Z\, (\approx h Z)$   & 15.2845 & 14.1038 & 14.7896 & 
                                  & 12.0972 & 14.6641 & 12.3536 \\
$e^+e^- \to h_3 Z\, (\approx H Z)$   &  0.0221 &  0.0174 & 0.0245 &
                                  &  0.0251 &  0.0275 & 0.0181 \\
$e^+e^- \to h_1 h_2\, (\approx h A)$ &  0.0262 &  0.0242 & 0.0165 &
                                  &  0.0220 &  0.0253 & 0.0292 \\
$e^+e^- \to h_2 h_3\, (\approx A H)$ &  6.1456 &  7.0250 & 6.7694 &  
                                  &  5.8913 &  6.8347 & 6.0994 \\
\bottomrule
\end{tabular}
\vspace{1em}
\end{table}

\item
In \citere{BeFeReVe2005} the processes $e^+e^- \to HA, hA$ 
(and $e^+e^- \to H^+ H^-$) have been calculated in the rMSSM.
Unfortunately, in \citere{BeFeReVe2005} the numerical evaluation
(shown in their Fig.~2) are only tree-level results, although the paper
deals with the respective one-loop corrections. For the comparison 
with \citere{BeFeReVe2005} we successfully reproduced their lower Fig.~2.

\item
In \citere{DjHaZe1996} the processes $e^+e^- \to AH,HZ,Ah$ have been 
calculated in the rMSSM at tree-level.  As input parameters we used
their parameters as far as possible.  For the comparison with 
\citere{DjHaZe1996} we successfully reproduced their upper Fig.~2.

\item
In \citere{HiggsProd-cMSSM-tree} a tree-level evaluation of the channels 
(\ref{eq:eehh}) and (\ref{eq:eehZ}) in the cMSSM was presented, where 
higher-order corrections were included via ($\CP$ violating) effective 
couplings. Unfortunately, no numbers are given in
\citere{HiggsProd-cMSSM-tree}, but only two-dimensional parameter scan
plots, which we could not reasonably compare to our results.
Consequently we omitted a comparison with \citere{HiggsProd-cMSSM-tree}.

\item
We performed a comparison with \citere{hff} for $e^+ e^- \to h_i Z, h_i h_j$ 
($i,j = 1,2,3$) at $\order{\alpha}$ in the cMSSM.  In \citere{hff} only
self-energy and vertex corrections involving $t, \stop, b, \sbot$
were included, and the numerical evaluation was performed in the
CPX scenario~\cite{cpx} (with $\MHp$ chosen to yield $\mh1 = 40\gev$) 
which is \textit{extremely} sensitive to the chosen input parameters.
Nevertheless, using their input parameters as far as possible, we found 
qualitative agreement for $\TB < 15$ with their Fig.~20.

\begin{figure}[t!]
\begin{center}
\begin{tabular}{c}
\includegraphics[width=0.48\textwidth,height=6cm]{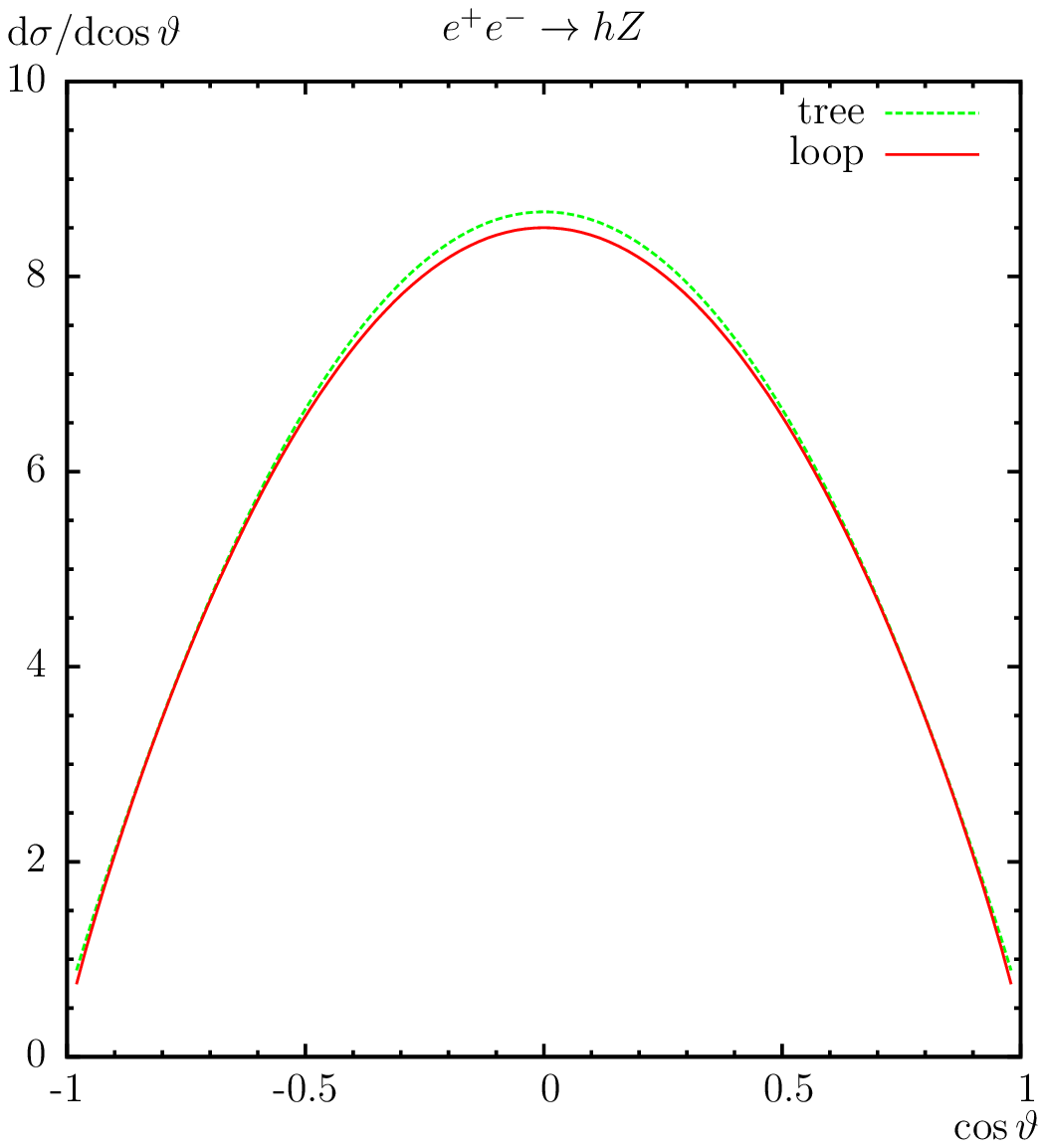}
\includegraphics[width=0.48\textwidth,height=6cm]{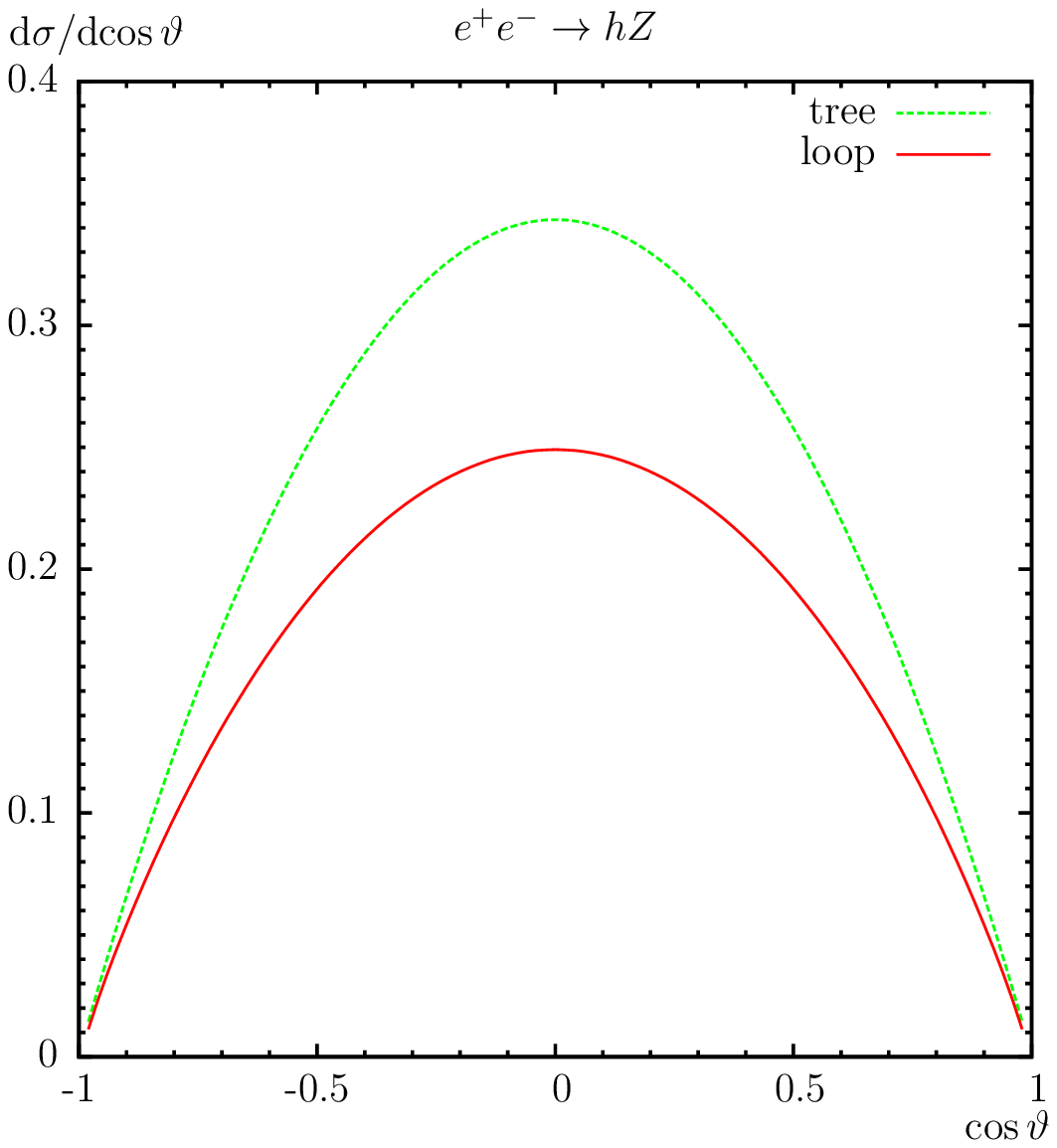}
\\[1em]
\includegraphics[width=0.48\textwidth,height=6cm]{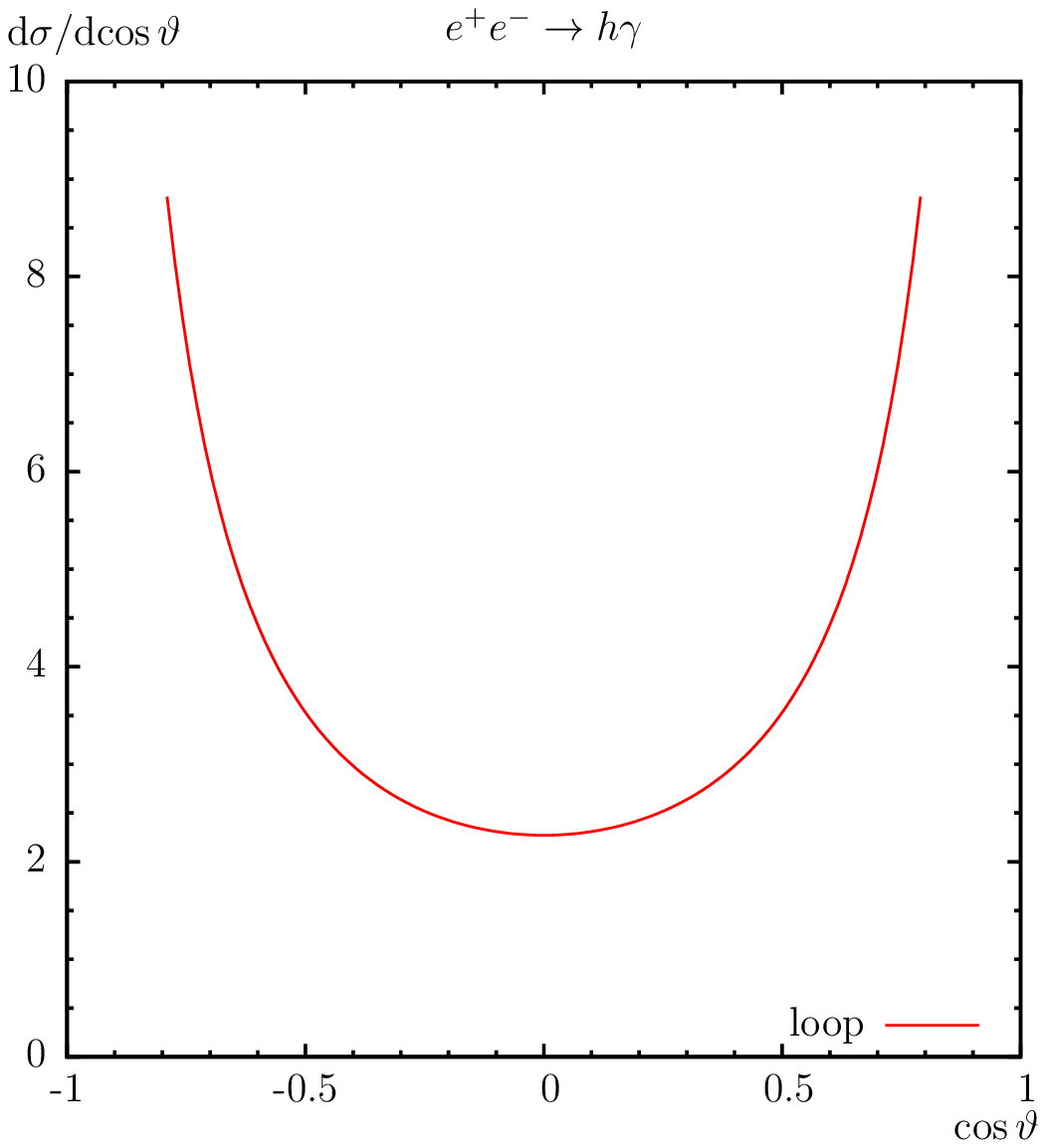}
\includegraphics[width=0.48\textwidth,height=6cm]{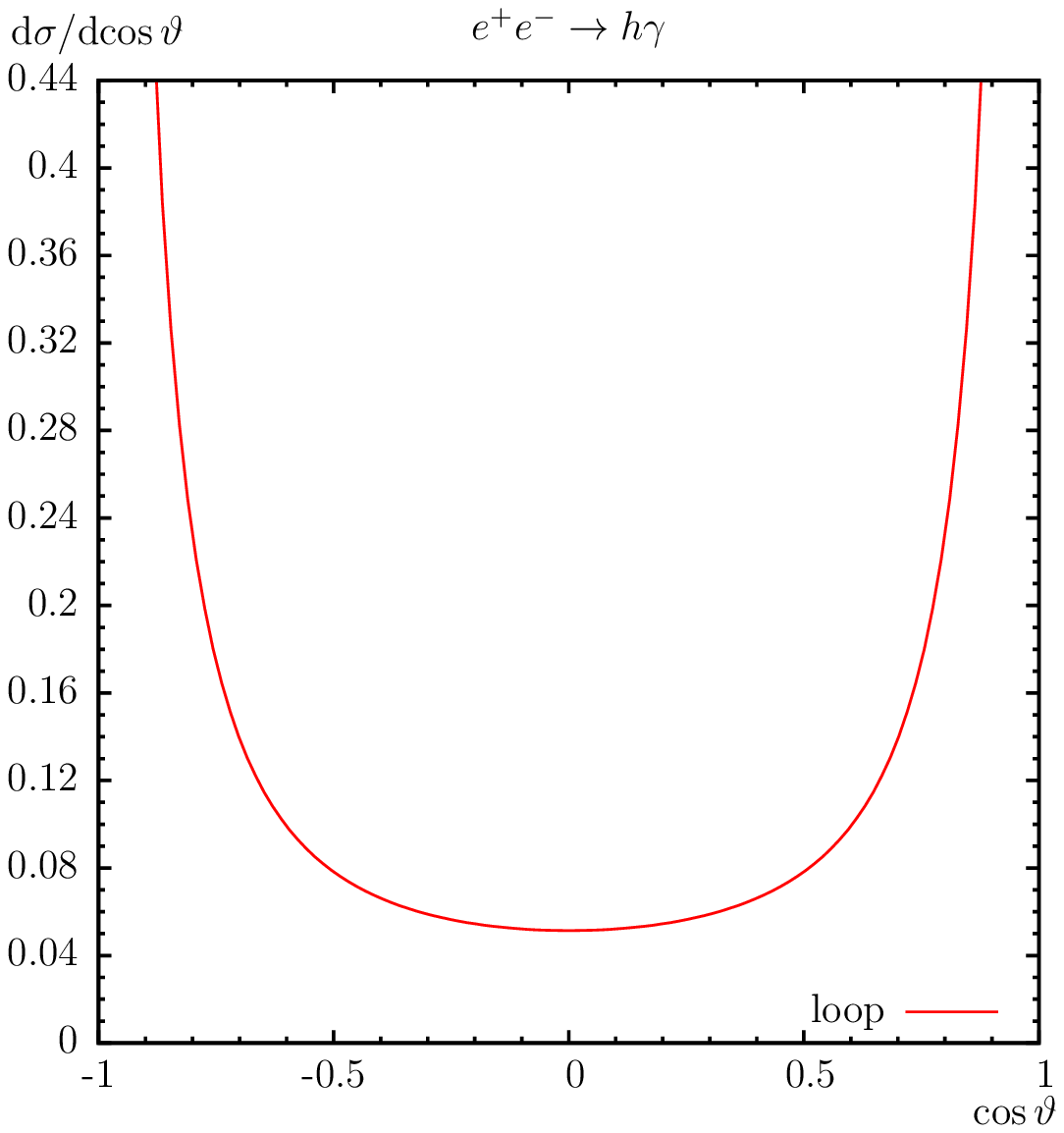}
\end{tabular}
\caption{\label{fig:supersim}
  $\sigma(e^+e^- \to h Z, h \gamma)$.
  The left (right) plots show the differential cross section with 
  $\sqrt{s} = 1\tev$ ($5\tev$) and $\cos\vartheta$ varied.
  Upper row: Tree-level and one-loop corrected differential cross sections 
  (in fb) are shown with parameters chosen according to S1 of 
  \citere{supersimple1}.
  Lower row: Loop induced differential cross sections (in ab) are shown 
  with parameters chosen according to S1 of \citere{supersimple2}.
}
\end{center}
\end{figure}

\begin{figure}[t!]
\begin{center}
\begin{tabular}{c}
\includegraphics[width=0.48\textwidth,height=6cm]{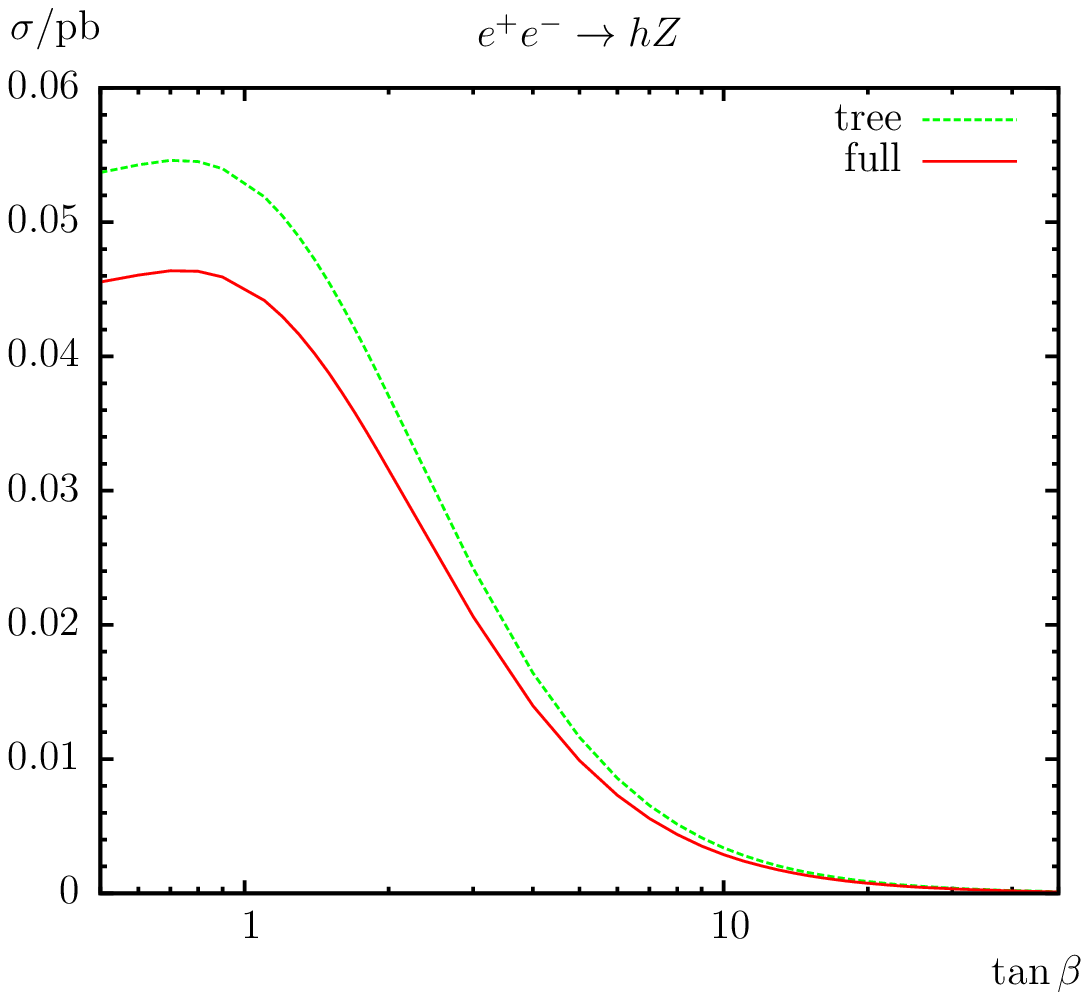}
\includegraphics[width=0.48\textwidth,height=6cm]{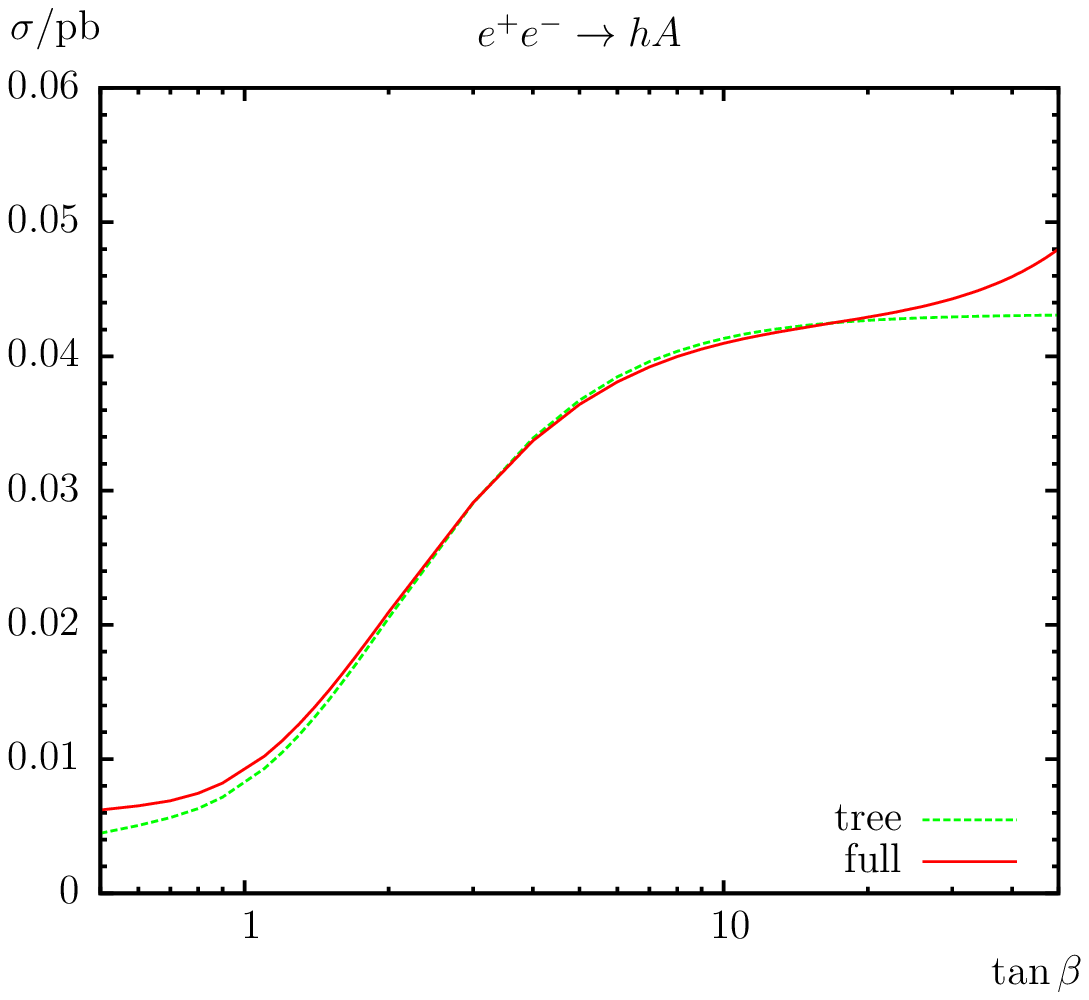}
\\[1em]
\includegraphics[width=0.48\textwidth,height=6cm]{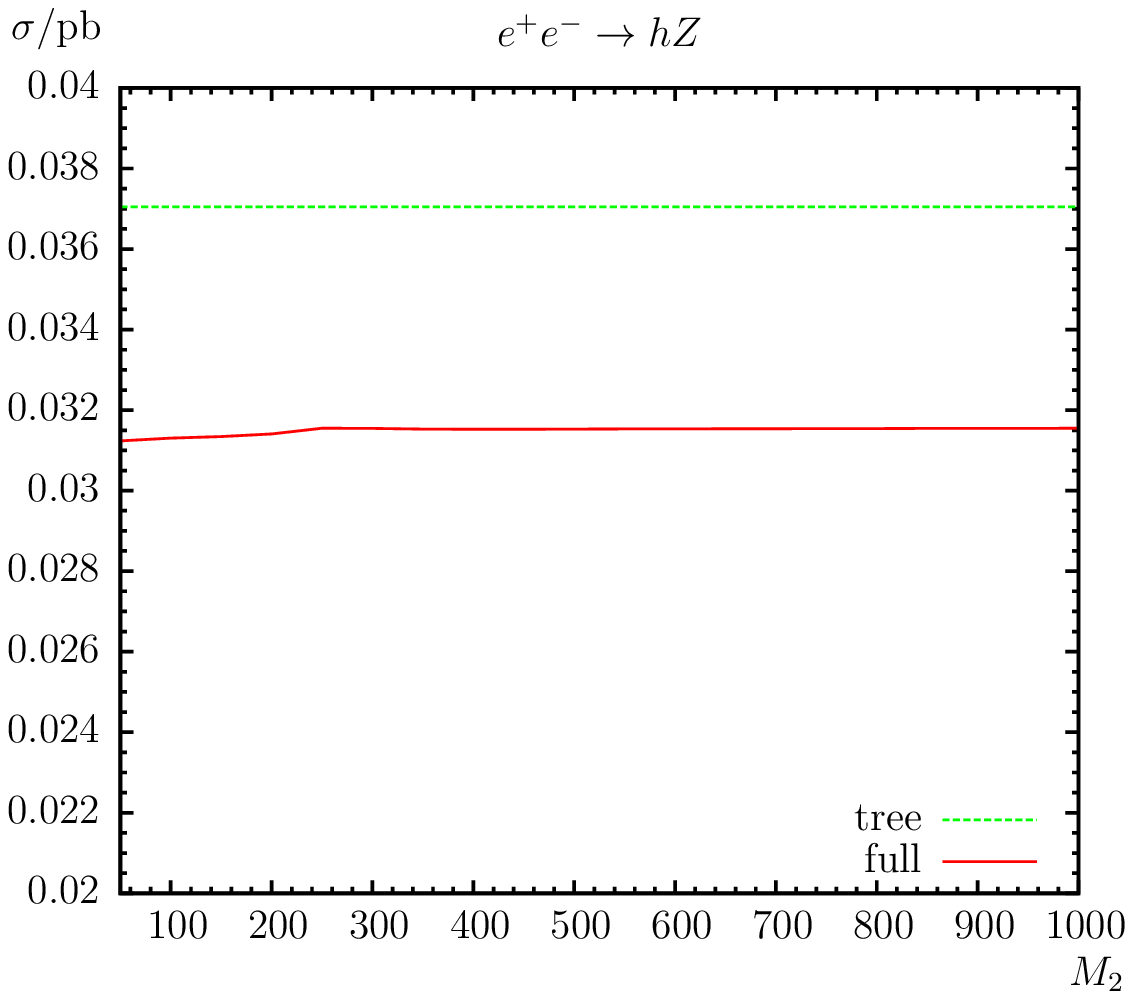}
\includegraphics[width=0.48\textwidth,height=6cm]{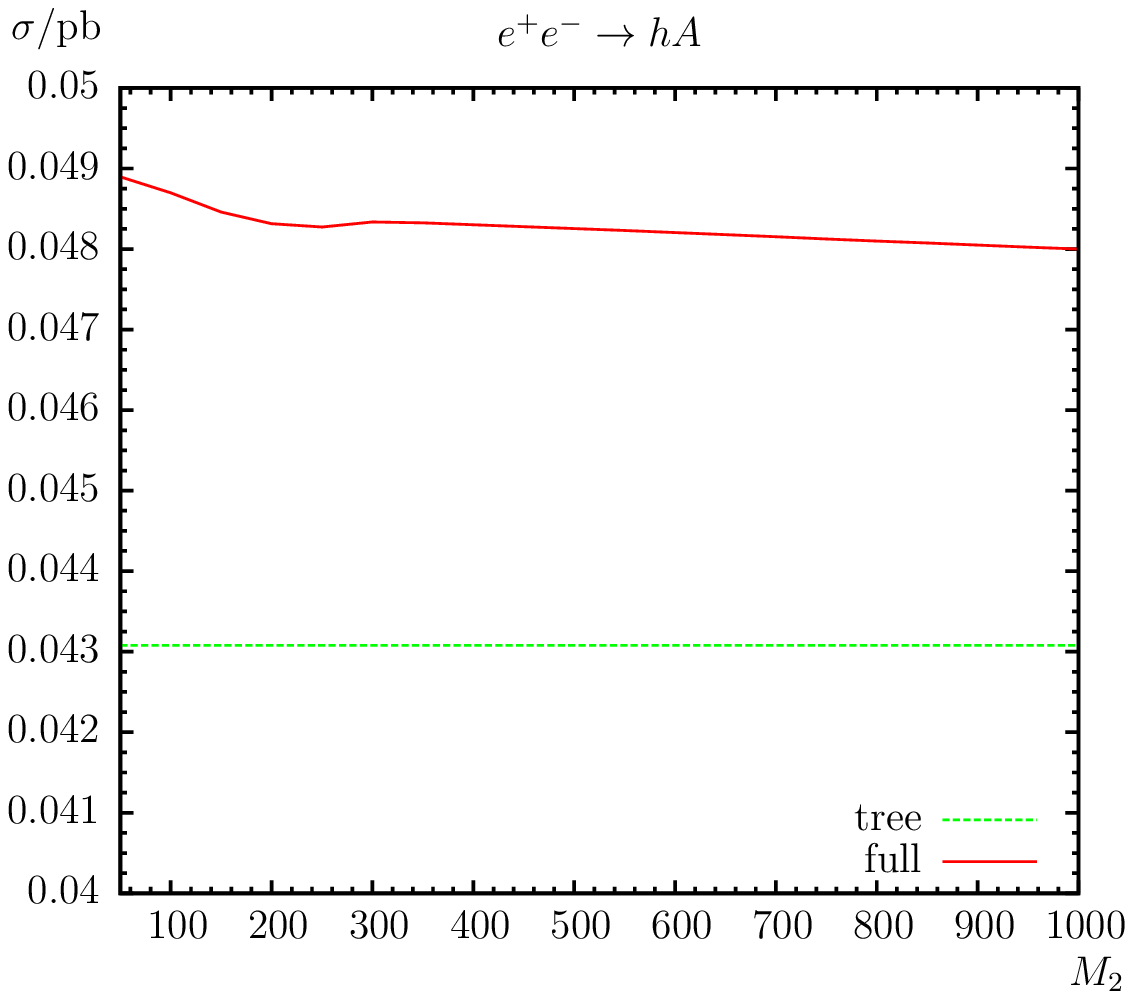}
\end{tabular}
\caption{\label{fig:fda}
  $\sigma(e^+e^- \to h Z, h A)$.
  One-loop corrected cross sections (in fb) are shown with parameters 
  chosen according to \citere{DrHoRo1996}.
  The upper left (right) plot shows the cross section for $e^+e^- \to h Z$
  ($e^+e^- \to h A$) with $\TB$ varied at $\sqrt{s} = 500\gev$.
  The lower left (right) plot shows the cross section for $e^+e^- \to h Z$
  ($e^+e^- \to h A$) with $M_2$ varied at $\sqrt{s} = 500\gev$.}
\end{center}
\end{figure}

\item
In \citeres{supersimple1,supersimple2} ``supersimple'' expressions have 
been derived for the processes $e^+e^- \to hZ, h\gamma$ in the rMSSM.
We successfully reproduced Fig.~4 (right panels) of \citere{supersimple1} in
the upper plots of our \reffi{fig:supersim} and Fig.~5 (right panels) of 
\citere{supersimple2} in the lower plots of our \reffi{fig:supersim}. 
As input parameters we used their (SUSY) parameter set S1. 
The small differences in the differential cross sections are caused by the 
SM input parameters (where we have used our parameters; 
see \refse{sec:paraset} below) and the slightly different renormalization 
schemes and treatment of the Higgs boson masses.

\item
$e^+e^- \to hZ$ at the full one-loop level (including hard and soft photon
bremsstrahlung) in the rMSSM has been analyzed in \citere{HiggsProd-1L}.
They also used \FA, \FC\ and \LT\ to generate and simplify their code. 
Unfortunately no numbers are given in \citere{HiggsProd-1L}, but only 
two-dimensional parameter scan plots, which we could not reasonably 
compare to our results.  Consequently, we omitted a comparison with 
\citere{HiggsProd-1L}.

\item
The Higgsstrahlung process $e^+e^- \to hZ$ with the expected leading
corrections in ``Natural SUSY'' models (\ie a one-loop calculation with 
third-generation (s)quarks) has been computed in \citere{Higgsstrahlung}.
They also used \FA, \FC\ and \LT\ to generate and simplify their code.
Unfortunately, again no numbers are given in \citere{Higgsstrahlung}, 
but mostly two-dimensional parameter scan plots, which we could not 
reasonably compare to our results.  Only in the left plot of their 
Fig.~4 they show (fractional) corrections to the Higgsstrahlung cross 
section.  However, the MSSM input parameters are not given in detail,
rendering a comparison again impossible.

\item
In \citere{DrHoRo1996} the processes $e^+e^- \to hZ, hA$ are computed
within a complete one-loop calculation.  Only the QED (including photon 
bremsstrahlung) has been neglected.  We used their input parameters
as far as possible and (more or less successfully) reproduced Fig.~5 and 
Fig.~6 (upper rows, solid lines) of \citere{DrHoRo1996} qualitatively in 
our \reffi{fig:fda}.  The differences are mainly due to different Higgs 
boson masses and the use of Higgs boson wave function corrections in
\citere{DrHoRo1996}, while we used an effective mixing angle
$\al_{\text{eff}}$.  In order to facilitate the comparison we used the same 
simple formulas for our Higgs boson masses and $\al_{\text{eff}}$ as in 
their Eqs.~(4)-(7). Therefore our $\sigma_{\text{tree}}$ correspond rather 
to their $\sigma^{\epsilon}$ and our $\sigma_{\text{full}}$ rather to their 
$\sigma^{\text{FDC}}$. 
It should be noted that the code of \citere{DrHoRo1996} is also part 
of the code from \citere{HiggsProd-org}.  Using \FHXS\ with the input 
parameters of \citere{DrHoRo1996} (as far as possible) gave also only 
qualitative agreement with the Figs. of \citere{DrHoRo1996}.

\begin{figure}[tp]
\begin{center}
\begin{tabular}{c}
\includegraphics[width=0.48\textwidth,height=6cm]{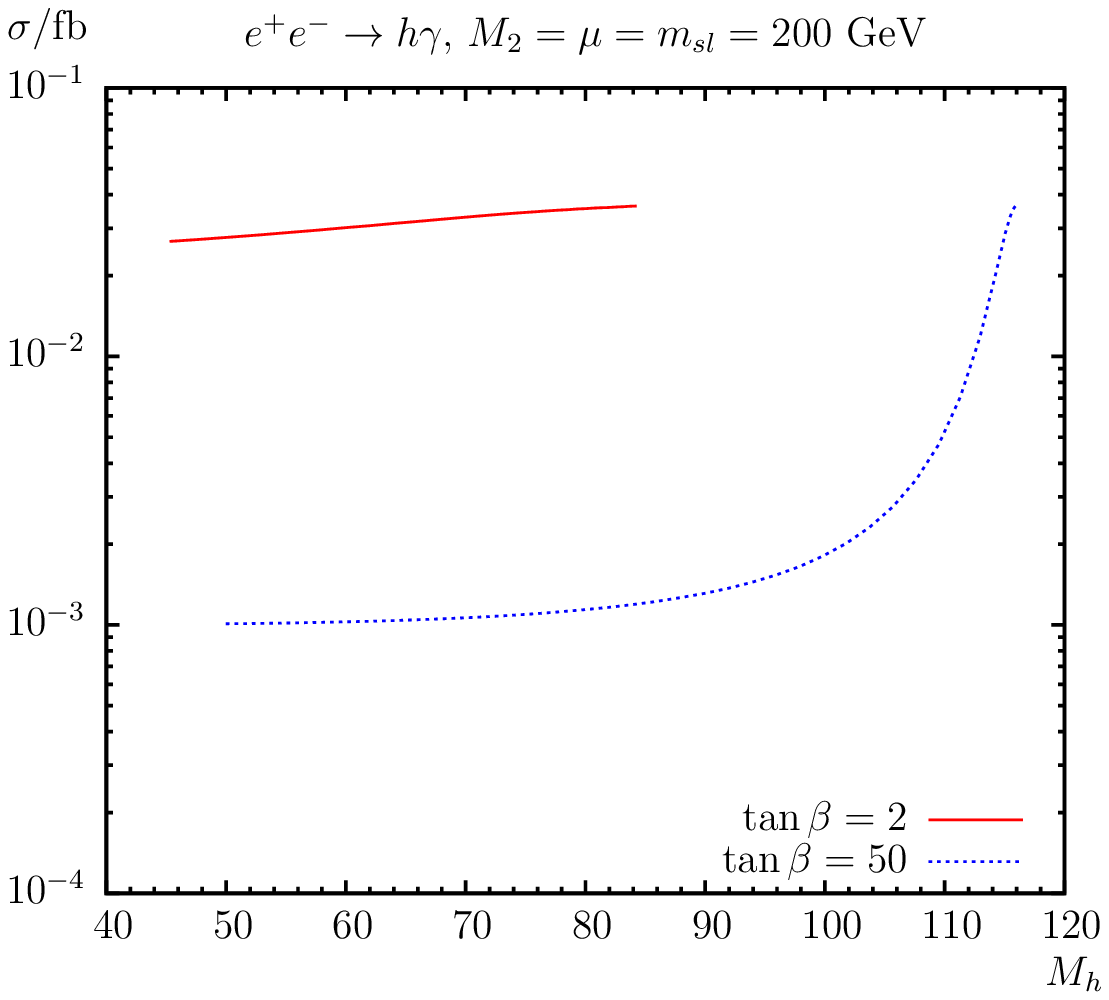}
\includegraphics[width=0.48\textwidth,height=6cm]{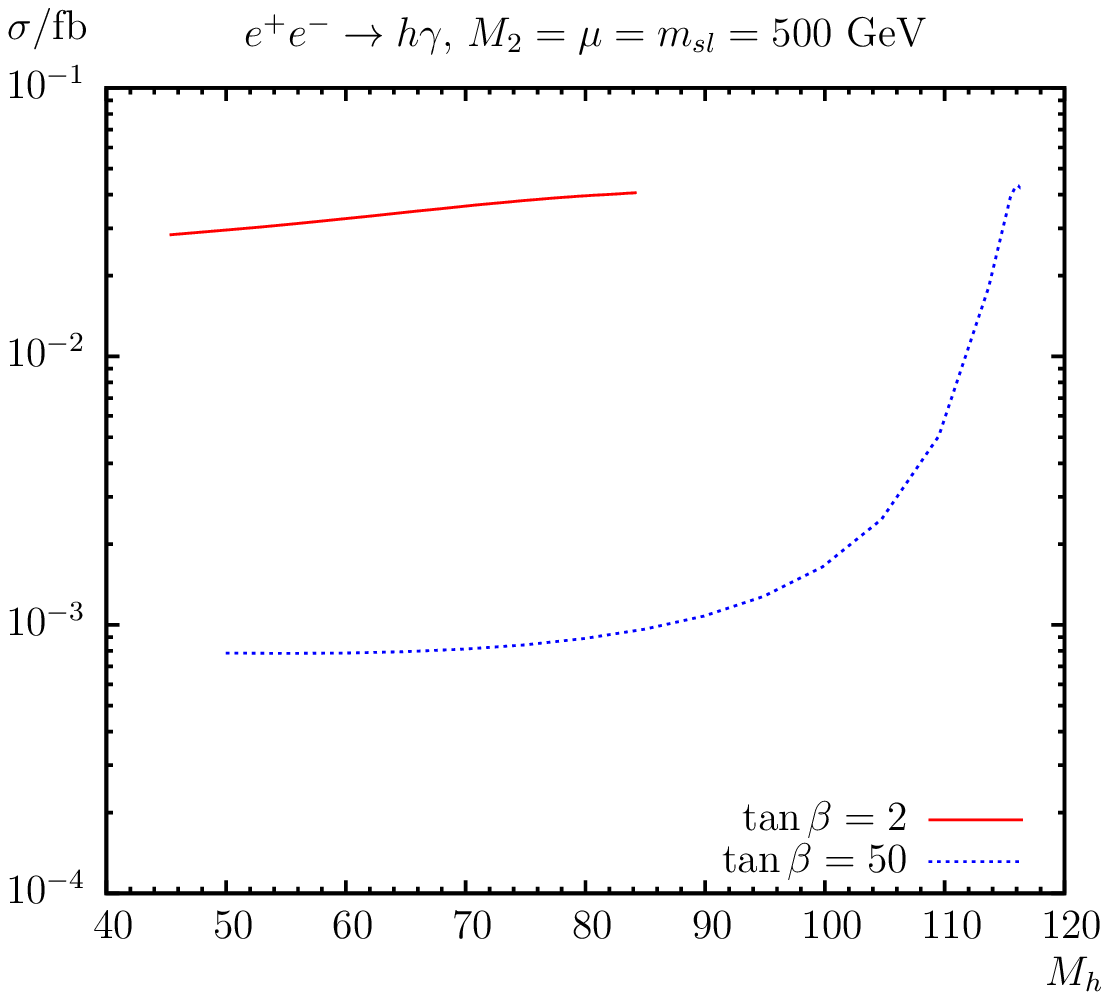}
\\[1em]
\includegraphics[width=0.48\textwidth,height=6cm]{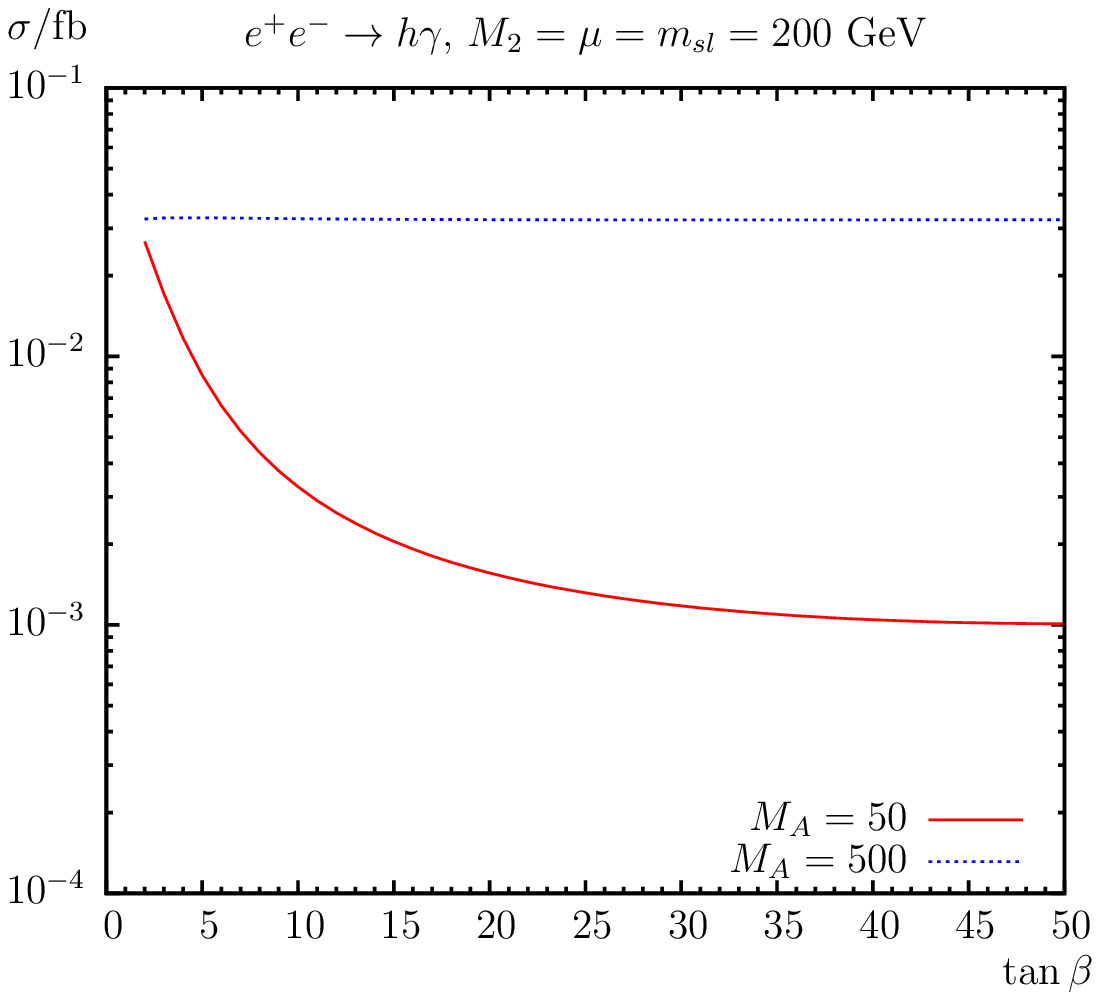}
\includegraphics[width=0.48\textwidth,height=6cm]{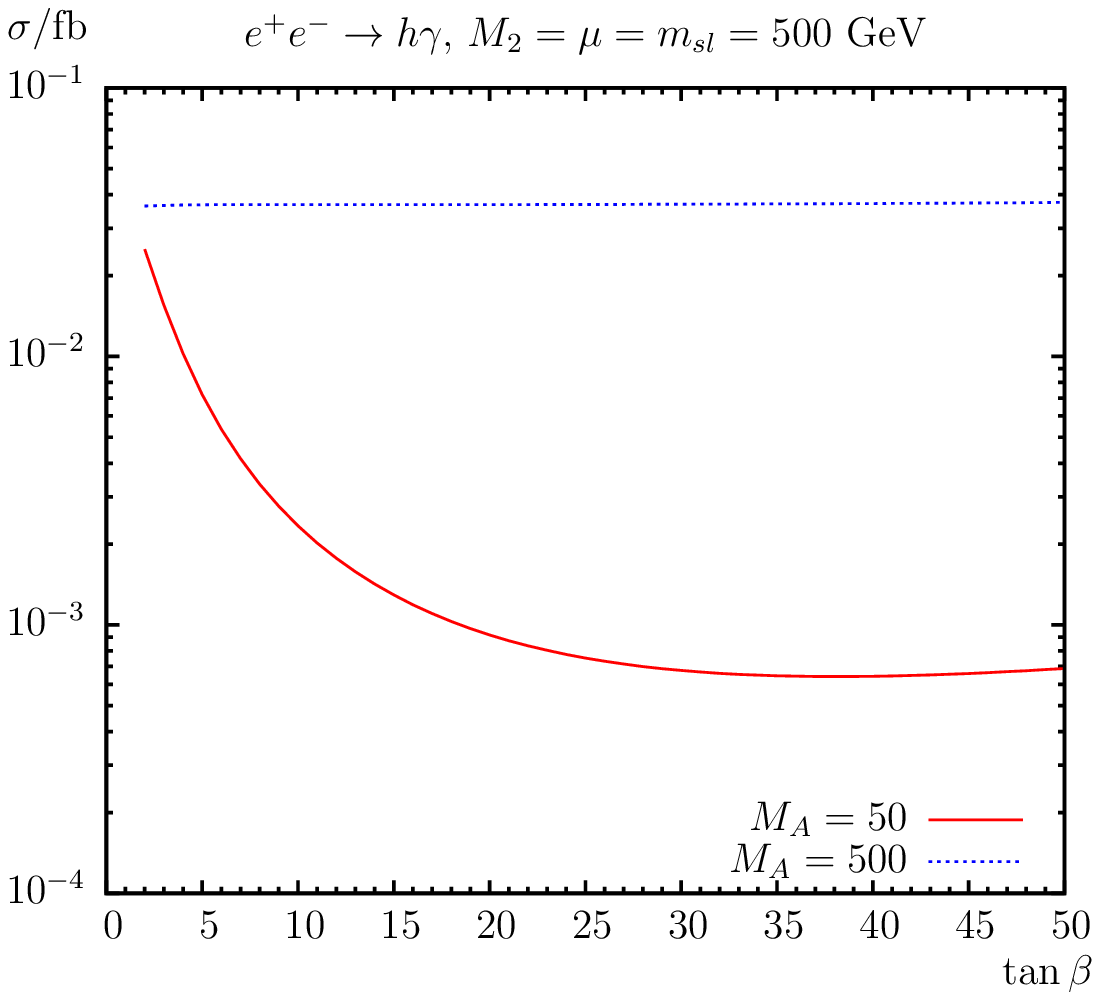}
\\[1em]
\includegraphics[width=0.48\textwidth,height=6cm]{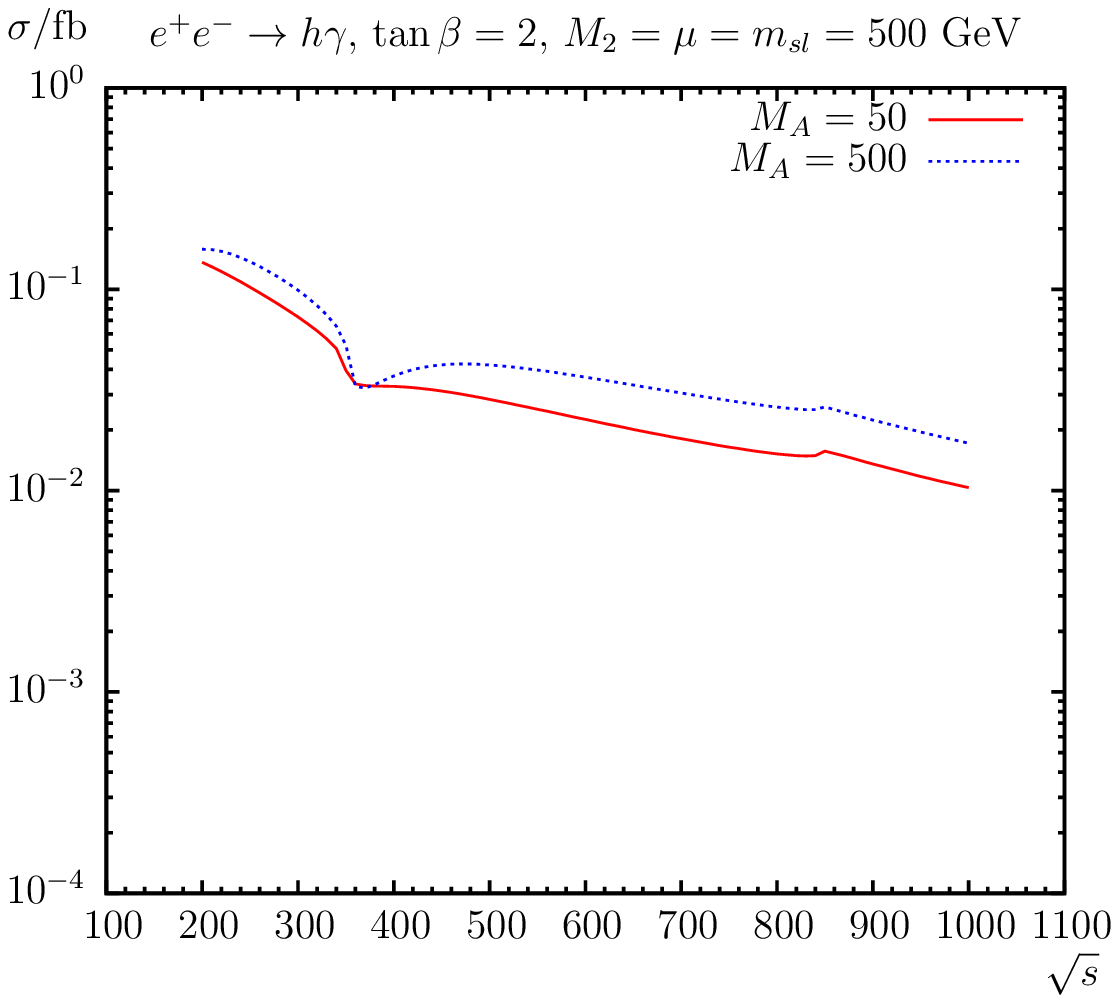}
\includegraphics[width=0.48\textwidth,height=6cm]{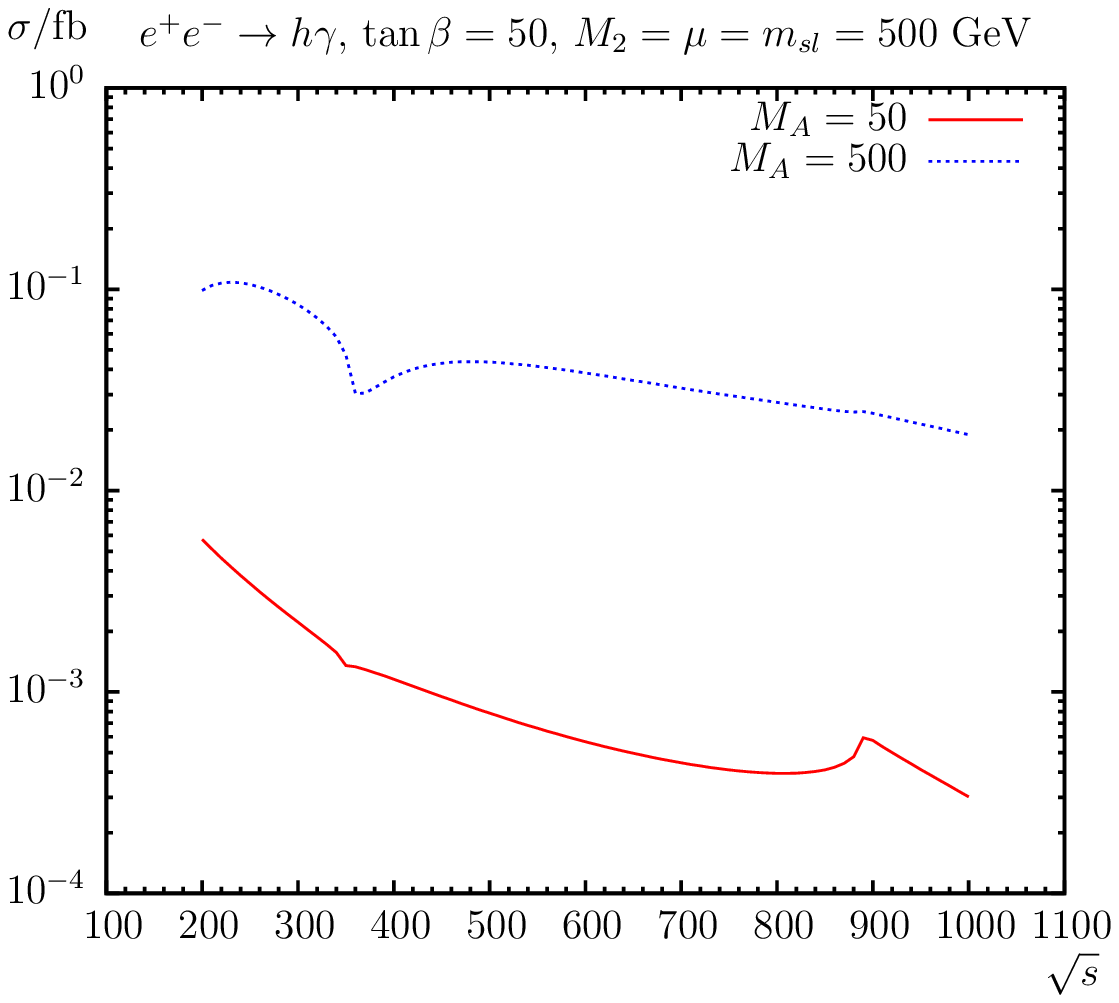}
\end{tabular}
\caption{\label{fig:hHAga}
  $\sigma(e^+e^- \to h \ga)$.
  Loop induced cross sections (in fb) are shown with parameters 
  chosen according to \citere{DjDrHoRo1997}.
  The upper plots show the cross section with $\Mh$ varied at $\sqrt{s} = 500$.
  The middle plots show the cross section for $\TB$ varied at $\sqrt{s} = 500$.
  The lower plots show the cross section for $\sqrt{s}$ varied.}
\end{center}
\end{figure}

\item
In \citere{DjDrHoRo1997} the loop induced processes 
$e^+e^- \to h\gamma, H\gamma, A\gamma$ have been computed.  We used the same 
simple formulas for our Higgs boson masses and $\alpha_{\text{eff}}$ as
in their Eqs. (3.48)-(3.50). We also used their input parameters as far as 
possible, but unfortunately they forgot to specify the trilinear parameters 
$A_f$.  Therefore we chose arbitrarily $A_f = 0$ for our comparison.  In view 
of this problem the comparison is acceptable; see our \reffi{fig:hHAga} vs. 
Fig.~4, Fig.~5 and Fig.~7 of \citere{DjDrHoRo1997}.  It should be noted 
that the code of \citere{DjDrHoRo1997} is also part of the code from 
\citere{HiggsProd-org} and \citere{DrHoRo1996}.

\end{itemize}

\noindent
A final comment is in order.  We argue that the problems in the comparison 
with \citere{HiggsProd-org} (\ie \FHXS), \citere{DrHoRo1996} and 
\citere{DjDrHoRo1997} are due to the fact that all three papers are based 
(effectively) on the same calculation/source.  
Therefore, these three papers should be considered as one rather than
three independent comparisons, and thus do not disprove the reliability 
of our calculation.  It should also be kept in mind that our calculational 
method/code has already been successfully tested and compared with quite 
a few other programs, see  
\citeres{HiggsDecaySferm,HiggsDecayIno,MSSMCT,SbotRen,Stop2decay,%
Gluinodecay,Stau2decay,LHCxC,LHCxN}.


\section{Numerical analysis}
\label{sec:numeval}

In this section we present our numerical analysis of neutral Higgs boson 
production at $e^+e^-$ colliders in the cMSSM. 
In the various figures below we show the cross sections at the tree-level 
(``tree'') and at the full one-loop level (``full'').  
In case of extremely small tree-level cross sections 
we also show results including the corresponding purely loop induced 
contributions (``loop'').  These leading two-loop contributions are 
$\propto |\cMl|^2$, where $\cMl$ denotes the one-loop matrix element of the 
appropriate process.


\subsection{Parameter settings}
\label{sec:paraset}

The renormalization scale $\mu_R$ has been set to the center-of-mass energy, 
$\sqrt{s}$.  The SM parameters are chosen as follows; see also \cite{pdg}:
\begin{itemize}

\item Fermion masses (on-shell masses, if not indicated differently):
\begin{align}
m_e    &= 0.510998928\mev\,, & m_{\nu_e}    &= 0\,, \notag \\
m_\mu  &= 105.65837515\mev\,, & m_{\nu_{\mu}} &= 0\,, \notag \\
m_\tau &= 1776.82\mev\,,      & m_{\nu_{\tau}} &= 0\,, \notag \\
m_u &= 68.7\mev\,,           & m_d         &= 68.7\mev\,, \notag \\ 
m_c &= 1.275\gev\,,          & m_s         &= 95.0\mev\,, \notag \\
m_t &= 173.21\gev\,,         & m_b         &= 4.18\gev\,.
\end{align}
According to \citere{pdg}, $m_s$ is an estimate of a so-called 
"current quark mass" in the \MSbar\ scheme at the scale 
$\mu \approx 2\gev$.  $m_c \equiv m_c(m_c)$ and $m_b \equiv m_b(m_b)$ 
are the "running" masses in the \MSbar\ scheme.  $m_u$ and $m_d$ are 
effective parameters, calculated through the hadronic contributions to
\begin{align}
\Delta\alpha_{\text{had}}^{(5)}(M_Z) &= 
      \frac{\alpha}{\pi}\sum_{f = u,c,d,s,b}
      Q_f^2 \Bigl(\ln\frac{M_Z^2}{m_f^2} - \frac 53\Bigr) \approx 0.027723\,.
\end{align}

\item Gauge boson masses\index{gaugebosonmasses}:
\begin{align}
M_Z = 91.1876\gev\,, \qquad M_W = 80.385\gev\,.
\end{align}

\item Coupling constant\index{couplingconstants}:
\begin{align}
\alpha(0) = 1/137.0359895\,.
\end{align}
\end{itemize}

The Higgs sector quantities (masses, mixings, $\hat{Z}$~factors, etc.) 
have been evaluated using \FH\ (version 2.11.0)
\cite{feynhiggs,mhiggslong,mhiggsAEC,mhcMSSMlong,Mh-logresum}.

\begin{table}[t!]
\caption{\label{tab:para}
  MSSM default parameters for the numerical investigation; all parameters 
  (except of $\TB$) are in GeV (calculated masses are rounded to 1 MeV). 
  The values for the trilinear sfermion Higgs couplings, $A_{t,b,\tau}$ are 
  chosen such that charge- and/or color-breaking minima are avoided 
  \cite{ccb}, and $A_{b,\tau}$ are chosen to be real.  It should be noted 
  that for the first and second generation of sfermions we chose instead 
  $A_f = 0$, $M_{\tilde Q, \tilde U, \tilde D} = 1500\gev$ and 
  $M_{\tilde L, \tilde E} = 500\gev$.
}
\centering
\begin{tabular}{lrrrrrrrrrr}
\toprule
Scen. & $\sqrt{s}$ & $\TB$ & $\mu$ & $\MHp$ & $M_{\tilde Q, \tilde U, \tilde D}$ & 
$M_{\tilde L, \tilde E}$ & $|A_{t,b,\tau}|$ & $M_1$ & $M_2$ & $M_3$ \\ 
\midrule
\Scs & 1000 & 7 & 200 & 300 & 1000 & 500 & $1500 + \mu/\TB$ & 100 & 200 & 1500 \\
\bottomrule
\end{tabular}

\vspace{0.5em}

\begin{tabular}{rrr}
\toprule
$\mh1$  & $\mh2$  & $\mh3$ \\
\midrule
123.404 & 288.762 & 290.588 \\
\bottomrule
\end{tabular}
\end{table}

The SUSY parameters are chosen according to the scenario \Scs, shown in 
\refta{tab:para}, unless otherwise noted. 
This scenario constitutes a viable scenario for the various cMSSM Higgs
production modes, \ie not picking specific parameters for each cross 
section.  The only variation will be the choice of $\sqrt{s} = 500\gev$ 
for production cross sections involving the light Higgs boson.%
\footnote{
  In a recent re-evaluation of ILC running strategies the first stage 
  was advocated to be at $\sqrt{s} = 500\gev$~\cite{ILCstages}.
}
This will be clearly indicated below.  We do not strictly demand that 
the lightest Higgs boson has a mass around $\sim 125\gev$, although 
for most of the parameter space this is given.  We will show the 
variation with $\sqrt{s}$, $\MHp$, $\TB$ and $\phiAt$, the phase of $\At$.

Concerning the complex parameters, some more comments are in order.
No complex parameter enters into the tree-level production cross
sections. Therefore, the largest effects are expected from the complex
phases entering via the $t/\stop$~sector, i.e.\ from $\phiAt$,
motivating our choice of $\phiAt$ as parameter to be varied.
Here the following should be kept in mind.
When performing an analysis involving complex parameters it should be 
noted that the results for physical observables are affected only by 
certain combinations of the complex phases of the parameters $\mu$, 
the trilinear couplings $A_f$ and the gaugino mass parameters 
$M_{1,2,3}$~\cite{MSSMcomplphasen,SUSYphases}.
It is possible, for instance, to rotate the phase $\phiMz$ away.
Experimental constraints on the (combinations of) complex phases 
arise, in particular, from their contributions to electric dipole 
moments of the electron and the neutron (see \citeres{EDMrev2,EDMPilaftsis} 
and references therein), of the deuteron~\cite{EDMRitz} and of heavy 
quarks~\cite{EDMDoink}.
While SM contributions enter only at the three-loop level, due to its
complex phases the MSSM can contribute already at one-loop order.
Large phases in the first two generations of sfermions can only be 
accommodated if these generations are assumed to be very heavy 
\cite{EDMheavy} or large cancellations occur~\cite{EDMmiracle};
see, however, the discussion in \citere{EDMrev1}. 
A review can be found in \citere{EDMrev3}.
Accordingly (using the convention that $\phiMz = 0$, as done in this paper), 
in particular, the phase $\phimu$ is tightly constrained~\cite{plehnix}, 
while the bounds on the phases of the third generation trilinear couplings 
are much weaker.  Setting $\phimu = 0$ and $\varphi_{A_{f \neq t}} = 0$ 
leaves us with $\phiAt$ as the only complex valued parameter. 

Since now the complex trilinear coupling $A_t$ can appear in the 
couplings, contributions from absorptive parts of self-energy type 
corrections on external legs can arise.  The corresponding formulas 
for an inclusion of these absorptive contributions via finite wave 
function correction factors can be found in \citeres{MSSMCT,Stop2decay}.

The numerical results shown in the next subsections are of course 
dependent on the choice of the SUSY parameters.  Nevertheless, they 
give an idea of the relevance of the full one-loop corrections.


\subsection{Full one-loop results for varying \boldmath{$\sqrt{s}$}, 
  \boldmath{$\MHp$}, \boldmath{$\TB$}, and \boldmath{$\phiAt$}}
\label{sec:full1L}

The results shown in this and the following subsections consist of 
``tree'', which denotes the tree-level value and of ``full'', which 
is the cross section including \textit{all} one-loop corrections 
as described in \refse{sec:calc}. 

We begin the numerical analysis with the cross sections of $\eehh$
($i,j = 1,2,3$) evaluated as a function of $\sqrt{s}$ (up to $3\tev$,
shown in the upper left plot of the respective figures), $\MHp$ 
(starting at $\MHp = 160\gev$ up to $\MHp = 500\gev$, shown in the 
upper right plots), $\TB$ (from 4 to 50, lower left plots) and 
$\phiAt$ (between $0^{\circ}$ and $360^{\circ}$, lower right plots).
Then we turn to the processes $\eehZ$ and $\eehga$ ($i = 1,2,3$). 
All these processes are of particular interest for ILC and CLIC 
analyses~\cite{ILC-TDR,teslatdr,ilc,CLIC} 
(as emphasized in \refse{sec:intro}).


\subsubsection{The process \boldmath{$\eehh$}}
\label{sec:eehh}

\begin{figure}[t!]
\begin{center}
\begin{tabular}{c}
\includegraphics[width=0.48\textwidth,height=6cm]{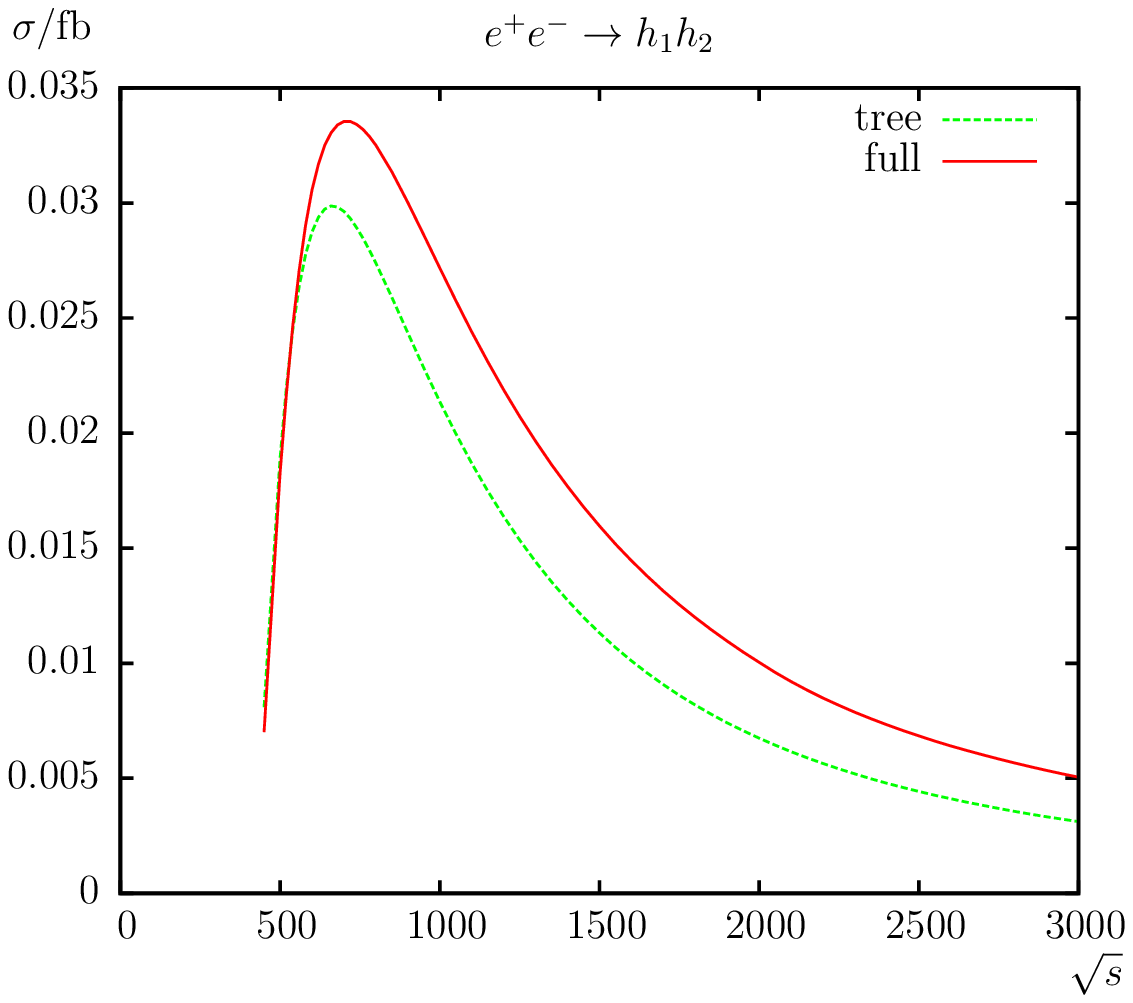}
\includegraphics[width=0.48\textwidth,height=6cm]{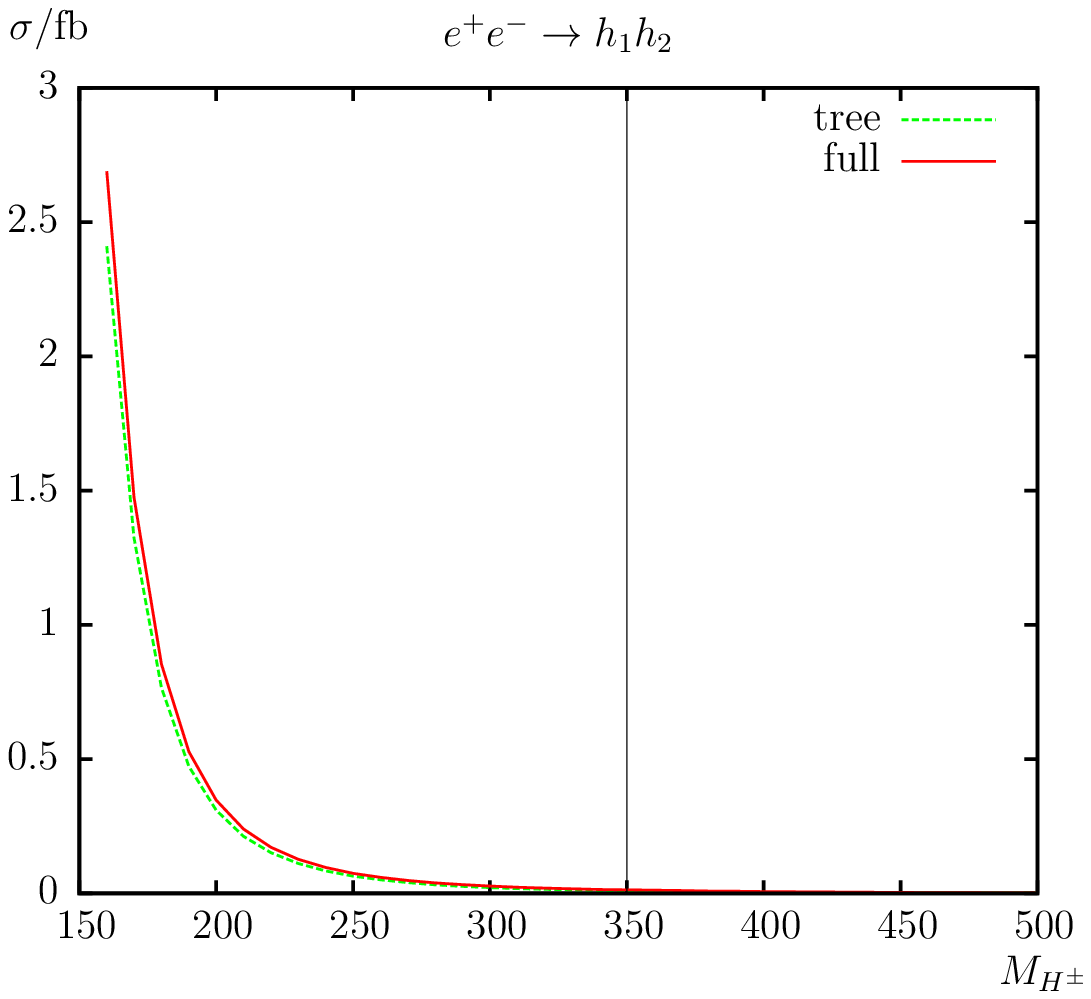}
\\[1em]
\includegraphics[width=0.48\textwidth,height=6cm]{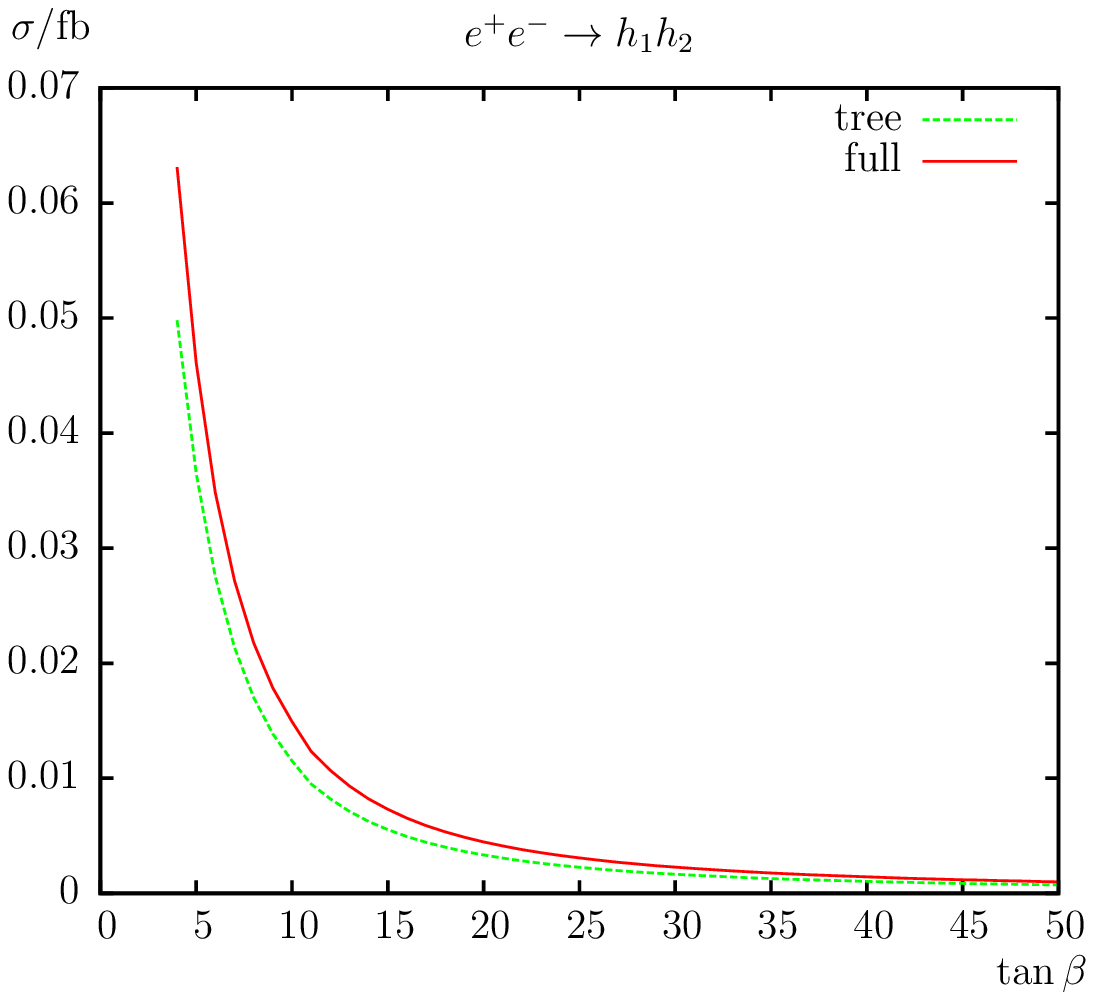}
\includegraphics[width=0.48\textwidth,height=6cm]{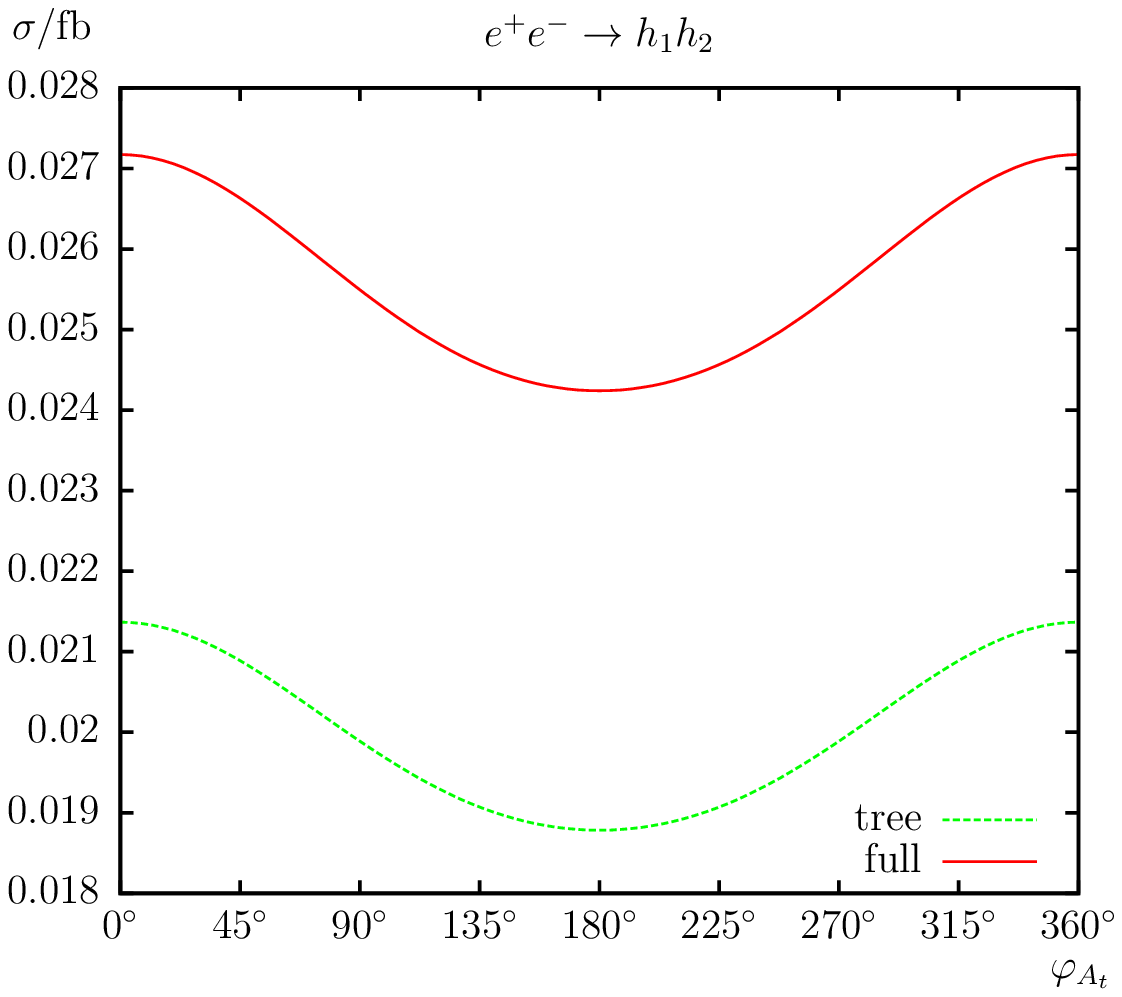}
\end{tabular}
\caption{\label{fig:eeh1h2}
  $\sigma(e^+e^- \to h_1 h_2)$.
  Tree-level and full one-loop corrected cross sections are shown  
  with parameters chosen according to \Scs; see \refta{tab:para}.
  The upper plots show the cross sections with $\sqrt{s}$ (left) 
  and $\MHp$ (right) varied;  the lower plots show $\TB$ (left) and 
  $\phiAt$ (right) varied.
}
\end{center}
\end{figure}

\begin{figure}[t!]
\begin{center}
\begin{tabular}{c}
\includegraphics[width=0.48\textwidth,height=6cm]{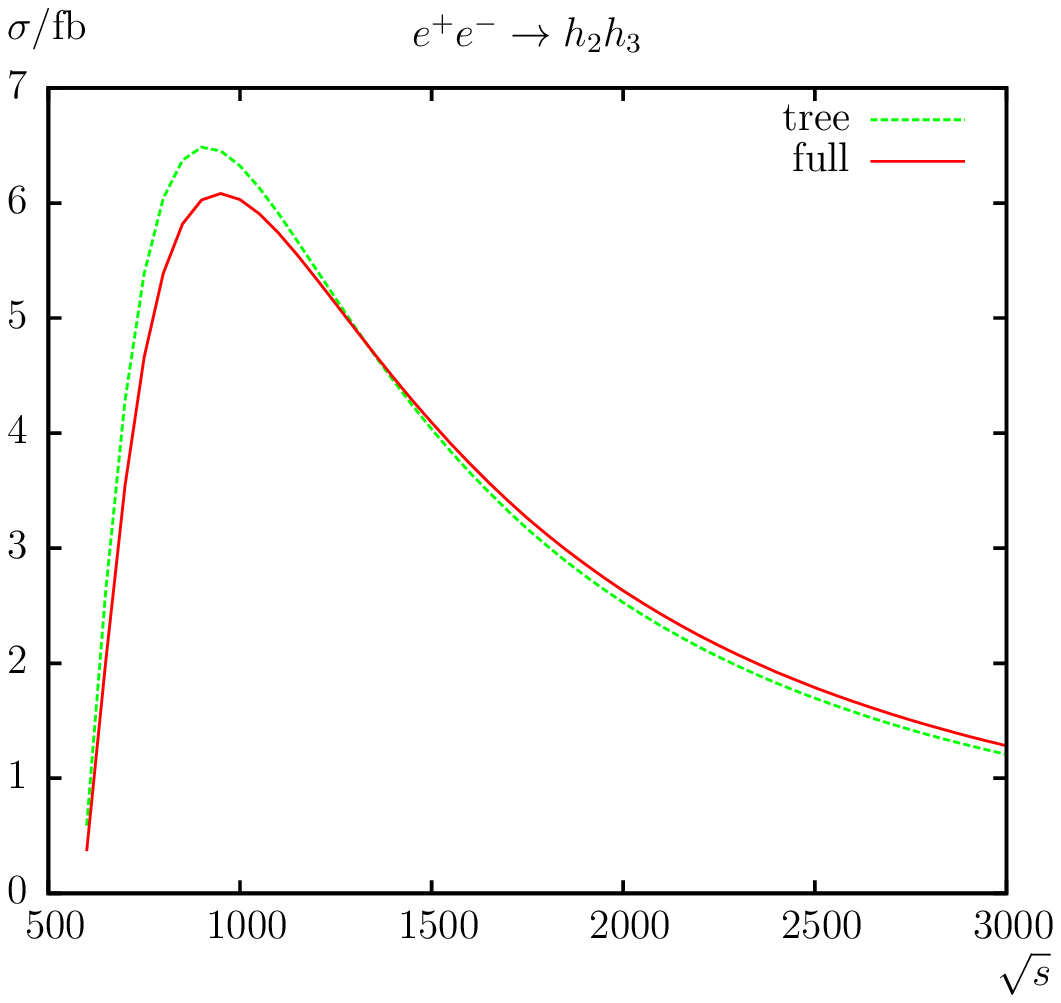}
\includegraphics[width=0.48\textwidth,height=6cm]{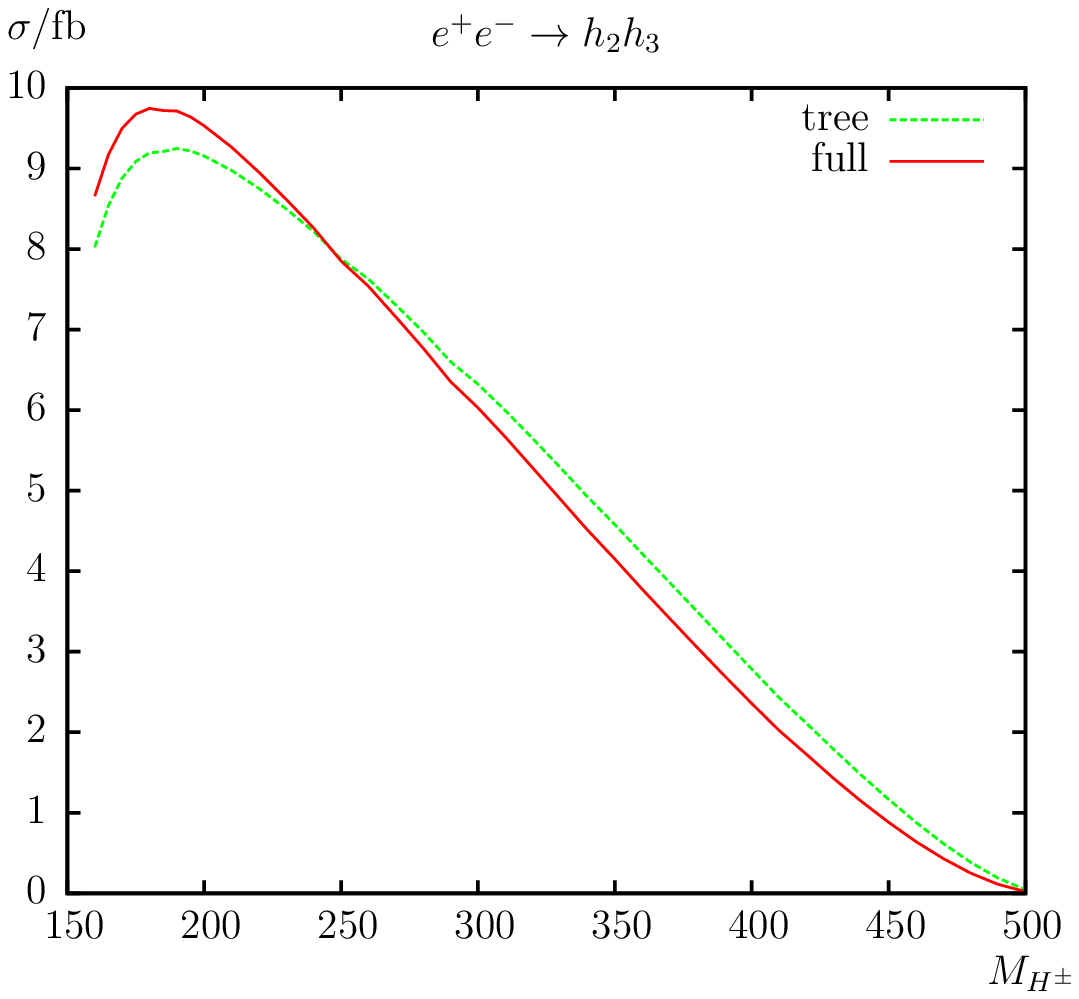}
\\[1em]
\includegraphics[width=0.48\textwidth,height=6cm]{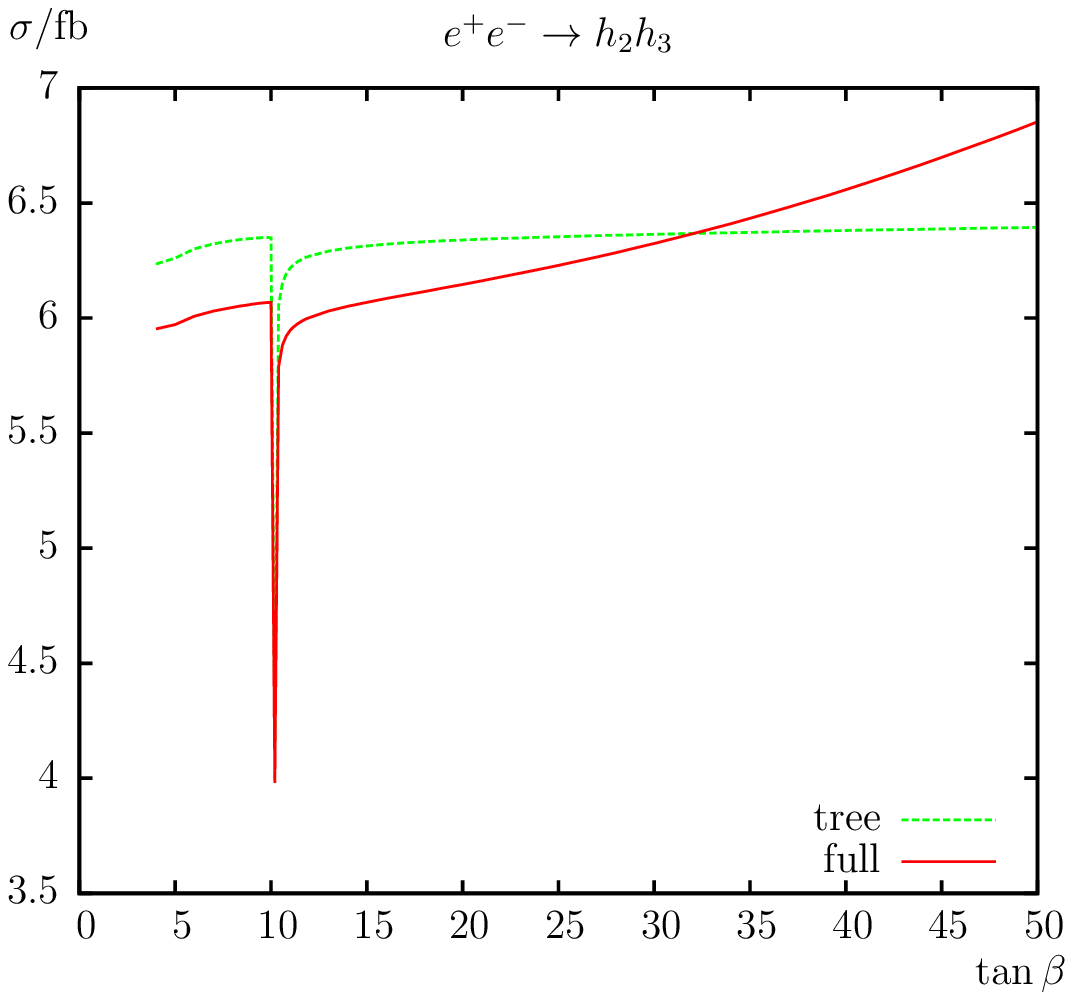}
\includegraphics[width=0.48\textwidth,height=6cm]{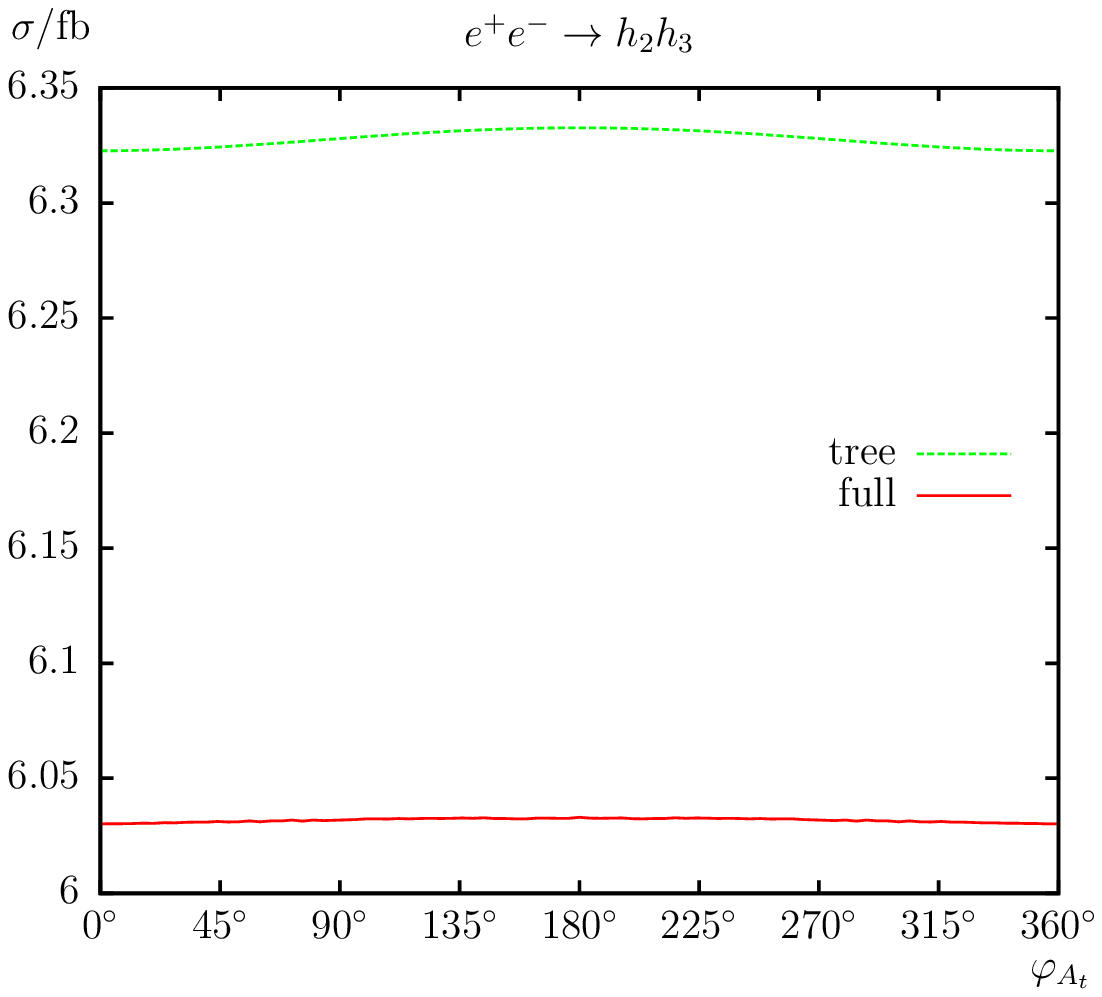}
\end{tabular}
\caption{\label{fig:eeh2h3}
  $\sigma(e^+e^- \to h_2 h_3)$.
  Tree-level and full one-loop corrected cross sections are shown  
  with parameters chosen according to \Scs; see \refta{tab:para}.
  The upper plots show the cross sections with $\sqrt{s}$ (left) 
  and $\MHp$ (right) varied;  the lower plots show $\TB$ (left) and 
  $\phiAt$ (right) varied.
}
\end{center}
\end{figure}

\begin{figure}[t!]
\begin{center}
\begin{tabular}{c}
\includegraphics[width=0.48\textwidth,height=6cm]{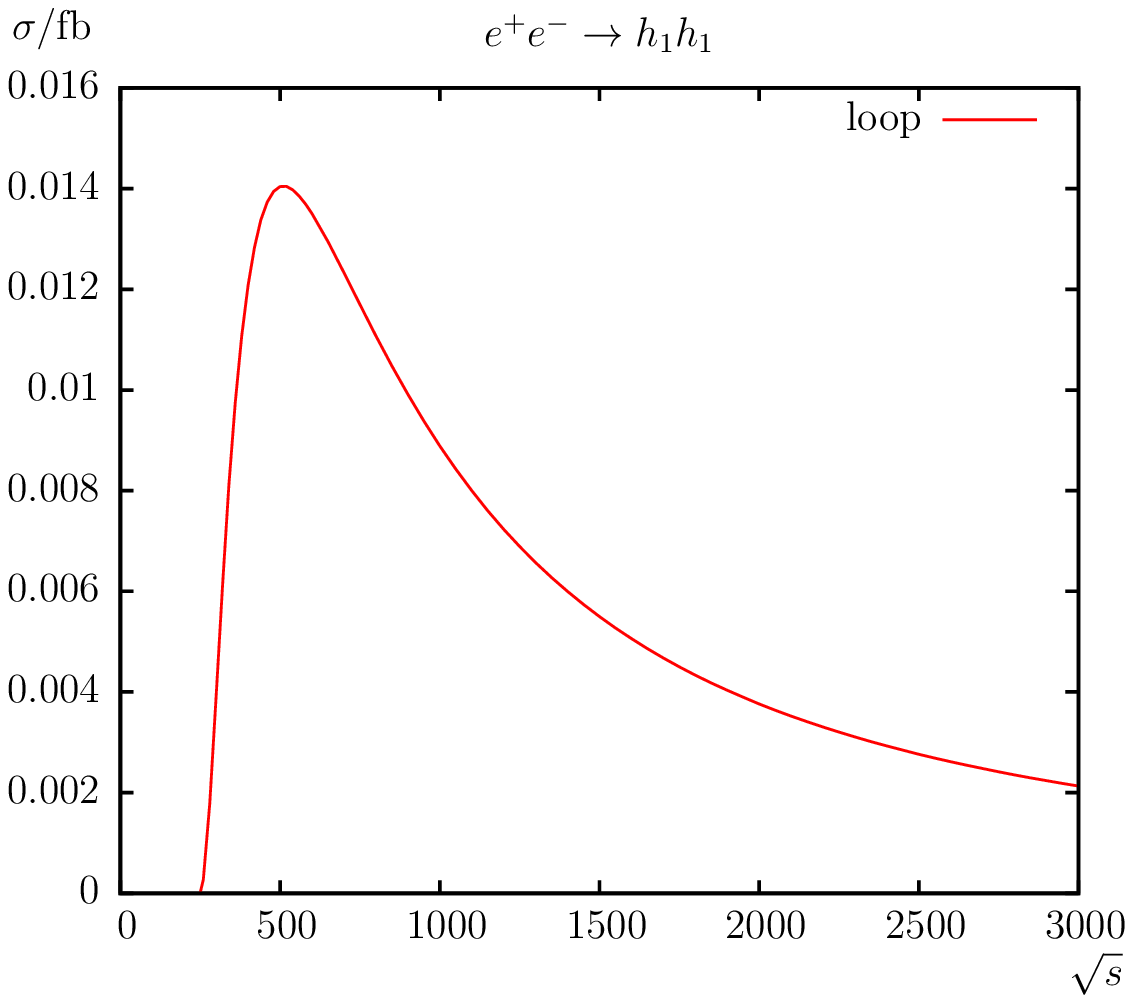}
\includegraphics[width=0.48\textwidth,height=6cm]{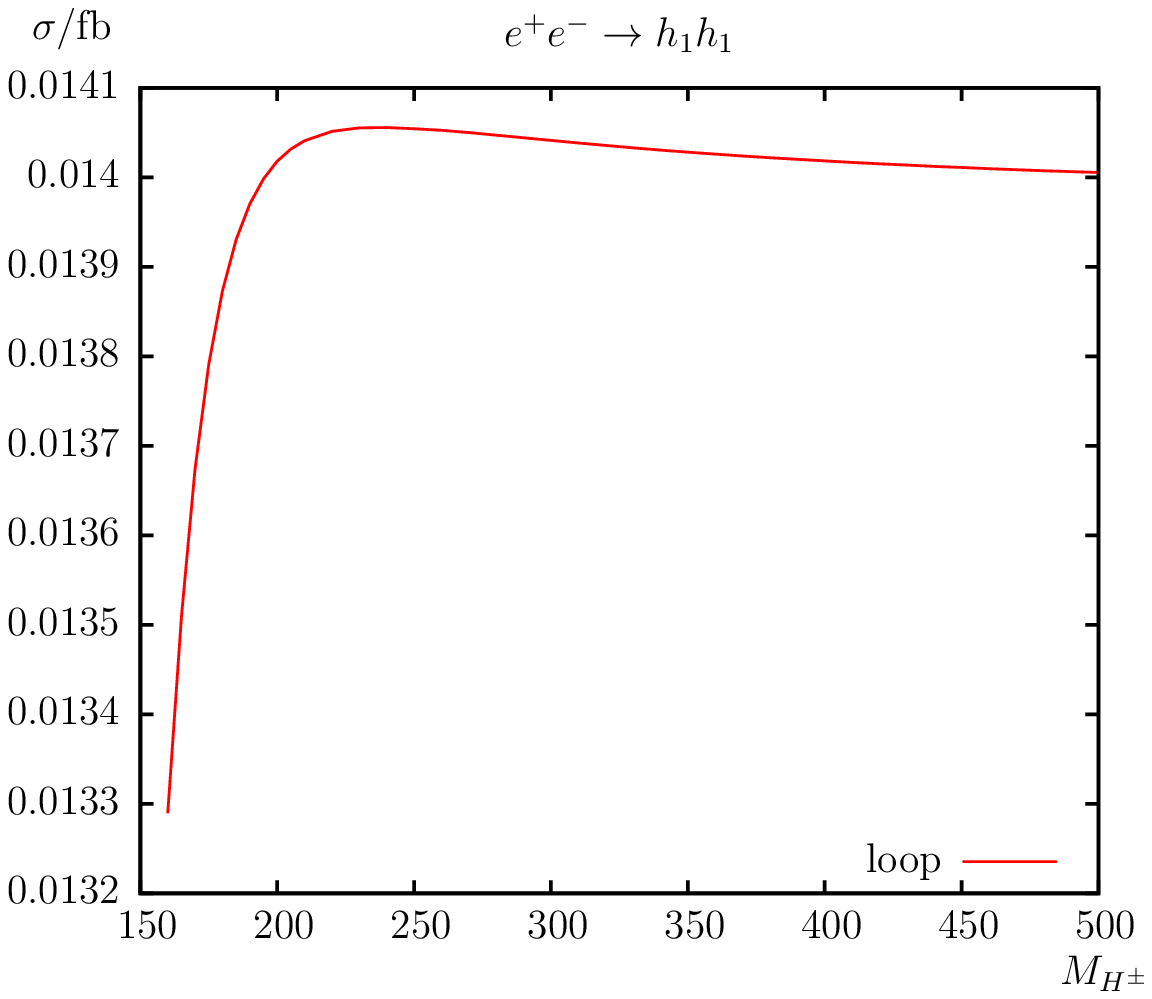}
\\[1em]
\includegraphics[width=0.48\textwidth,height=6cm]{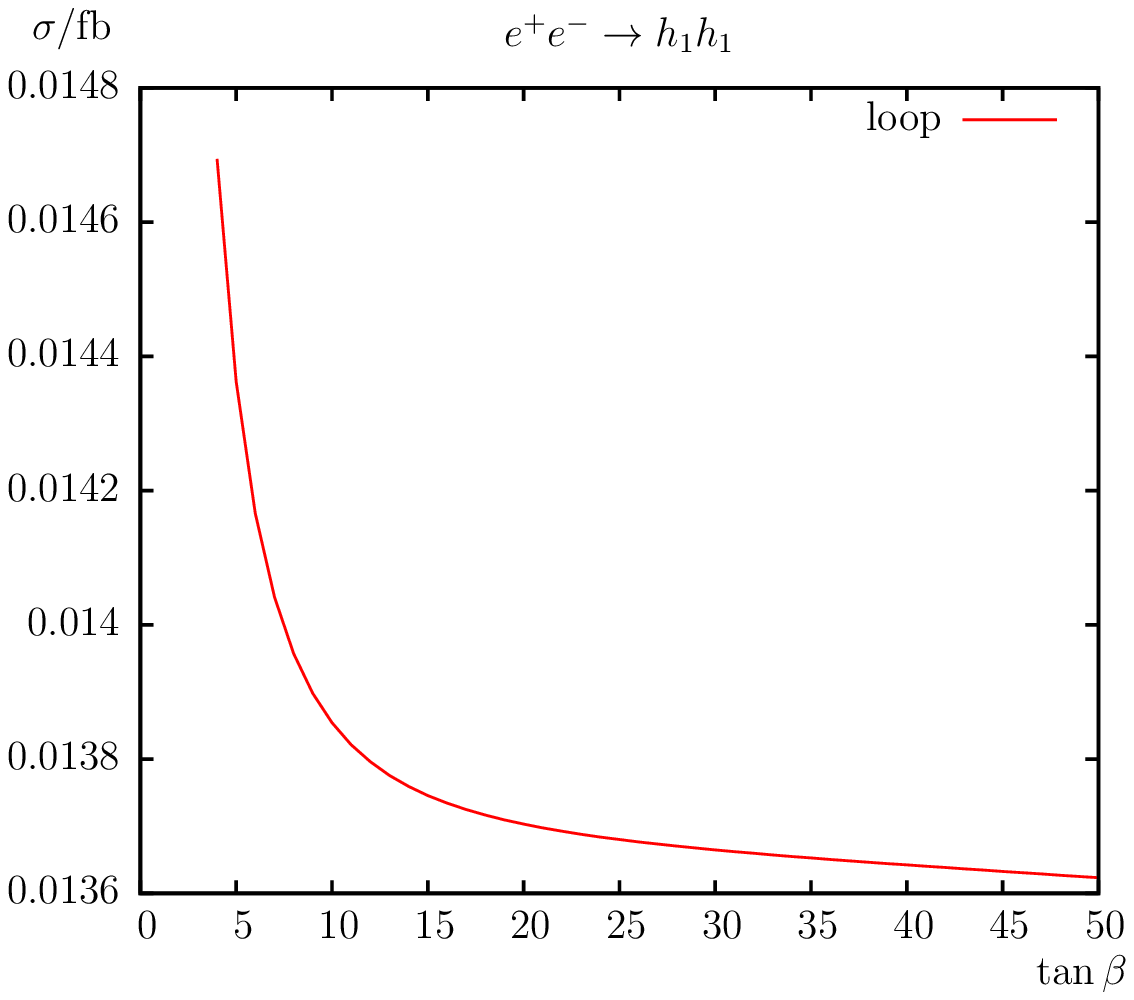}
\includegraphics[width=0.48\textwidth,height=6cm]{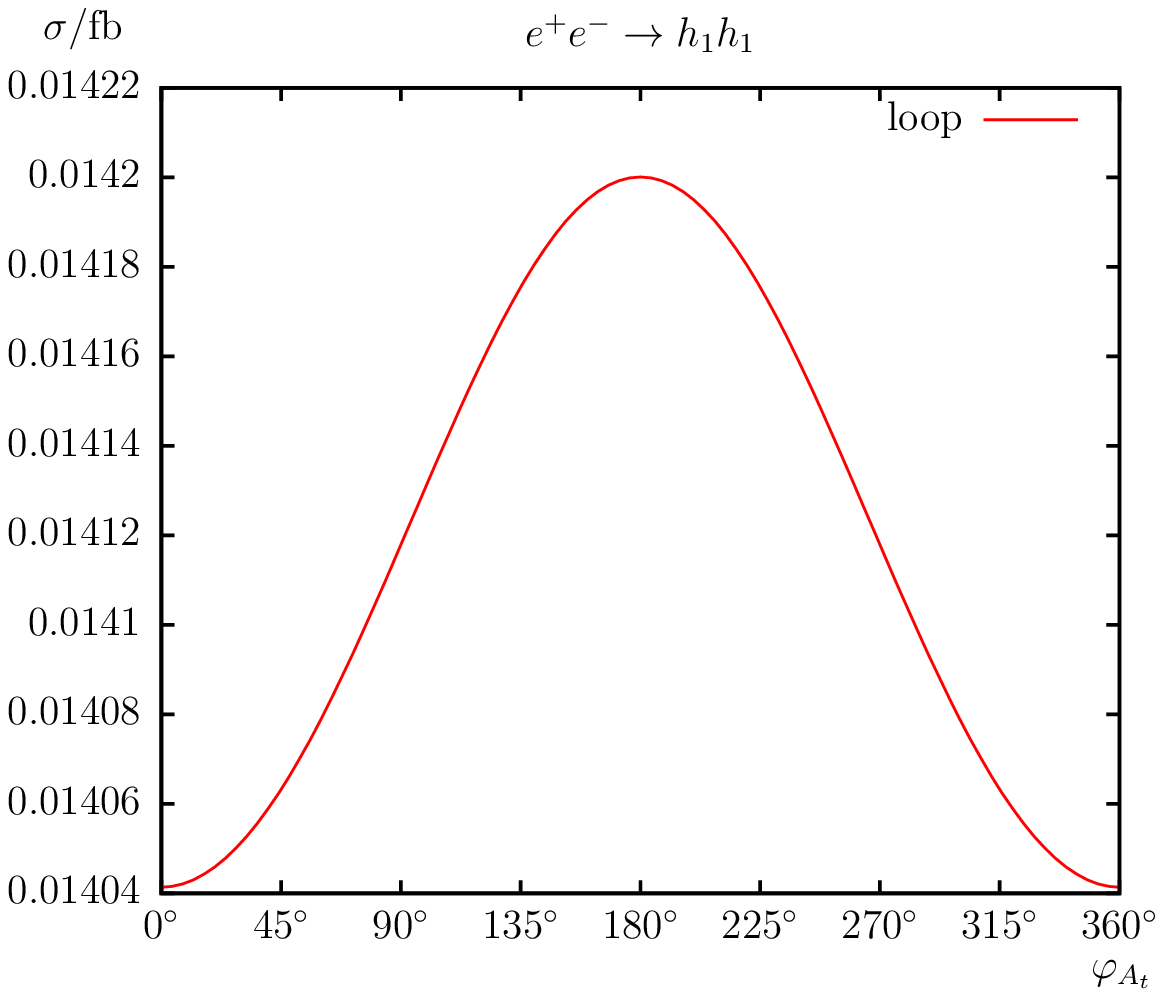}
\end{tabular}
\caption{\label{fig:eeh1h1}
  $\sigma(e^+e^- \to h_1 h_1)$.
  Loop induced (\ie leading two-loop corrected) cross sections are 
  shown with parameters chosen according to \Scs\ (see \refta{tab:para}), 
  but with $\sqrt{s} = 500\gev$. 
  The upper plots show the cross sections with $\sqrt{s}$ (left) 
  and $\MHp$ (right) varied;  the lower plots show $\TB$ (left) and 
  $\phiAt$ (right) varied.
}
\end{center}
\end{figure}

\begin{figure}[t!]
\begin{center}
\begin{tabular}{c}
\includegraphics[width=0.48\textwidth,height=6cm]{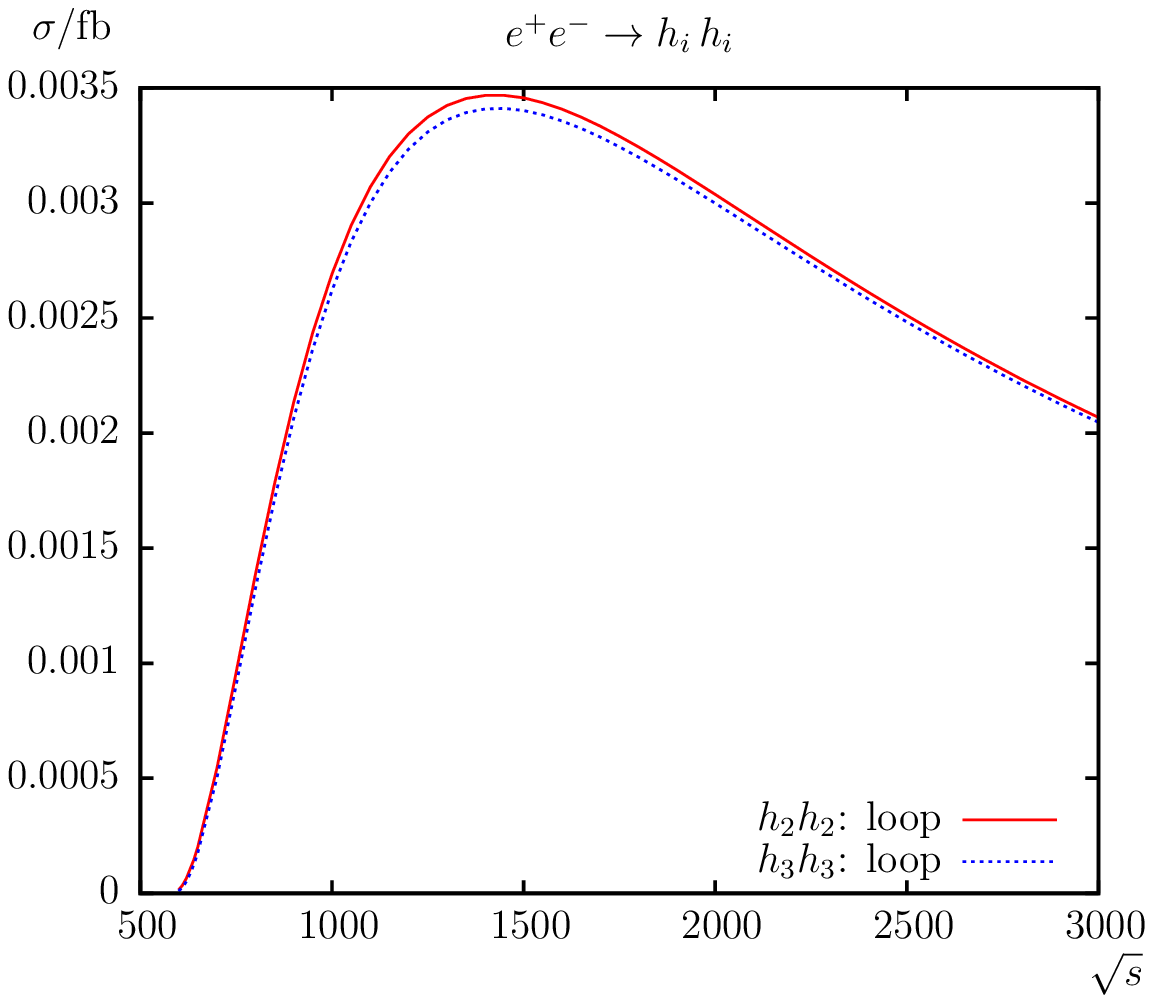}
\includegraphics[width=0.48\textwidth,height=6cm]{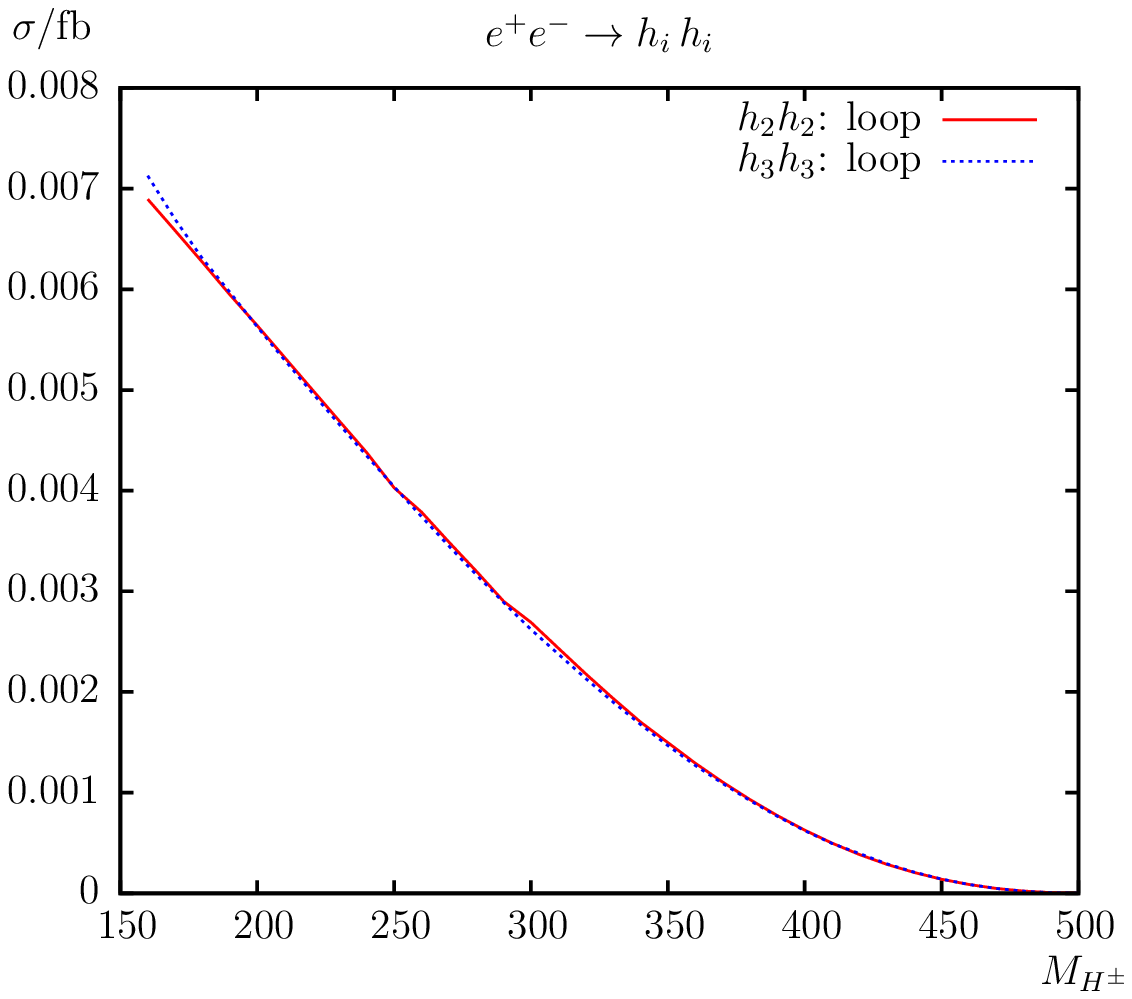}
\\[1em]
\includegraphics[width=0.48\textwidth,height=6cm]{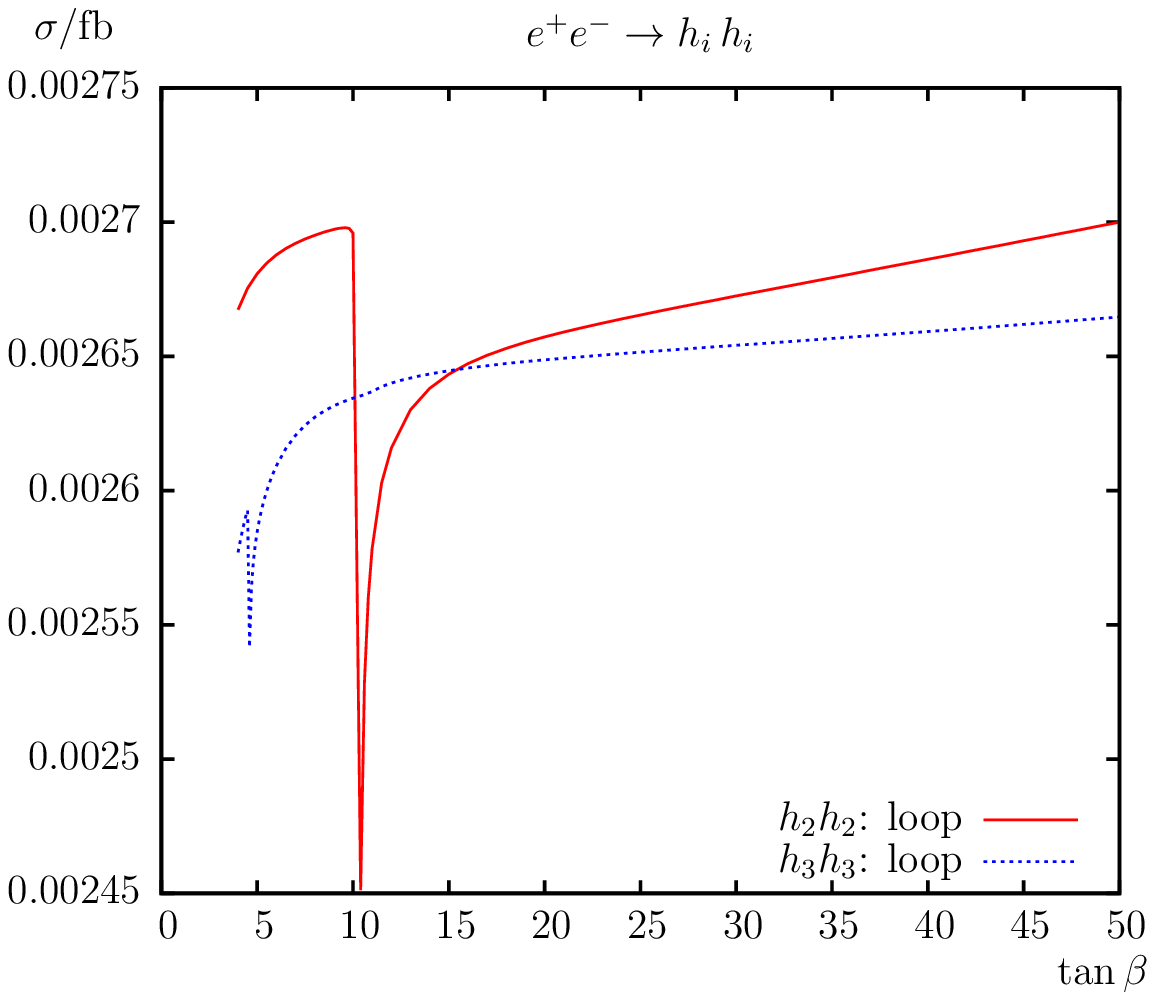}
\includegraphics[width=0.48\textwidth,height=6cm]{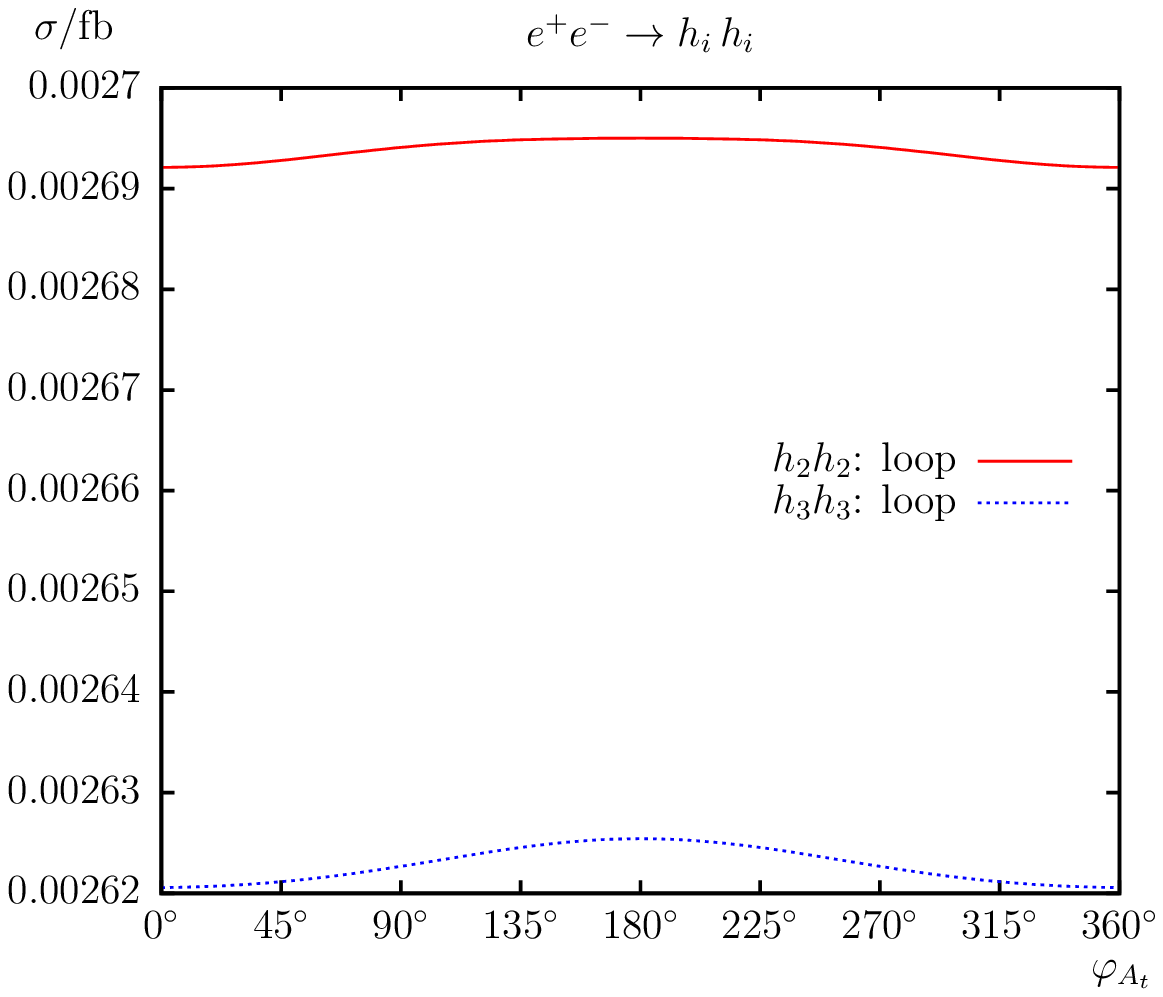}
\end{tabular}
\caption{\label{fig:eehihi}
  $\sigma(e^+e^- \to h_2 h_2, h_3 h_3)$.
  Loop induced (\ie leading two-loop corrected) cross sections are 
  shown with parameters chosen according to \Scs; see \refta{tab:para}. 
  The upper plots show the cross sections with $\sqrt{s}$ (left) 
  and $\MHp$ (right) varied;  the lower plots show $\TB$ (left) and 
  $\phiAt$ (right) varied.
}
\end{center}
\end{figure}

We start our analysis with the production modes $\eehh$ ($i,j = 1,2,3$). 
Results are shown in the \reffis{fig:eeh1h2} -- \ref{fig:eehihi}.
It should be noted that there are no $hHZ$ couplings in the rMSSM
(see \citere{feynarts-mf}).  For real parameters this leads to vanishing 
tree-level cross sections if $h_i \sim h$ and $h_j \sim H$ (or vice versa).
Furthermore there are no $hhZ$, $HHZ$ and $AAZ$ couplings in the rMSSM,
but also in the complex case the tree couplings $h_i h_i Z$ ($i = 1,2,3$) 
are exactly zero (see \citere{feynarts-mf}).
In the following analysis $e^+ e^- \to h_i h_i$ ($i = 1,2,3$) are loop 
induced via (only) box diagrams and therefore $\propto |\cMl|^2$.

\medskip

We begin with the process $e^+e^- \to h_1 h_2$ as shown in \reffi{fig:eeh1h2}. 
As a general comment it should be noted that in \Scs\ one finds that
$h_1 \sim h$, $h_2 \sim A$ and $h_3 \sim H$.  The $hAZ$ coupling is 
$\propto \CBA$ which goes to zero in the 
decoupling limit~\cite{decoupling}, and consequently relatively small cross 
sections are found.
In the analysis of the production cross section as a function of
$\sqrt{s}$ (upper left plot) we find the expected behavior: a strong
rise close to the production threshold, 
followed by a decrease with increasing $\sqrt{s}$. We find a relative
correction of $\sim -15\%$ around the production threshold.
Away from the production threshold, 
loop corrections of $\sim +27\%$ at $\sqrt{s} = 1000\gev$ are found 
in \Scs\ (see \refta{tab:para}).  The relative size of loop corrections
decrease with 
increasing $\sqrt{s}$ and reach $\sim +61\%$ at $\sqrt{s} = 3000\gev$ 
where the tree-level becomes very small.
With increasing $\MHp$ in \Scs\ (upper right plot) we find a strong
decrease of the production cross section, as can be expected from
kinematics, but in particular from the decoupling limit discussed above.
The loop corrections reach $\sim +27\%$ at $\MHp = 300\gev$ and 
$\sim +62\%$ at $\MHp = 500\gev$. These large loop corrections 
are again due to the (relative) smallness of the tree-level results.
It should be noted that at $\MHp \approx 350\gev$ the limit of $0.01$ 
fb is reached, corresponding to 10 events at an integrated luminosity of
$\cL = 1\, \iab$.
The cross sections decrease with increasing $\TB$ (lower left plot), 
and the loop corrections reach the maximum of $\sim +38\%$ at 
$\TB = 36$ while the minimum of $\sim +26\%$ is at $\TB = 5$.
The phase dependence $\phiAt$ of the cross section in \Scs\ (lower right
plot) is at the 10\% level at tree-level. 
The loop corrections are nearly constant, $\sim +28\%$ for all $\phiAt$
values and do not change the overall dependence of the cross section on
the complex phase.

\medskip

Not shown is the process $e^+e^- \to h_1 h_3$.
In this case, for our parameter set \Scs\ (see \refta{tab:para}), one
finds $h_3 \sim H$. Due to the absence of the $hHZ$ coupling in the MSSM 
(see \citere{feynarts-mf}) this leads to vanishing tree-level cross sections 
in the case of real parameters.  For complex parameters (\ie $\phiAt$) the 
tree-level results stay below $10^{-5}$~fb.
Also the loop induced cross sections $\propto |\cMl|^2$ (where only 
the vertex and box diagrams contribute in the case of real parameters) 
stay below $10^{-5}$~fb for our parameter set \Scs.
Consequently, in this case we omit showing plots to the process 
$e^+e^- \to h_1 h_3$.

\medskip

In \reffi{fig:eeh2h3} we present the cross section $e^+e^- \to h_2 h_3$
with $h_2 \sim A$ and $h_3 \sim H$ in \Scs.   The $HAZ$ coupling is 
$\propto \SBA$, which goes to one in the decoupling limit, and 
consequently relatively large cross sections are found.
In the analysis as a function of $\sqrt{s}$ (upper left plot) we find 
relative corrections of $\sim -37\%$ around the production 
threshold, $\sim -5\%$ at $\sqrt{s} = 1000\gev$ (\ie \Scs),
and $\sim +6\%$ at $\sqrt{s} = 3000\gev$.
The dependence on $\MHp$ (upper right plot) is nearly linear above 
$\MHp \gsim 250\gev$, and mostly due to kinematics.
The loop corrections are $\sim -8\%$ at $\MHp = 160\gev$,
$\sim -5\%$ at $\MHp = 300\gev$ (\ie \Scs), and 
$\sim -52\%$ at $\MHp = 500\gev$ where the tree-level goes to zero.
As a function of $\TB$ (lower left plot) the tree-level cross section 
is rather flat, apart from a dip at $\TB \approx 10$, corresponding 
to the threshold $\mcha1 + \mcha1 = m_{h_2}$.  
This threshold enter into the tree-level and the loop corrections only 
via the $\matr{\hat{Z}}$~matrix contribution (calculated by \FH). 
The relative corrections increase from $\sim -5\%$ at $\TB = 7$
to $\sim +7\%$ at $\TB = 50$.
The dependence on $\phiAt$ (lower right plot) is very small, 
below the percent level.  The loop corrections are found to be nearly 
independent of $\phiAt$ at the level of $\sim -4.6\%$.

\medskip

We now turn to the processes with equal indices.  The tree couplings 
$h_i h_i Z$ ($i = 1,2,3$) are exactly zero; see \citere{feynarts-mf}.
Therefore, in this case we show the pure loop induced cross sections
$\propto |\cMl|^2$ (labeled as ``loop'') where only the box diagrams 
contribute.  These box diagrams are UV and IR finite.

In \reffi{fig:eeh1h1} we show the results for $e^+e^- \to h_1 h_1$.
This process might have some special interest, since it is the lowest
energy process in which triple Higgs boson couplings play a role, which 
could be relevant at a high-luminosity collider operating above the two 
Higgs boson production threshold.  In our numerical analysis, as a 
function of $\sqrt{s}$ we find a maximum of $\sim 0.014$~fb, at 
$\sqrt{s} = 500\gev$, decreasing to $\sim 0.002$~fb at 
$\sqrt{s} = 3\tev$.
The dependence on $\MHp$ is rather small, as is the dependence on $\TB$
and $\phiAt$ in \Scs. However, with cross sections found at the level 
of up to $0.015$~fb this process could potentially be observable at 
the ILC running at $\sqrt{s} = 500\gev$ or below (depending on the
integrated luminosity).

\medskip

We finish the $e^+e^- \to h_i h_i$ analysis in \reffi{fig:eehihi} 
in which the results for $i = 2,3$ are displayed. 
Both production processes have rather similar (purely loop-induced)
production cross sections.  As a function of $\sqrt{s}$ we find a maximum
of $\sim 0.0035$~fb at $\sqrt{s} = 1.4 \tev$. In \Scs, but with $\MHp$
varied we find the highest values of $\sim 0.007$~fb at the lowest 
mass scales, going down below $0.001$~fb at around $\MHp \sim 380\gev$.
The production cross sections depend only very weakly on $\TB$ and
$\phiAt$, where in \Scs\ values of $\sim 0.0026$~fb are found, 
leading only to about 5~events for an integrated luminosity of 
$\cL = 2\, \iab$. Furthermore, due to the similar decay patterns of 
$h_2 \sim A$ and $h_3 \sim H$ and the similar masses of the two states it 
might be difficult to disentangle it from $e^+e^- \to h_2 h_3$, and a more
dedicated analysis (beyond the scope of our paper) will be necessary to 
determine its observability.
The large dip at $\TB \approx 10$ (red solid line) is the threshold 
$\mcha1 + \mcha1 = m_{h_2}$ in $e^+e^- \to h_2 h_2$.  
The dip at $\TB \approx 5$ (blue dotted line) is the threshold 
$\mneu1 + \mneu1 = m_{h_3}$ in $e^+e^- \to h_3 h_3$.
These thresholds enter into the loop corrections only via the 
$\matr{\hat{Z}}$~matrix contribution (calculated by \FH).
The cross sections are quite similar and very small for the parameter 
set chosen; see \refta{tab:para}.

\medskip

Overall, for the neutral Higgs boson pair production we observed 
an increasing cross section $\propto 1/s$ for $s \to \infty$;
see \refeq{eehhTree}.  
The full one-loop corrections reach a level of $10\%$ -- $20\%$ or higher 
for cross sections of $0.01$ -- $10$~fb.  The variation with $\phiAt$ is 
found rather small, except for $e^+e^- \to h_1 h_2$, where it is at the level 
of $10\%$.  The results for $h_i h_i$ production turn out to be small 
(but not necessarily hopeless) for $i = 1$, and negligible for $i = 2,3$ 
for Higgs boson masses above $\sim 200\gev$.


\subsubsection{The process \boldmath{$\eehZ$}}
\label{eehZ}

\begin{figure}[t!]
\begin{center}
\begin{tabular}{c}
\includegraphics[width=0.48\textwidth,height=6cm]{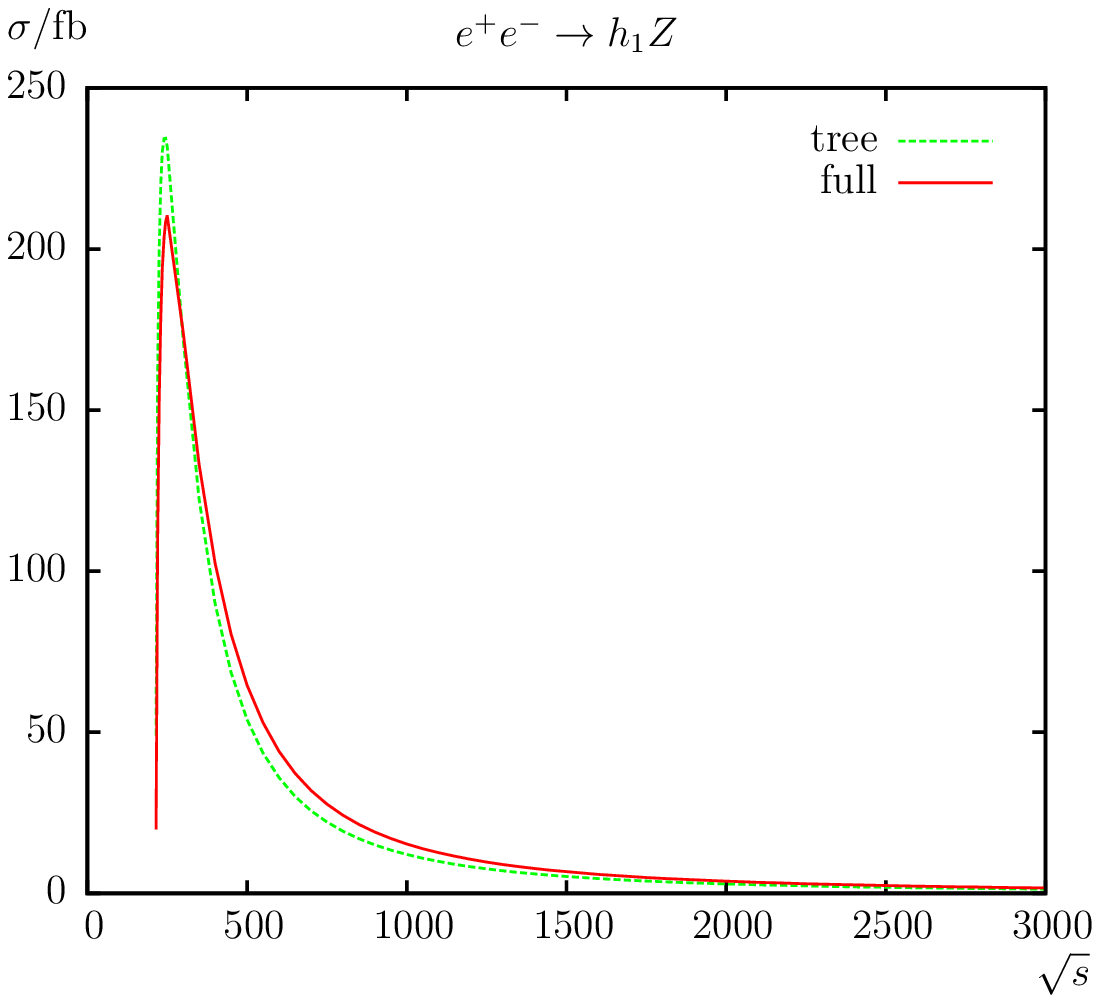}
\includegraphics[width=0.48\textwidth,height=6cm]{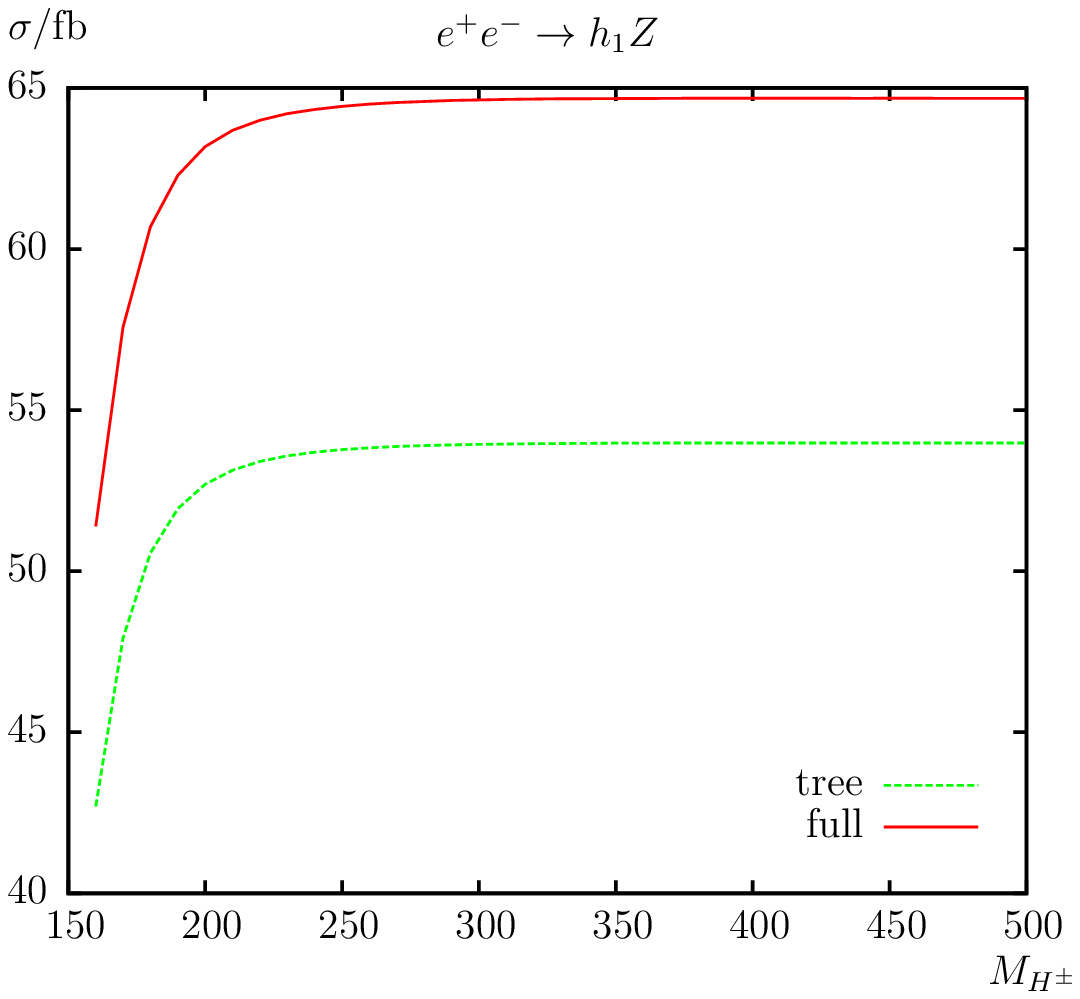}
\\[1em]
\includegraphics[width=0.48\textwidth,height=6cm]{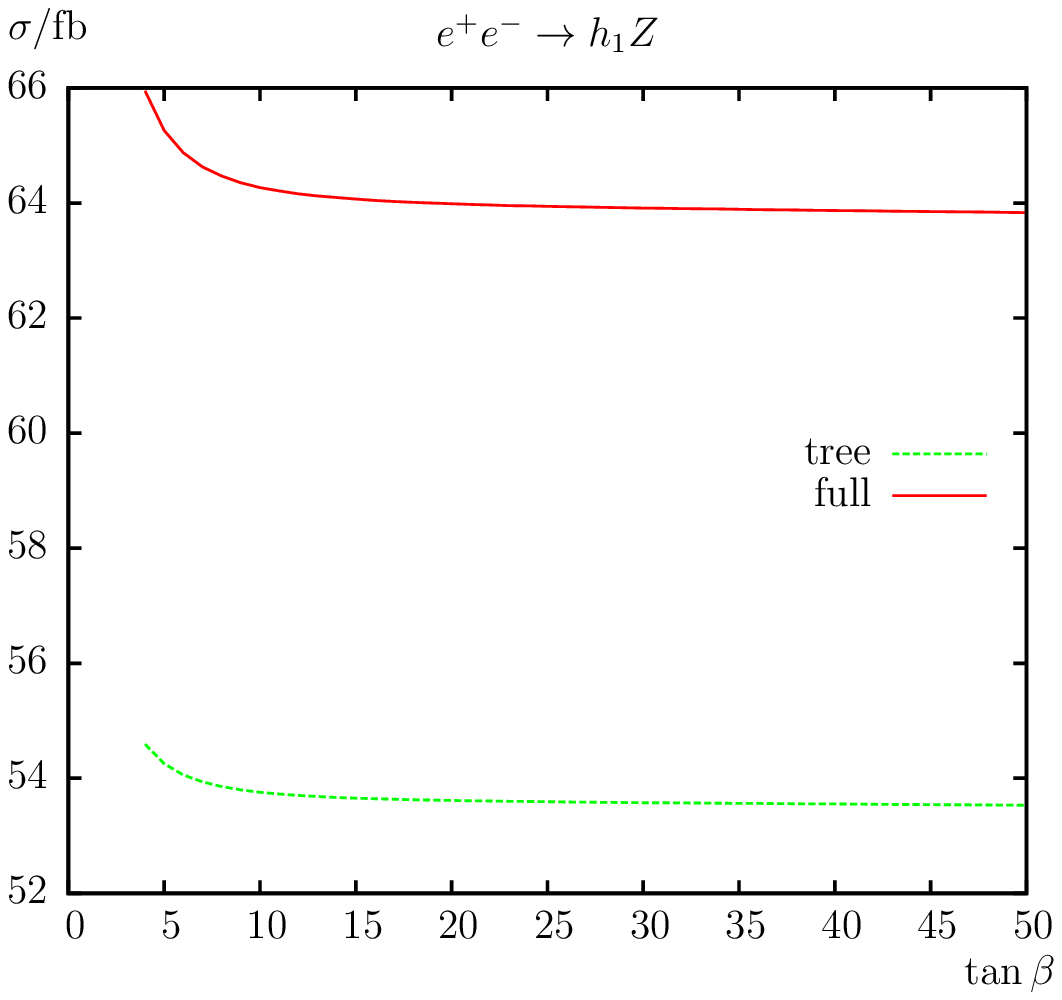}
\includegraphics[width=0.48\textwidth,height=6cm]{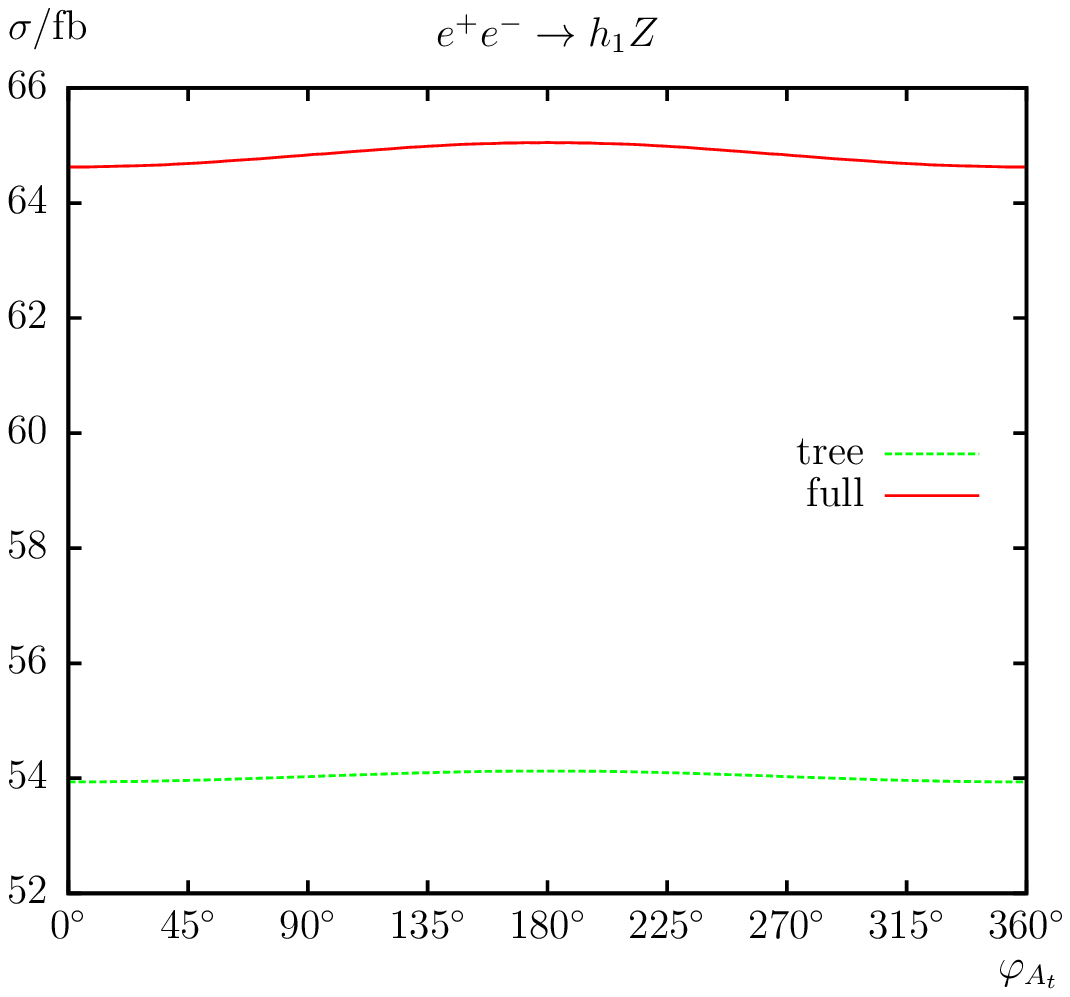}
\end{tabular}
\caption{\label{fig:eeh1Z}
  $\sigma(e^+e^- \to h_1 Z)$. 
  Tree-level and full one-loop corrected cross sections are shown  
  with parameters chosen according to \Scs\ (see \refta{tab:para}), 
  but with $\sqrt{s} = 500\gev$. 
  The upper plots show the cross sections with $\sqrt{s}$ (left) 
  and $\MHp$ (right) varied;  the lower plots show $\TB$ (left) and 
  $\phiAt$ (right) varied.
}
\end{center}
\end{figure}

\begin{figure}[t!]
\begin{center}
\begin{tabular}{c}
\includegraphics[width=0.48\textwidth,height=6cm]{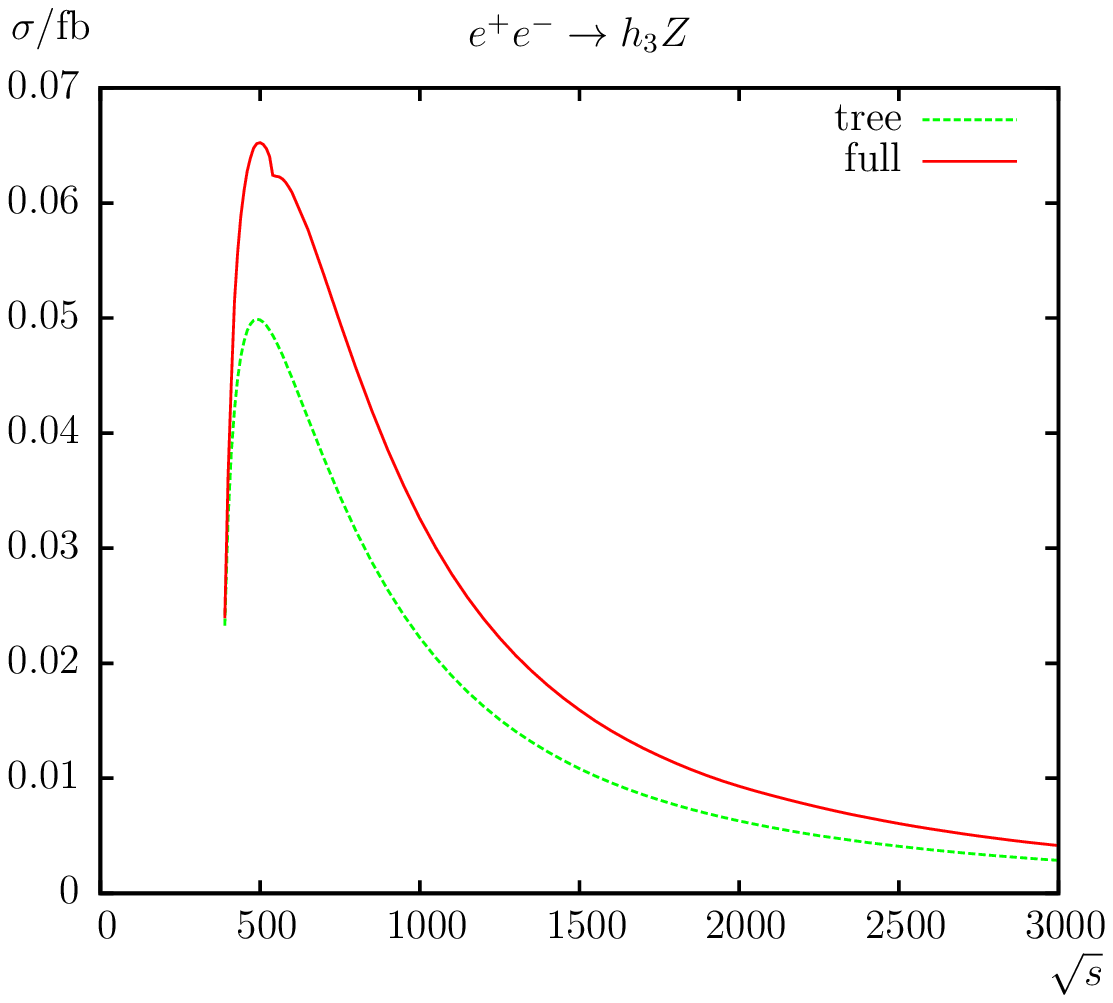}
\includegraphics[width=0.48\textwidth,height=6cm]{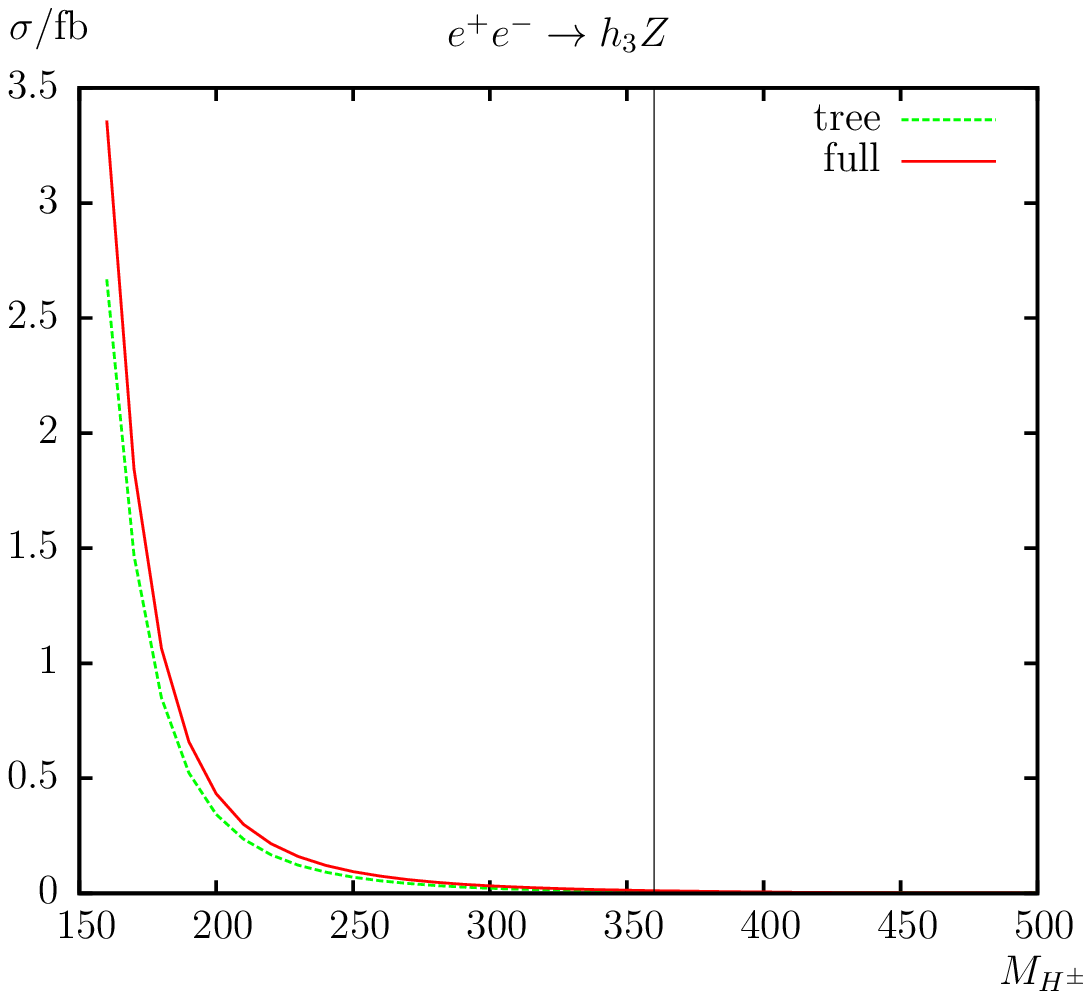}
\\[1em]
\includegraphics[width=0.48\textwidth,height=6cm]{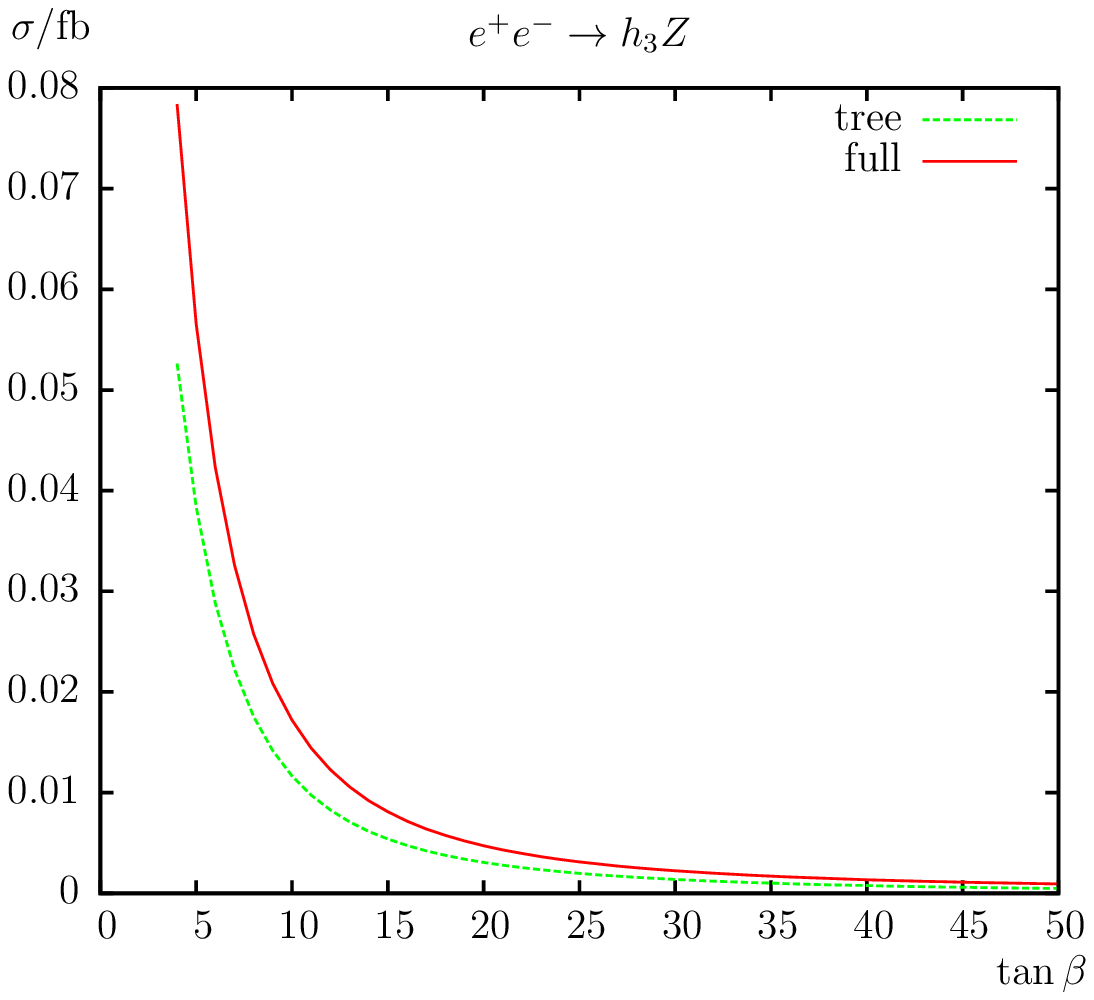}
\includegraphics[width=0.48\textwidth,height=6cm]{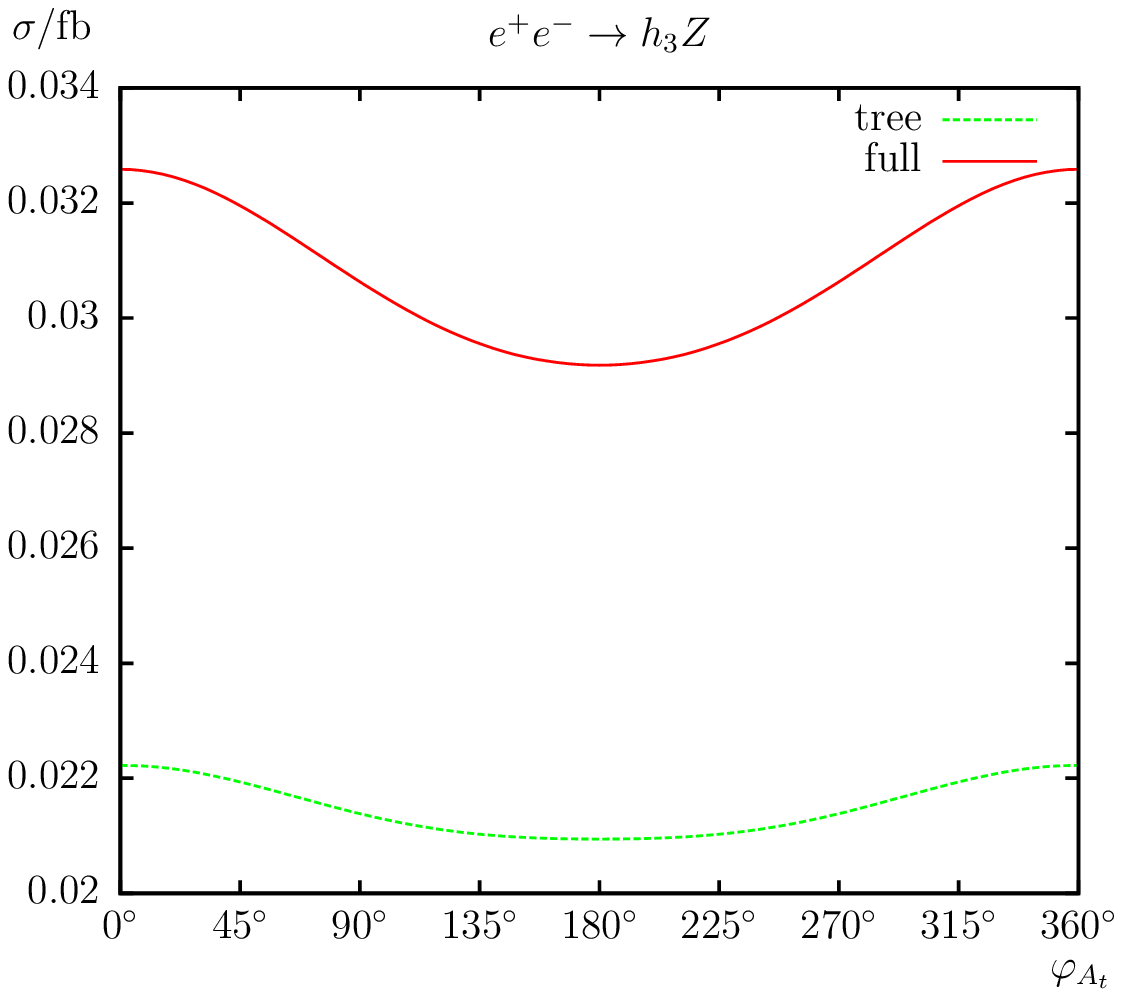}
\end{tabular}
\caption{\label{fig:eeh3Z}
  $\sigma(e^+e^- \to h_3 Z)$. 
  Tree-level and full one-loop corrected cross sections are shown  
  with parameters chosen according to \Scs; see \refta{tab:para}.  
  The upper plots show the cross sections with $\sqrt{s}$ (left) 
  and $\MHp$ (right) varied;  the lower plots show $\TB$ (left) and 
  $\phiAt$ (right) varied.
}
\end{center}
\end{figure}

In \reffis{fig:eeh1Z} and \ref{fig:eeh3Z} we show the results for the 
processes $\eehZ$, as before as a function of $\sqrt{s}$, $\MHp$, $\TB$ 
and $\phiAt$. 
It should be noted that there are no $AZZ$ couplings in the MSSM
(see \cite{feynarts-mf}).  In the case of real parameters this 
leads to vanishing tree-level cross sections if $h_i \sim A$.

\medskip

We start with the process $e^+e^- \to h_1 Z$ shown in \reffi{fig:eeh1Z}.
In \Scs\ one finds $h_1 \sim h$, and since the $ZZh$ coupling is 
$\propto \SBA \to 1$ in the decoupling limit, relative large cross 
sections are found. 
As a function of $\sqrt{s}$ (upper left plot) a maxium of more than
$200$~fb is found at $\sqrt{s} \sim 250\gev$ with a decrease for
increasing $\sqrt{s}$. 
The size of the corrections of the cross section can be especially 
large very close to the production threshold%
\footnote{
  It should be noted that a calculation very close to the production 
  threshold requires the inclusion of additional (nonrelativistic) 
  contributions, which is beyond the scope of this paper. 
  Consequently, very close to the production threshold our calculation 
  (at the tree- and loop-level) does not provide a very accurate 
  description of the cross section.
}
from which on the considered process is kinematically possible.
At the production threshold we found relative corrections of 
$\sim -60\%$.  Away from the production threshold, loop corrections of 
$\sim +20\%$ at $\sqrt{s} = 500\gev$ are found, increasing to 
$\sim +30\%$ at $\sqrt{s} = 3000\gev$.
In the following plots we assume, deviating from the definition of \Scs,
$\sqrt{s} = 500\gev$. 
As a function of $\MHp$ (upper right plot) the cross sections strongly 
increases up to $\MHp \lsim 250\gev$, corresponding to $\SBA \to 1$ 
in the decoupling limit discussed above.  For higher $\MHp$ values 
it is nearly constant, and the loop corrections are $\sim +20\%$ for 
$160\gev < \MHp < 500\gev$.
Hardly any variation is found for the production cross section as a
function of $\TB$ or $\phiAt$. In both cases the one-loop corrections 
are found at the level of $\sim +20\%$.

\medskip

Not shown is the process $e^+e^- \to h_2 Z$.
In this case, for our parameter set \Scs\ (see \refta{tab:para}), one finds
$h_2 \sim A$.  Because there are no $AZZ$ couplings in the MSSM 
(see \cite{feynarts-mf}) this leads to vanishing tree-level cross sections 
in the case of real parameters.  For complex parameters (\ie $\phiAt$) the 
tree-level results stay below $10^{-5}$~fb.
Also the loop induced cross sections $\propto |\cMl|^2$ (where only 
the vertex and box diagrams contribute in the case of real parameters) 
are below $10^{-3}$~fb for our parameter set \Scs.
Consequently, in this case we omit showing plots to the process 
$e^+e^- \to h_2 Z$.

\medskip

We finish the $\eehZ$ analysis in \reffi{fig:eeh3Z} in which the results 
for $e^+e^- \to h_3 Z$ are shown. In \Scs\ one has $h_3 \sim H$, 
and with the $ZZH$ coupling being proportional to $\CBA \to 0$ in the 
decoupling limit relatively small production cross sections are found 
for $\MHp$ not too small.
As a function of $\sqrt{s}$ (upper left plot) a dip can be seen at 
$\sqrt{s} \approx 540\gev$, due to the threshold 
$\mcha2 + \mcha2 = \sqrt{s}$.  Around the production threshold we found 
relative corrections of $\sim 3\%$. The maxium production cross section
is found at $\sqrt{s} \sim 500\gev$ of about $0.065$~fb including loop
corrections, rendering this process observable with an accumulated
luminosity $\cL \lsim 1\, \iab$. Away from the production 
threshold, one-loop corrections of $\sim 47\%$ at $\sqrt{s} = 1000\gev$ 
are found in \Scs\ (see \refta{tab:para}), with a cross section of about
$0.03$~fb. The cross section further decreases
with increasing $\sqrt{s}$ and the loop corrections reach $\sim 45\%$ 
at $\sqrt{s} = 3000\gev$, where it drops below the level of
$0.0025$~fb. 
As a function of $\MHp$ we find the afore mentioned decoupling behavior
with increasing $\MHp$. The
the loop corrections reach $\sim 26\%$ at $\MHp = 160\gev$,
$\sim 47\%$ at $\MHp = 300\gev$ and $\sim +56\%$ at 
$\MHp = 500\gev$.  These large loop corrections ($> 50\%$) are again 
due to the (relative) smallness of the tree-level results. 
It should be noted that at $\MHp \approx 360\gev$ the limit of $0.01$~fb
is reached; see the line in the upper right plot.
The production cross section decreases strongly with $\TB$ (lower right
plot). The loop corrections reach the maximum of $\sim +95\%$ at 
$\TB = 50$ do to the very small tree-level result, while the minimum 
of $\sim +47\%$ is found at $\TB = 7$.
The phase dependence $\phiAt$ of the cross section (lower right plot)
is at the level of $5\%$ at tree-level, but increases to about $10\%$ 
including loop corrections. Those are found to vary from $\sim +47\%$ 
at $\phiAt = 0^{\circ}, 360^{\circ}$ to $\sim +39\%$ at $\phiAt = 180^{\circ}$.

\medskip

Overall, for the $Z$ Higgs boson production we observed an increasing 
cross section $\propto 1/s$ for $s \to \infty$; see \refeq{eehZTree}. 
The full one-loop corrections reach a level of $20\%$ ($50\%$) for cross 
sections of $60$~fb ($0.03$~fb). 
The variation with $\phiAt$ is found to be small, reaching up to $10\%$
for $e^+e^- \to h_3 Z$, after including the loop corrections.


\subsubsection{The process \boldmath{$\eehga$}}
\label{sec:eehga}

\begin{figure}[t!]
\begin{center}
\begin{tabular}{c}
\includegraphics[width=0.48\textwidth,height=6cm]{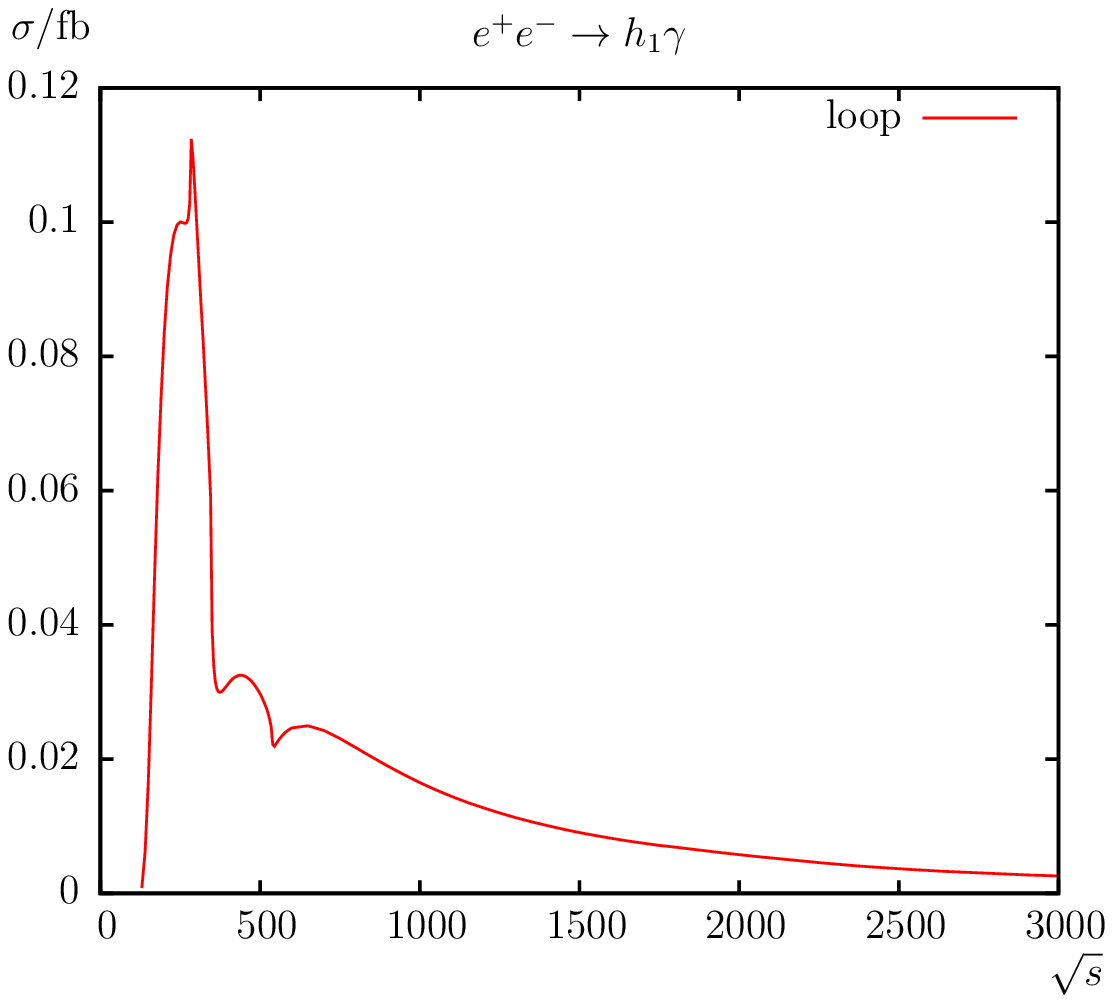}
\includegraphics[width=0.48\textwidth,height=6cm]{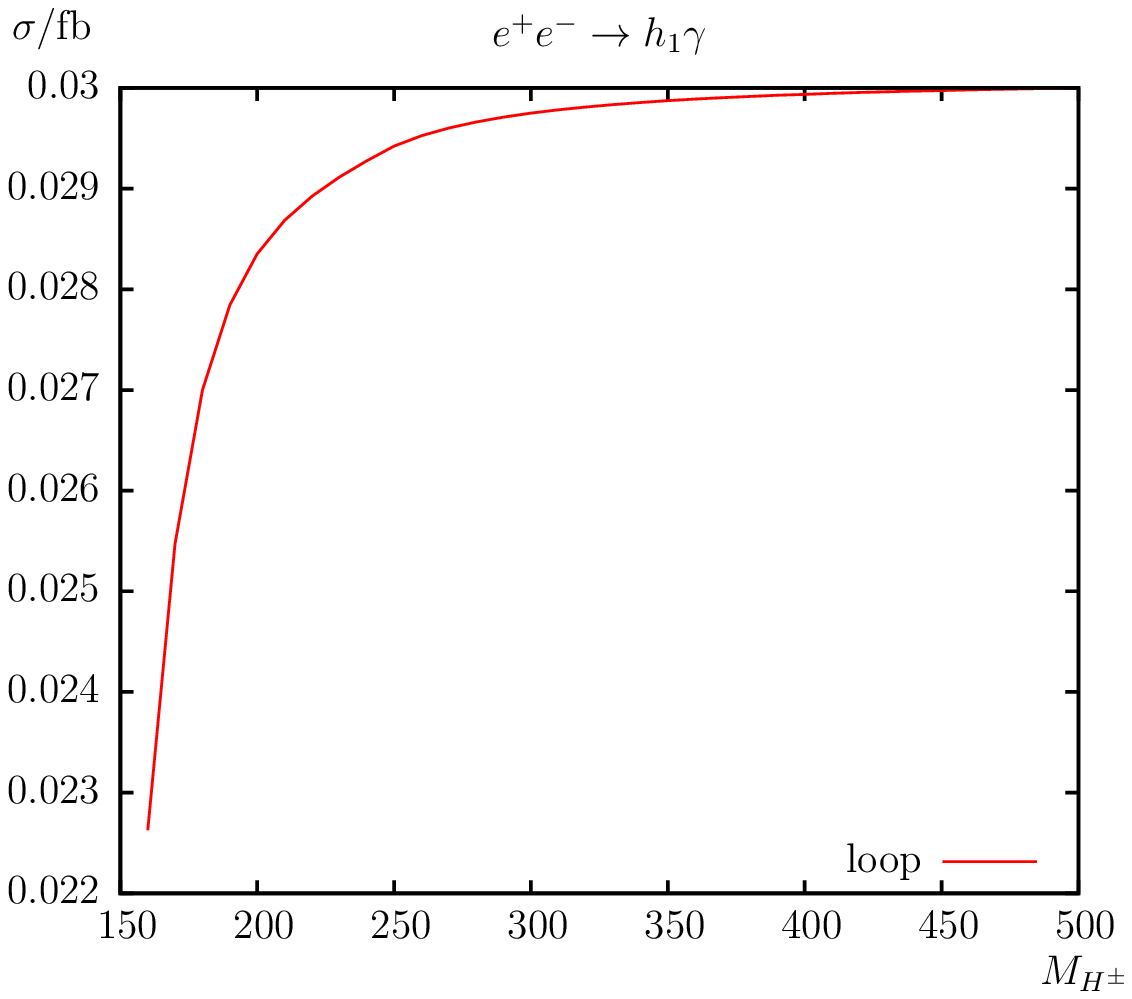}
\\[1em]
\includegraphics[width=0.48\textwidth,height=6cm]{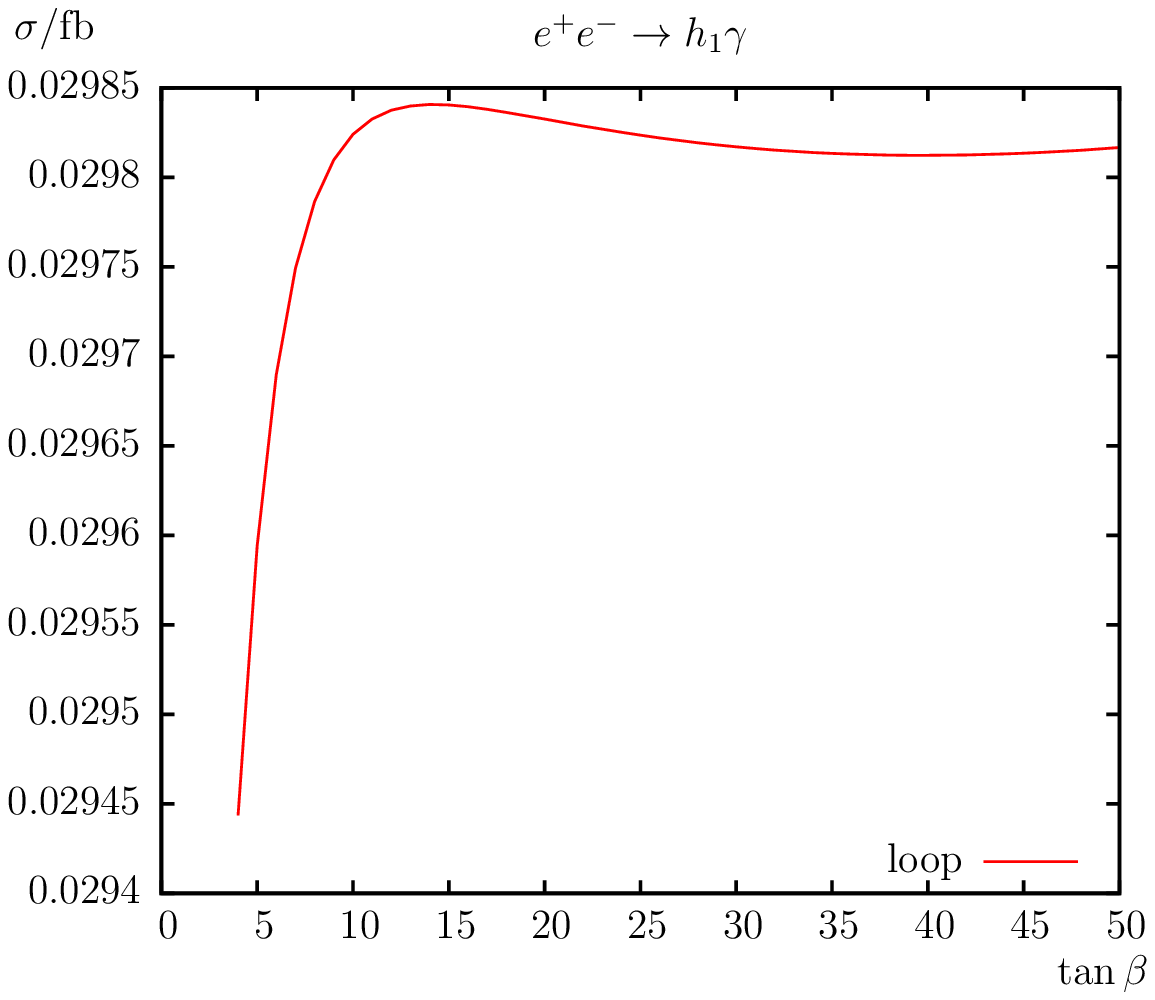}
\includegraphics[width=0.48\textwidth,height=6cm]{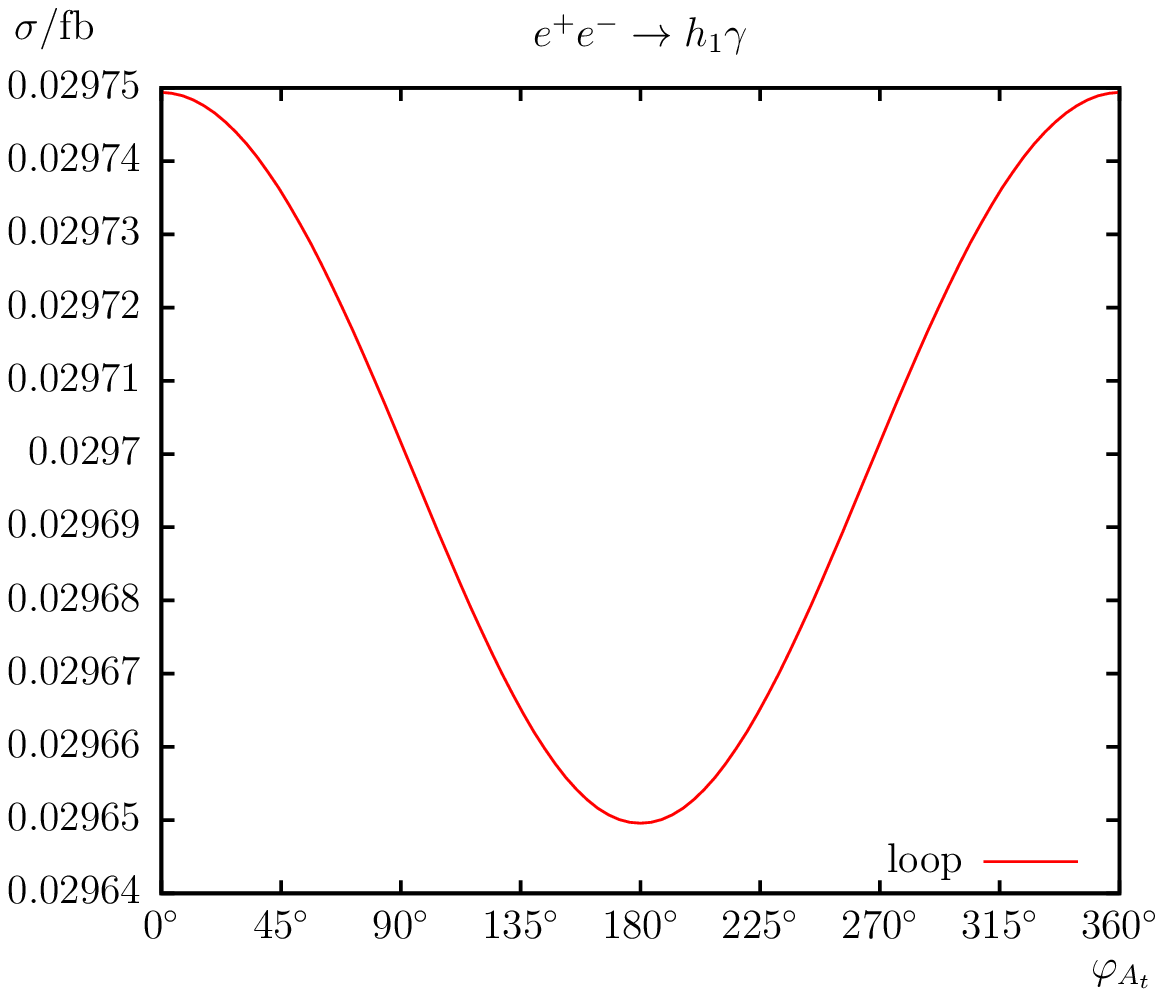}
\end{tabular}
\caption{\label{fig:eeh1ga}
  $\sigma(e^+e^- \to h_1 \ga)$.
  Loop induced (\ie leading two-loop corrected) cross sections are 
  shown with parameters chosen according to \Scs\ (see \refta{tab:para}), 
  but with $\sqrt{s} = 500\gev$.  
  The upper plots show the cross sections with $\sqrt{s}$ (left) 
  and $\MHp$ (right) varied;  the lower plots show $\TB$ (left) and 
  $\phiAt$ (right) varied.
}
\end{center}
\end{figure}

\begin{figure}[t!]
\begin{center}
\begin{tabular}{c}
\includegraphics[width=0.48\textwidth,height=6cm]{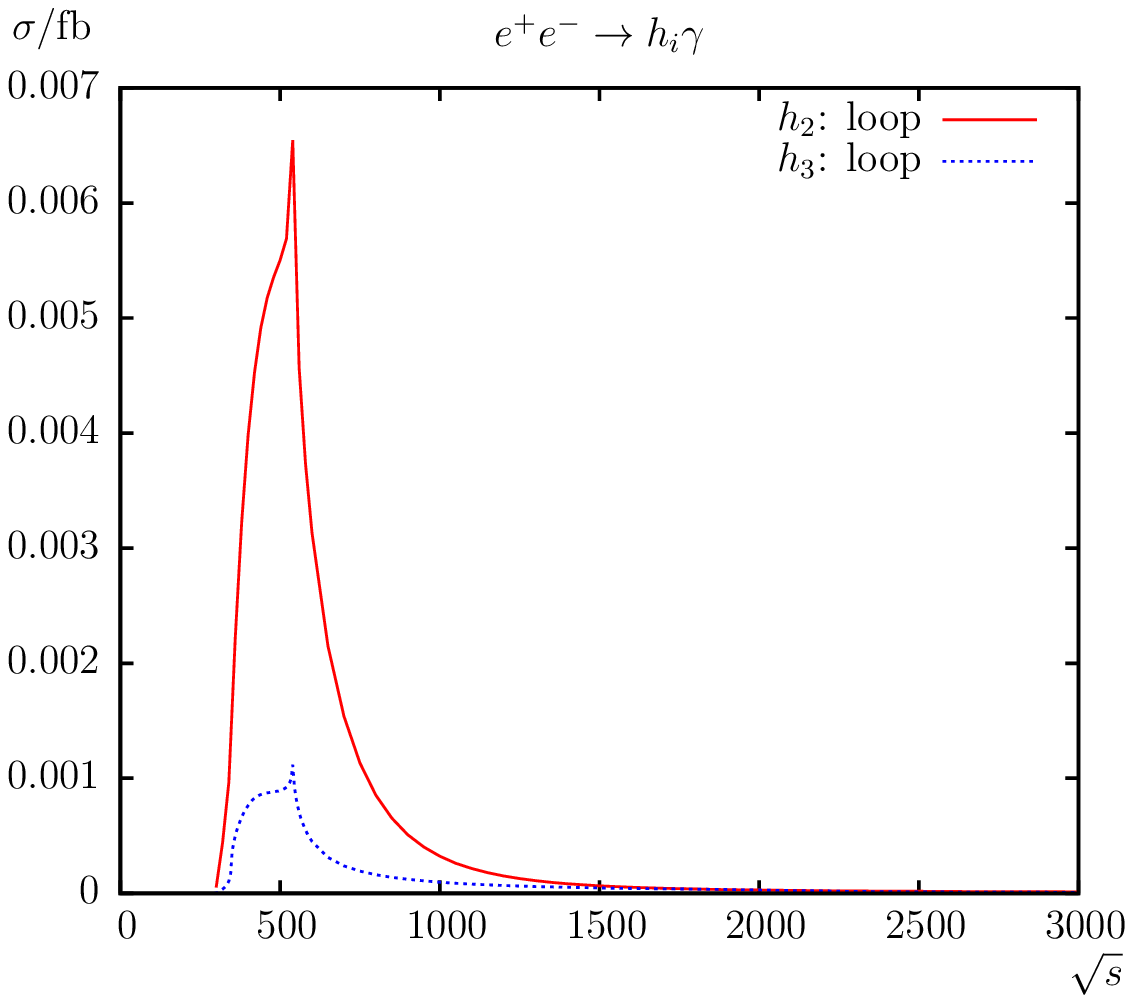}
\includegraphics[width=0.48\textwidth,height=6cm]{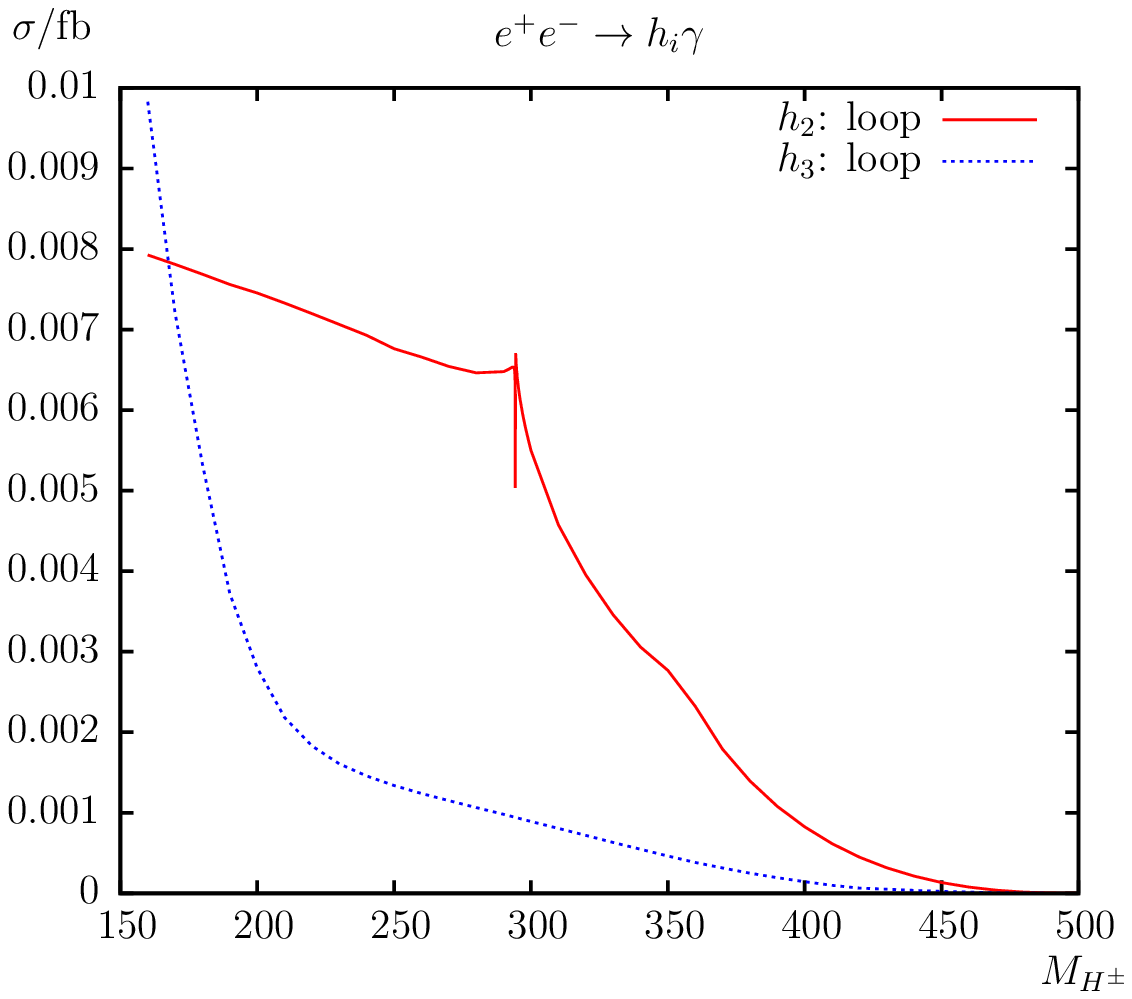}
\\[1em]
\includegraphics[width=0.48\textwidth,height=6cm]{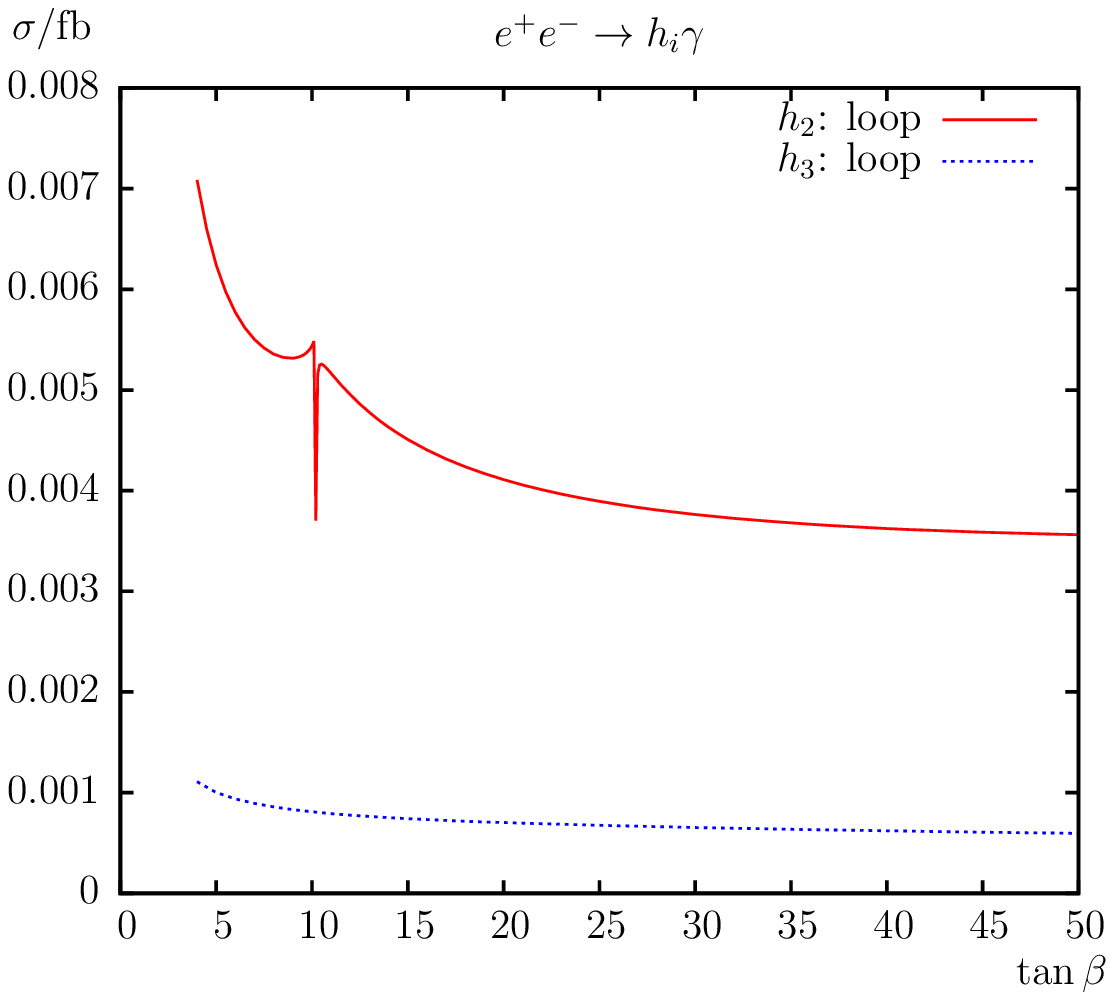}
\includegraphics[width=0.48\textwidth,height=6cm]{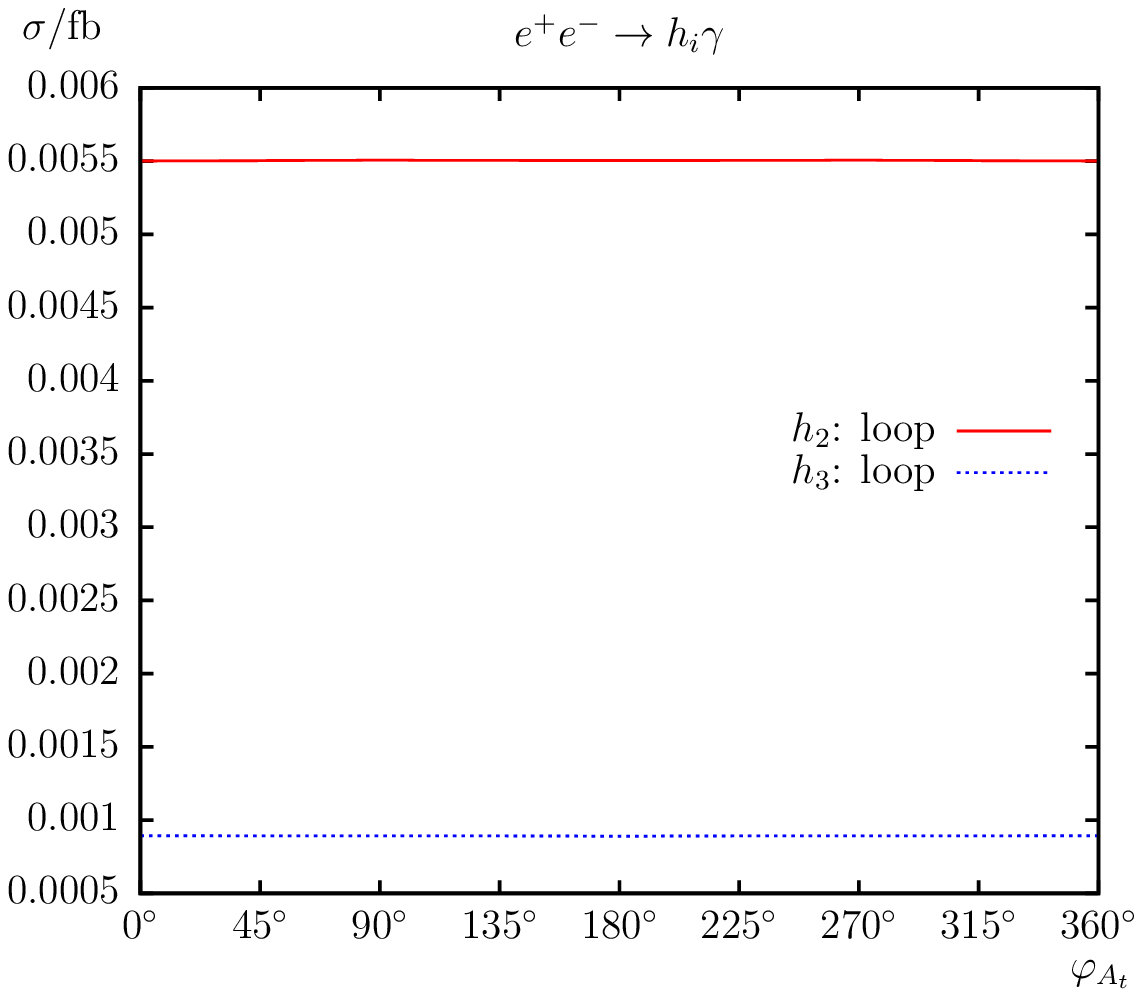}
\end{tabular}
\caption{\label{fig:eehiga}
  $\sigma(e^+e^- \to h_i \ga)$ ($i = 2,3$).
  Loop induced (\ie leading two-loop corrected) cross sections are 
  shown with parameters chosen according to \Scs\ (see \refta{tab:para}), 
  but with $\sqrt{s} = 500\gev$. 
  The upper plots show the cross sections with $\sqrt{s}$ (left) 
  and $\MHp$ (right) varied;  the lower plots show $\TB$ (left) and 
  $\phiAt$ (right) varied.
}
\end{center}
\end{figure}

In \reffis{fig:eeh1ga} and \ref{fig:eehiga} we show the results for the 
processes $\eehga$ as before as a function of $\sqrt{s}$, $\MHp$, $\TB$ 
and $\phiAt$. 
It should be noted that there are no $h_i Z \ga$ or $h_i \ga \ga$ 
($i = 1,2,3$) couplings in the MSSM; see \citere{feynarts-mf}.
In the following analysis $e^+ e^- \to h_i \ga$ ($i = 1,2,3$) are purely
loop induced processes (via vertex and box diagrams) and therefore 
$\propto |\cMl|^2$.

\medskip

We start with the process $e^+e^- \to h_1 \ga$ shown in
\reffi{fig:eeh1ga}. The largest contributions are expected from loops
involving top quarks and SM gauge bosons.
The cross section is rather small for the parameter set chosen; 
see \refta{tab:para}. 
As a function of $\sqrt{s}$ (upper left plot) a maxium of $\sim 0.1$~fb
is reached around $\sqrt{s} \sim 250\gev$, where several thresholds and
dip effects overlap. The first peak is found at 
$\sqrt{s} \approx 283\gev$, due to the threshold $\mcha1 + \mcha1 = \sqrt{s}$.
A dip can be found at $\mt + \mt = \sqrt{s} \approx 346\gev$. 
The next dip at $\sqrt{s} \approx 540\gev$ is the threshold 
$\mcha2 + \mcha2 = \sqrt{s}$.
The loop corrections for $\sqrt{s}$ vary between $0.1$~fb at 
$\sqrt{s} \approx 250\gev$, $0.03$~fb at
$\sqrt{s} \approx 500\gev$ and $0.003$~fb at 
$\sqrt{s} \approx 3000\gev$.
Consequently, this process could be observable for larger ranges of
$\sqrt{s}$. In particular in the initial phase with 
$\sqrt{s} = 500\gev$~\cite{ILCstages} 30 events could be produced with
an integrated luminosity of $\cL = 1\, \iab$. 
As a function of $\MHp$ (upper right plot) we find an increase in \Scs\ 
(but with $\sqrt{s} = 500\gev$), increasing the production cross
sections from $0.023$~fb at $\MHp \approx 160\gev$ to about $0.03$~fb 
in the decoupling regime. This dependence shows the relevance of the SM
gauge boson loops in the production cross section, indicating that the
top quark loops dominate this production cross section.
The variation with $\TB$ and $\phiAt$ (lower row) is rather small, and
values of $0.03$~fb are found in \Scs.

\medskip

We finish the $\eehga$ analysis in \reffi{fig:eehiga} in which the 
results for $e^+e^- \to h_i \ga$ ($i = 2,3$) are displayed, where 
$e^+e^- \to h_2 \ga\ (h_3 \ga)$ is shown as solid red (dashed blue) line.
In \Scs, as dicussed above, one finds $h_2 \sim A$ and $h_3 \sim H$. 
While both Higgs bosons have reduced (enhanced) couplings to top
(bottom) quarks, only the $H$~can have a non-negligible coupling to SM
gauge bosons. 
As function of $\sqrt{s}$ (upper left plot) we find that for the
$h_2\ga$ ($h_1\ga$) production maximum values of about 
$0.006\ (0.001)$~fb are found. However, due to the similar decay pattern
and masses (for not too small $\MHp$, $300\gev$ here) it will be
difficult to disentangle those to production cross sections, and the
effective cross section is given roughly by the sum of the two.  This
renders these loop-induced processes at the border of observability. 
The peaks oberved are found at $\sqrt{s} \approx 540\gev$
due to the threshold $\mcha2 + \mcha2 = \sqrt{s}$ for both production
cross sections. They drop to the unobservable level for 
$\sqrt{s} \gsim 1 \tev$. 
As a function of $\MHp$ (upper right plot) one can observe the
decoupling of $h_3 \sim H$ of the SM gauge bosons with increasing
$\MHp$, lowering the cross section for larger values.
The ``knee'' at $\MHp \approx 294\gev$ is the threshold 
$\mcha1 + \mcha1 = m_{h_2}$.
This threshold enter into the loop corrections only via 
the $\matr{\hat{Z}}$~matrix contribution (calculated by \FH).
The loop corrections vary between $0.008$~fb at 
$\MHp \approx 160\gev$ and far below $0.001$~fb at 
$\MHp \approx 500\gev$.
The dependence on $\TB$ (lower left plot) is rather strong for the
$h_2\ga$ production going from $0.007$~fb at $\TB = 4$ down to
$0.0035$~fb at $\TB = 50$. 
The dip at $\TB \approx 10$ is the threshold $\mcha1 + \mcha1 = m_{h_2}$.
This threshold enter into the loop corrections again only via the 
$\matr{\hat{Z}}$~matrix contribution (calculated by \FH).
For the $h_3\ga$ production the cross section stays at the very low
level of $0.001$~fb for all $\TB$ values. 
The dependence on the phase $\phiAt$ of the cross sections (lower right plot)
is very small in \Scs, with no visible variation in the plot.

\medskip

Overall, for the $\ga$ Higgs boson production the leading order 
corrections can reach a level of $0.1$~fb, depending on the SUSY
parameters. This renders these loop-induced processes in principle
observable at an $e^+e^-$ collider. The variation with $\phiAt$ is
found to be extremely small.


\section{Conclusions}
\label{sec:conclusions}

We evaluated all neutral MSSM Higgs boson production modes at $e^+e^-$
colliders with a two-particle final state, i.e.\ 
$e^+e^- \to h_i h_j, h_i Z, h_i \ga$ ($i,j = 1,2,3$), 
allowing for complex parameters. 
In the case of a discovery of additional Higgs bosons a subsequent
precision measurement of their properties will be crucial determine
their nature and the underlying (SUSY) parameters. 
In order to yield a sufficient accuracy, one-loop corrections to the 
various Higgs boson production modes have to be considered. 
This is particularly the case for the high anticipated accuracy of the
Higgs boson property determination at $e^+e^-$
colliders~\cite{LCreport}. 

The evaluation of the processes (\ref{eq:eehh}) -- (\ref{eq:eehga})
is based on a full one-loop calculation, also including hard and soft 
QED radiation.  The renormalization is chosen to be identical as for 
the various Higgs boson decay calculations; see, e.g.,
\citeres{HiggsDecaySferm,HiggsDecayIno}. 

We first very briefly reviewed the relevant sectors including some
details on the one-loop renormalization procedure of the cMSSM, which are 
relevant for our calculation. In most cases we follow \citere{MSSMCT}. 
We have discussed the calculation of the one-loop diagrams, the
treatment of UV, IR and collinear divergences that are canceled by the 
inclusion of (hard, soft and collinear) QED radiation. 
We have checked our result against the literature as far as possible, 
and in most cases we found accaptable or qualitative agreement,
where parts of the differences can be attributed to problems with
input parameters (conversions) and/or special szenarios.  
Once our set-up was changed successfully to the one used in the existing 
analyses we found good agreement.

For the analysis we have chosen a parameter set that allows
simultaneously a maximum number of production processes.
In this scenario (see \refta{tab:para}) we have $h_1 \sim h$, 
$h_2 \sim A$ and $h_3 \sim H$. 
In the analysis we investigated the variation of the various production
cross sections with the center-of-mass energy $\sqrt{s}$, the charged
Higgs boson mass $\MHp$, the ratio of the vacuum expectation values 
$\TB$ and the phase of the trilinear Higgs-top squark coupling, $\phiAt$. 
For light (heavy) Higgs production cross sections we have chosen 
$\sqrt{s} = 500\, (1000)\gev$. 

In our numerical scenarios we compared the tree-level production cross
sections with the full one-loop corrected cross sections. 
In certain cases the tree-level cross sections are identical zero (due to
the symmetries of the model), and in those cases we have evaluated the 
one-loop squared amplitude, $\sigma_{\text{loop}} \propto |\cMl|^2$. 

We found sizable corrections of $\sim 10 - 20\%$ in the $h_i h_j$
production cross sections. Substantially larger corrections are 
found in cases where the tree-level result is (accidentally) small and 
thus the production mode likely is not observable. 
The purely loop-induced processes of $e^+e^- \to h_ih_i$ could be
observable, in particular in the case of $h_1 h_1$ production. 
For the $h_i Z$ modes corrections around $10-20\%$, but going up to 
$\sim 50\%$, are found. The purely loop-induced processes of $h_i\ga$
production appear observable for $h_1\ga$, but very challenging for
$h_{2,3}\ga$. 

Only in very few cases a relevant dependence on $\phiAt$ was
found. Examples are $e^+e^- \to h_1 h_2$ and $e^+e^- \to h_3 Z$, where a
variation, after the inclusion of the loop corrections, of up to $10\%$
with $\phiAt$ was found. In those cases
neglecting the phase dependence could lead to a wrong impression of 
the relative size of the various cross sections.

The numerical results we have shown are, of course, dependent on the choice 
of the SUSY parameters. Nevertheless, they give an idea of the relevance
of the full one-loop corrections. 
Following our analysis it is evident that the full one-loop corrections
are mandatory for a precise prediction of the various cMSSM Higgs boson
production processes.
The full one-loop corrections must be taken into account in any precise 
determination of (SUSY) parameters from the production of cMSSM Higgs
bosons at $e^+e^-$ linear colliders. 
There are plans to implement the evaluation of the Higgs boson 
production into the public code \FH.


\subsection*{Acknowledgements}

We thank G.~Gounaris, T.~Hahn, F.~Renard, J.~Rosiek and K.~Williams
for helpful discussions. 
The work of S.H.\ is supported in part by CICYT (grant FPA 2013-40715-P) 
and by the Spanish MICINN's Consolider-Ingenio 2010 Program under grant 
MultiDark CSD2009-00064.


\newcommand\jnl[1]{\textit{\frenchspacing #1}}
\newcommand\vol[1]{\textbf{#1}}


\begin{thebibliography}{99}
\bibitem{feynarts-mf}
The couplings can be found in \texttt{HMix.ps.gz} and 
\texttt{MSSM.ps.gz} as part of the \FA\ package~\cite{feynarts}.
\end{thebibliography}
\end{document}